\documentclass[11pt]{article}
\usepackage{epsfig}
\usepackage{amsfonts}
\usepackage{slashed}
\usepackage{amsmath,amssymb}
\usepackage{bbm,bm}
\usepackage{cite}
\allowdisplaybreaks[1]

\usepackage{tikz}
\usetikzlibrary{shapes,arrows,shadows,automata,positioning}
\newdimen\nodeDist
\usepackage{comment}
\usepackage{cellspace} 
\usepackage[normalem]{ulem} 

\usepackage[
  colorlinks=true,
  linkcolor={red!50!black},
  citecolor={blue!50!black},
  filecolor=black,
  urlcolor=black,
  breaklinks=true
  ]{hyperref}
\usepackage{orcidlink}

\newcommand{\nn}{\nonumber \\}
\renewcommand{\vec}[1]{{\bf #1}}
\newcommand{\rmi}[1]{{\mbox{\scriptsize #1}}}
\newcommand{\rmii}[1]{{\mbox{\tiny\rm{#1}}}}

\newcommand{\bmu}{\bar{\mu}}

\newcommand{\LamD}{\bmu}
\newcommand{\LamDref}{\LamD_\text{ref}}
\newcommand{\LamDInput}{\LamD_0}

\newcommand{\gammaE}{{\gamma_\rmii{E}}}
\newcommand{\Veff}{V_{\text{eff}}}
\newcommand{\Seff}{{S_{\text{eff}}}}

\newcommand{\Afluc}{A_\text{fluc}}
\newcommand{\Astat}{A_\text{stat}}
\newcommand{\Adyn}{A_\text{dyn}}

\newcommand\newDimSix[1]{{\color{red!50!black}#1}}
\newcommand{\alphaFR}{\alpha}
\newcommand{\threedEFT}{3d~EFT}
\newcommand{\alphaF}{\widehat\alpha}
\newcommand{\fluctO}{\mathcal{O}} 

\newcommand{\Tc}{T_{\rm c}}

\newcommand{\Tn}{T_{\rm n}}
\newcommand{\Tp}{T_{\rm p}}
\newcommand{\vw}{v_w}

\newcommand{\mD}{m_\rmii{D}}
\newcommand{\mF}{m_{\fv}}

\newcommand{\mX}{m_\rmii{$X$}}

\newcommand{\mXzero}{m_{\rmii{$X_0$}}}

\newcommand{\Mpl}{M_\rmii{Pl}}

\newcommand{\WX}{W_\rmii{$X$}}

\newcommand{\REFT}{R_\rmii{EFT}}
\newcommand{\GammaT}{\Gamma_\rmii{$T$}}

\newcommand{\tr}{{\rm Tr\,}}
\renewcommand{\det}{\mbox{det}}
\newcommand{\detS}{\det_\rmii{$S$}}
\newcommand{\detV}{\det_\rmii{$V$}}
\newcommand{\detXzero}{\det_\rmii{$X_0$}}
\newcommand{\detXT}{\det_\rmii{$X_T$}}
\newcommand{\detXG}{\det_\rmii{$XG$}}
\newcommand{\detg}{\det_\rmi{gh}}

\newcommand{\Vol}{\mathcal{V}}
\newcommand{\Tstar}{T_\star}
\newcommand{\Hstar}{H_\star}
\newcommand{\Hv}{H_v}

\newcommand{\geff}{g_*}

\newcommand{\PhiField}{\Phi}
\newcommand{\phiField}{\phi}
\newcommand{\phibg}{\varphi}
\newcommand{\bnc}{b} 
\newcommand{\fv}{\mathchoice{\mbox{\rm F}}{\mbox{\rm F}}{\rmii{F}}{\rmii{F}}} 
\newcommand{\phib}{\phibg_{\bnc}}
\newcommand{\phiF}{\phibg_{\fv}}

\newcommand{\Lb}{L_b}

\usepackage{pifont}

\makeatletter \@addtoreset{equation}{section} \makeatother
\renewcommand{\theequation}{\arabic{section}.\arabic{equation}}
\makeatletter
\renewcommand\section{\@startsection{section}{1}{\z@}%
  {-5.5ex \@plus -1ex \@minus -.2ex}
  {2.3ex \@plus.2ex}%
  {\normalfont\large\bfseries}}
\renewcommand\subsection{\@startsection{subsection}{2}{\z@}%
  {-3.25ex\@plus -1ex \@minus -.2ex}%
  {1.5ex \@plus .2ex}%
  {\normalfont\normalsize\bfseries}}
\renewcommand\thesection{\@arabic\c@section}
\renewcommand\thesubsection{\thesection.\@arabic\c@subsection}
\renewcommand{\@seccntformat}[1]{%
  \csname the#1\endcsname.\hspace{1.0em}}
\makeatother

\begin{document}

\flushbottom

\begin{titlepage}

\begin{flushright}
MITP-26-053
\end{flushright}
\begin{centering}

\vfill

{\Large{\bf
  Bubble nucleation
  with thermal higher-dimensional operators
}}

\vspace{0.8cm}

\renewcommand{\thefootnote}{\fnsymbol{footnote}}
\hypersetup{hidelinks}
Fabio Bernardo%
\orcidlink{0009-0008-0719-3219}%
,$^{\rm a,}$%
\footnote{\href{mailto:fabio.bernardo@unige.ch}{fabio.bernardo@unige.ch}}
Maciej Kierkla%
\orcidlink{0000-0002-2785-5370}%
,$^{\rm b,}$%
\footnote{\href{mailto:maciej.kierkla@physics.uu.se}{maciej.kierkla@physics.uu.se}}
Nicholas Leister%
\orcidlink{0009-0007-6113-313X}%
,$^{\rm c,}$%
\footnote{\href{mailto:nleister@uni-mainz.de}{nleister@uni-mainz.de}}
\\
Philipp Schicho%
\orcidlink{0000-0001-5869-7611}%
,$^{\rm a,}$%
\footnote{\href{mailto:philipp.schicho@unige.ch}{philipp.schicho@unige.ch}}
and
Pedro Schwaller%
\orcidlink{0000-0002-6307-3069}%
$^{\rm c,}$%
\footnote{\href{mailto:pedro.schwaller@uni-mainz.de}{pedro.schwaller@uni-mainz.de}}

\vspace{0.8cm}

$^\rmi{a}$%
{\em
  D\'epartement de Physique Th\'eorique, Universit\'e de Gen\`eve,\\
  24 quai Ernest Ansermet, CH-1211 Gen\`eve 4, Switzerland\\}
\vspace{0.3cm}

$^\rmi{b}$%
{\em
Department of Physics and Astronomy,
Uppsala University,\\
Box 516, SE-751 20 Uppsala, Sweden\\}
\vspace{0.3cm}

$^\rmi{c}$%
{\em
PRISMA$^{++}$ Cluster of Excellence \& Mainz Institute for Theoretical Physics,\\
Johannes Gutenberg University, 55099 Mainz, Germany\\}
\vspace{0.3cm}

\vspace*{0.8cm}

\mbox{\bf Abstract}

\end{centering}

\vspace*{0.3cm}

\noindent
We quantify the impact of higher-dimensional operators in
the effective theory for bubble nucleation,
focusing on
supercooled phase transitions in
the classically conformal Abelian-Higgs model.
While high-temperature dimensional reduction organizes
hard thermal corrections into a tower of operators in a three-dimensional
effective field theory,
recent results show that
dimension-six terms can
dominate over higher-loop corrections to
the equilibrium thermodynamics of the strongest transitions.
To assess their effect on
the nucleation rate and
the derived near-equilibrium phase-transition observables,
we develop a perturbative framework
that includes them in the full one-loop nucleation rate.
Including higher-dimensional operators both in the bounce action and in the fluctuation determinants, we solve the resulting Sturm--Liouville problem without a derivative expansion, using a modified Gel'fand--Yaglom method.
For strongly supercooled transitions,
we find that higher-dimensional operators
only marginally affect the nucleation rate and the resulting observables. As a byproduct, we delineate the regime of validity of the high-temperature nucleation effective theory
in supercooled phase transitions.

\vfill
\end{titlepage}

{\hypersetup{hidelinks}
\tableofcontents
}

\clearpage

\renewcommand{\thefootnote}{\arabic{footnote}}
\setcounter{footnote}{0}

%
\section{Introduction}
\label{sec:intro}

First-order phase transitions in hidden sectors are among
the most compelling early-Universe sources of a stochastic
gravitational-wave~(GW) background, including the nHz signal probed by
pulsar timing arrays~\cite{NANOGrav:2023gor,EPTA:2023fyk,Reardon:2023gzh,Xu:2023wog}.
Transitions that source such a nHz signal could also indirectly affect
big-bang nucleosynthesis and
thereby potentially alleviate tensions in
the observed light-element abundances~\cite{Elahi:2026ldn, Girmohanta:2026ogq, Tanaka:2026ssu}.
This can be seen as further motivation to study them.

Since the GW spectrum depends strongly on
the transition parameters,
such as the percolation temperature and $\beta/H$,
which follow from the bubble-nucleation rate,
a quantitatively controlled prediction of the transition
requires theoretical control not only of the bounce action, but also
of the corrections that enter the nucleation rate beyond leading
order~\cite{Croon:2020cgk,Gould:2021oba,Ekstedt:2021kyx,Ekstedt:2022ceo,Ekstedt:2022tqk,Ekstedt:2023sqc,Kierkla:2025qyz}.

Thermal transitions are most systematically described using
high-temperature dimensional reduction, which integrates out the
hard Matsubara modes in favor of a three-dimensional (3d) effective
field theory (EFT)
for the infrared bosonic
modes~\cite{Farakos:1994kx,Kajantie:1995dw,Braaten:1995cm}\footnote{
For alternative, recent approaches see e.g.~\cite{Navarrete:2025yxy, Klose:2026pux, Berthiaume:2026dkt, Carosi:2024lop}. 
}.
The thermal corrections are thereby organized into a tower of local
operators~\cite{Chakrabortty:2026swu}.
Most analyses truncate this tower at the
super-renormalizable level, which is adequate for weak transitions
driven by thermal loops~\cite{Bernardo:2026whs}.
For moderately strong transitions with
small supercooling, however,
dimension-six operators,
first appearing at $\mathcal{O}(g^6)$
in the gauge coupling,
can significantly modify
the transition strength
and the resulting GW predictions~\cite{Chala:2024xll,Bernardo:2025vkz,Bernardo:2026whs}.
In particular, a three-loop dimensional reduction of the Abelian-Higgs
model shows that one-loop dimension-six effects typically \emph{dominate}
over two- and three-loop corrections to super-renormalizable parameters
for the strongest of these transitions~\cite{Bernardo:2025vkz,Bernardo:2026whs}.

In strongly supercooled transitions, the escape point of the bounce configuration stays at field values much smaller than the vacuum expectation value.  
Hence, the high-temperature expansion was claimed
to converge along the bounce~\cite{Levi:2022bzt,Kierkla:2023von},
suggesting that higher-\hspace{0pt}dimensional operators are
suppressed~\cite{Bernardo:2026whs,Puchades-Ibanez:2026ksh}.
While previous treatments
implemented these operators as perturbative corrections to the nucleation rate~\cite{Chala:2024xll,Puchades-Ibanez:2026ksh},
their direct impact on
the full one-loop contributions to the nucleation rate is an open question.
We present the first systematic study of this impact,
including dimension-six operators in both
the bounce and the one-loop fluctuation determinants.

Quantifying this impact requires a consistent perturbative treatment
of the full one-loop nucleation rate.
Following Langer's theory, the thermal rate factorizes into a
dynamical and a statistical part~\cite{Langer:1967ax,Langer:1969bc,Gould:2021ccf},
\begin{equation}
  \label{eq:rate}
    \GammaT =
      \Adyn\times\Astat =
      \Adyn\times\Afluc\, e^{-\Seff}
    \,,
\end{equation}
with a dynamical prefactor $\Adyn \sim T$ and
a statistical fluctuation part $\Afluc \sim T^3$.
The dynamical part accounts for the non-equilibrium
evolution~\cite{Langer:1969bc,Hanggi:1990zz,Berera:2019uyp},
for which we adopt the no-damping
approximation~\cite{Affleck:1980ac},
with $\Adyn \simeq \sqrt{|\lambda_-|}/(2\pi)$, where
$\lambda_-$
is the negative eigenvalue of the scalar fluctuation operator,
as implemented in {\tt BubbleDet}~\cite{Ekstedt:2023sqc} and
in our in-house software for the nucleation
rate~\cite{Kierkla:2025qyz,Puchades-Ibanez:2026ksh}.%
\footnote{%
  This treatment neglects out-of-equilibrium and
  damping effects~\cite{Hirvonen:2024rfg,Hirvonen:2025hqn,Gould:2026mtl} that arise once the thermal bath is itself driven
  out of equilibrium by the nucleation process and back-reacts on
  the bubble dynamics.
  Such effects have recently been addressed from first principles
  using kinetic and Wigner-function descriptions of the
  plasma~\cite{Hirvonen:2026zaq}.
}
For the transitions considered
in this work, the statistical part dominates eq.~\eqref{eq:rate} and reduces to a zero-temperature
vacuum-decay problem in one lower dimension. It is controlled by the effective action $\Seff$ evaluated on the classical solution of the Euclidean equations of motion (the bounce)
and by the fluctuation prefactor $\Afluc$~\cite{%
  Langer:1969bc,Callan:1977pt,Ekstedt:2021kyx,Ekstedt:2022tqk, Baratella:2025dum, Baratella:2024hju}.
At one loop, the latter is given by the functional determinant of all the remaining light
fields at the nucleation scale ($\Psi_{\rm l}$),
\begin{equation}
  \label{eq:Afluc}
    \Afluc^{} =
      \prod_{\Psi_{\rm l}}\left|
          \frac{
            \det\!\left(-\nabla_d^2+\mathcal{M}_i^{}(\phib^{})\right)}{
            \det\!\left(-\nabla_d^2+\mathcal{M}_i^{}(\phiF^{})\right)}
      \right|^{-n_i^{}/2}
    \,,
\end{equation}
where
$\phib$ is the bounce,
$\phiF$ the false vacuum,
$\nabla_d^2$ the spatial Laplacian,
$n_i=1~(-2)$ for bosons (ghosts) and
$\mathcal{M}_i$ the background-dependent effective mass squared.
Higher-dimensional operators enter the rate by deforming
the bounce and its action at tree level, and
by modifying the fluctuation operators in
the determinants,
corresponding to a modification of the quadratic part of the action.

The determinants are often approximated
by a derivative expansion of the effective action.
Such an expansion is known to converge poorly or to break down
along the tail of the bounce,
especially for scale-shifting gauge modes~\cite{
  Gould:2021ccf,Kierkla:2023von,Kierkla:2025qyz,
  Kierkla:2025vwp,Kierkla:2026bnm,Puchades-Ibanez:2026ksh
}.
We avoid this expansion altogether and instead compute the
determinants exactly using the Gel'fand--Yaglom (Forman) theorem, which
reduces each partial-wave determinant to an initial-value
problem~\cite{Gelfand:1959nq,Forman:1987gha,Kirsten:2004qv,Kirsten:2010eg, Carosi:2026gpi}.
This is the same approach underlying~\cite{Baratella:2025dum} and
{\tt BubbleDet}~\cite{Ekstedt:2023sqc} and
its extension to the mixed Goldstone--gauge sector and
scale-shifting modes~\cite{%
  Gould:2021ccf,Kierkla:2025qyz,Kierkla:2026bnm,
  Puchades-Ibanez:2026ksh}.
Thus, the full one-loop correction is included,
without the pathologies of the
derivative expansion.

In this work,
we combine
dimension-six operators with exact one-loop determinants.
Using the classically conformal Abelian-Higgs model as
a simplified setup, we include the higher-dimensional operators of the
thermal EFT derived in~\cite{Bernardo:2025vkz,Bernardo:2026nyq}
consistently in the full one-loop nucleation rate.
We achieve this by computing, for the first time,
the thermal fluctuation determinants with higher-dimensional operators
using a modified Gel'fand--Yaglom method. 
This provides a perturbative framework for the nucleation rate
that allows us to separate 
the effect of higher-dimensional operators
from other higher-order corrections.
The framework applies to higher-dimensional operators of general EFTs.
In our case, these operators are determined via hard thermal matching.
Our analysis illustrates the validity of using
the high-temperature EFT for
bubble nucleation,
even in the case of supercooled transitions, and
we additionally assess the importance of higher-dimensional operators for
the percolation temperature and
the resulting phase-transition parameters, such as the average bubble radius. 

The remainder of this paper is organized as follows.
Section~\ref{sec:model} introduces the model and
discusses the renormalization-group (RG) improvement and the connection
between physical and Lagrangian parameters.
Section~\ref{sec:nucleation} develops the nucleation framework.
Section~\ref{sec:EFTvalidity}
derives a validity range for the EFT in supercooled phase transitions.
Section~\ref{sec:results} presents the numerical results.
Finally,
the appendices detail
the corresponding higher-dimensional operators and
their inclusion in the fluctuation determinants.

%
\section{Model}
\label{sec:model}

We consider a dark Abelian-Higgs model with one complex scalar
$\PhiField$ charged under
a dark ${\rm U}(1)_\rmii{D}$ gauge field
$A_\mu$ with gauge coupling $g$. 
In four-dimensional Euclidean space, the Lagrangian is
\begin{align}
  \label{eq:Lag4d}
    \mathcal{L}_\rmi{4d} &=
        \frac14 F_{\mu\nu}F_{\mu\nu}
      + (D_\mu\PhiField)^\dagger (D_\mu\PhiField)
      + m^2\,\PhiField^\dagger\PhiField
      + \lambda\,(\PhiField^\dagger\PhiField)^2
    \,,
\end{align}
with
covariant derivative
$D_\mu=\partial_\mu+i g A_\mu$ and
field strength
$F_{\mu\nu}=\partial_\mu A_\nu-\partial_\nu A_\mu$.
Around a real background $\phibg$,
the complex scalar is decomposed as
\begin{align}
  \label{eq:field:decomp}
    \PhiField &=
      \frac{1}{\sqrt{2}}\bigl(\phibg+\phiField+i\chi\bigr)
    \,,
\end{align}
with
the radial fluctuation $\phiField$ and
the Goldstone mode $\chi$.
Here,
$\phib$ denotes the bounce and
$\phiF$ the false vacuum.
The model is the simplest gauge-Higgs theory that captures the
essential features of more complex realizations, and it serves as
a controlled setup for
the higher-order thermal EFT
computations relevant to nucleation~\cite{Hirvonen:2021zej}.
See~\cite{Bernardo:2025vkz,Bernardo:2026whs}
for the same discussion in equilibrium.

We work in the classically conformal limit, imposing $m^2=0$ at tree
level so that the classical action contains no intrinsic mass scale.
The true vacuum is then generated radiatively through the
Coleman--Weinberg mechanism~\cite{Coleman:1973jx}.
There, the quartic coupling
$\lambda$ runs and turns negative at intermediate scales, producing a
non-trivial minimum.
At finite temperature,
the thermal mass stabilizes the symmetric phase,
while the running quartic coupling is negative at low scales.
Together with the thermal cubic term, they generate a barrier
far below the true vacuum that persists down to low temperatures,
leading to a strongly supercooled first-order
transition~\cite{Kierkla:2022odc, Kierkla:2023von, Kierkla:2026bnm, Ellis:2018mja, Ellis:2020nnr}.
Since the radiatively generated quartic is of order $g^4/(4\pi)^2$,
tunneling becomes efficient only far below the scale of the true vacuum,
and the relevant field values span many orders of magnitude,
from $g\phibg\lesssim\pi T$ to the zero-temperature
minimum~\cite{Christiansen:2025xhv}.
The resulting large logarithms
$\ln(g\phibg/\LamD)$ and $\ln(\pi T/\LamD)$
are resummed through RG improvement,
discussed
in sec.~\ref{sec:RG:improvement}.

%
\subsection{Thermal effective action}
\label{sec:thermal:action}

At high temperatures,
the theory is reduced to a \threedEFT{} for
the soft bosonic modes by integrating out the hard, non-zero Matsubara modes.
Beyond the super-renormalizable operators,
this matching generates a tower of higher-dimensional operators.
The leading ones are of dimension six and first appear at $\mathcal{O}(g^6)$.
The corresponding operator basis and Wilson coefficients are given in
appendix~\ref{sec:dim6:matching} and~\cite{Bernardo:2025vkz}.
These operators also modify
the bounce (see~\cite{Chala:2024xll,Puchades-Ibanez:2026ksh}) and
the fluctuation determinants,
which is the subject of the following sections.

The resulting 3d Lagrangian is
\begin{align}
  \label{eq:Lag3d:dim4}
    \mathcal{L}_0^{} &=
        (D_i^{}\PhiField)^\dagger(D_i^{}\PhiField)
      + \frac{1}{4}F_{ij}^{}F_{ij}^{}
      + \frac{1}{2}(\partial_i^{}B_0^{})^2
      + V(\PhiField)
      + \frac{1}{2}\mD^2B_0^2
      + h_3^{}\,B_0^2\PhiField^\dagger\PhiField
    \nn &
      + \frac{1}{2\xi}(
          \partial_i^{} B_i^{}
        + g\phibg\chi
        + f(\phibg)g\phibg\chi
      )^2
    \,,
\end{align}
where
$\PhiField$ is the scalar field in three dimensions,
$B_i$ and $B_0$ are the spatial and temporal components of
the gauge field,
$\mD$ is the Debye mass, and
$h_3$ the coupling between the temporal gauge mode and the scalar field.
Henceforth,
we set $\xi=1$,
which reduces to Feynman gauge in the absence of higher-dimensional operators.
In the last line of eq.~\eqref{eq:Lag3d:dim4},
we introduced
a background-field-dependent gauge fixing similar to~\cite{Martin:2018emo}.
The gauge freedom
allows for choosing the function $f(\phibg)$
to simplify the fluctuation operators
$-\nabla_d^2+\mathcal{M}_i^{}(\phibg)$
that enter the determinants in eq.~\eqref{eq:Afluc}.

The contributions to the effective action from the higher-dimensional operators
up to dimension six are given by~\cite{Bernardo:2025vkz}
\begin{align}
  \label{eq:Lag3d:dim6}
    \mathcal{L}_1^{} &=
        \alpha_{\phi^2F^2}^{}F_{ij}^{}F_{ij}^{}\PhiField^\dagger\PhiField
      + \alpha_{D^2\phi^4}^{}(D_i^{}\PhiField)^\dagger (D_i^{}\PhiField)(\PhiField^\dagger\PhiField)
    \nn &
      + \alpha_{D^2\phi^2B_0^2}^{}(D_i^{}\PhiField)^\dagger (D_i^{}\PhiField)B_0^2
      + \alpha_{\phi^4B_0^2}^{}(\PhiField^\dagger\PhiField)^2 B_0^2
      + \alpha_{\phi^6}^{}(\PhiField^\dagger\PhiField)^3
    \,,
\end{align}
where the Wilson coefficients $\alpha_i$ are functions of the running 4d couplings
and the temperature, and are given in appendix~\ref{sec:dim6:matching}.
Similar to
the dimension-four
$B_0^4$~\cite{Ekstedt:2021kyx},
operators of the form
$B_0^6$,
$B_0^4(\PhiField^\dagger\PhiField)$,
and $F_{ij}^{}F_{ij}^{}B_0^2$ are neglected,
since they are more than quadratic in the fluctuations once expanded around the
background solutions.
They would, however, contribute
to the effective potential at higher orders
than our matching target accuracy of
$\mathcal{O}(g^6)$%
\footnote{%
    At this order,
    also other contributions
    arise, such as the three-loop effective potential~\cite{Ekstedt:2024etx} or
    two-loop corrections to the rate beyond the one-loop fluctuation determinants,
    so far computed for vacuum decay in scalar theories~\cite{Bezuglov:2018qpq,Bezuglov:2019uxg}.
    Such contributions are not further investigated
    in this work.
}
or to scenarios
where
the temporal scalar $B_0$ obtains a
background~\cite{Bandyopadhyay:2026nrv,Biermann:2026qju,Li:2025kyo}.

To simplify the fluctuation operator once higher-dimensional operators are included,
we use the following gauge fixing throughout
\begin{align}
  \label{eq:gauge-fixing}
    f(\phibg) &=
      f_0(\phibg) \equiv
      \frac{\alpha_{D^2\phi^4}}{2} \phibg^2
    \,,
\end{align}
which
cancels the dimension-six mixing between
the Goldstone and
the longitudinal gauge mode in the fluctuation operator. 
For further details,
we refer to appendix~\ref{sec:bounce:higherdim}.

%
\subsection{Renormalization-group improvement and physical parameters}
\label{sec:RG:improvement}

Without a tree-level mass,
the potential is generated radiatively over field values
from $g\phibg\lesssim\pi T$ to
the zero-temperature minimum,
and no single renormalization scale minimizes all logarithms.
Four-dimensional computations therefore can use a field-dependent scale,
e.g.\ $\LamD=\max(g\phibg,\pi T)$~\cite{Kierkla:2022odc,Kierkla:2023von,Kierkla:2026bnm}.
In the high-temperature expansion, instead, the field-dependent logarithms of the hard
modes cancel~\cite{Kierkla:2023von},
and the soft ones are resummed within the \threedEFT{}~\cite{Gould:2021ccf}.
It thus suffices to fix the couplings at an input scale $\LamDInput$
and to run them to a thermal matching scale $\LamD\sim\pi T$.
Following~\cite{Christiansen:2025xhv}, we impose Gildener--Weinberg-like conditions~\cite{Gildener:1976ih}
\begin{align}
  \label{eq:RG:prescription}
    \lambda(\LamDInput) &= 0
    \,,&
    g(\LamDInput) &= g_0
    \,.
\end{align}
In the absence of a dimensionful parameter,
the input scale $\LamDInput$ merely sets the unit of measurement.
All dimensionful quantities are therefore understood 
in units of $\LamDInput$ and
the theory is specified by
the single dimensionless parameter,
the gauge coupling $g_0$ at the input scale.
In this context,
the quartic coupling $\lambda$ is not an independent input but
is generated radiatively at the running scale
$\LamD \neq \LamDInput$ via RG evolution.
Since $\beta_\lambda=6g^4/(4\pi)^2>0$ at $\lambda=0$,
$\lambda$
is negative for
$\LamD<\LamDInput$~\cite{Christiansen:2025xhv}.

For dimensional reduction,
the four-dimensional couplings are run to
the reference scale $\LamDref$,
which is the typical scale of the hard modes in the plasma, or simply the mass of the lowest non-zero Matsubara mode.
These running couplings directly enter
the effective parameters of
the \threedEFT.
Thus, the temperature dependence of the matched theory follows from RG
evolution and thermal matching, while $g_0$ remains the only physical
free parameter.
The higher-dimensional operators are likewise generated by this
matching.
Their Wilson coefficients are determined by
the running couplings (cf.\ appendix~\ref{sec:dim6:matching}) and
the temperature.
These coefficients
introduce no new free parameters.

We now connect the Lagrangian parameters to physical quantities.
At $\LamD=\LamDInput$,
minimizing the one-loop potential
gives~\cite{Christiansen:2025xhv}
\begin{equation}
  \label{eq:vev}
    v_\rmii{CW} =
      e^{\frac{1}{6}}\frac{\LamDInput}{g_0}
    \,,
\end{equation}
which identifies $\LamDInput$ as
the scale of dimensional transmutation.
The residual dependence on the running scale $\LamD$
cancels order by order once the running of the
couplings and of the field is consistently included.
The vacuum expectation value itself is gauge dependent and not an observable.
Physical, $\LamD$-independent quantities are instead
the (on-shell) 4d gauge-boson mass,
at leading order
$\mX^{\rmi{4d}}=g_0v_\rmii{CW}=e^{\frac16}\LamDInput$,
and vacuum-energy differences~\cite{Christiansen:2025xhv,Puchades-Ibanez:2026ksh}.
Both depend only on $\LamDInput$, as expected from dimensional transmutation.
The corresponding vacuum-energy difference is
\begin{equation}
  \label{eq:DeltaV0}
    \Delta V_0^{} =
      \frac{3}{8}\frac{g_0^4 v_\rmii{CW}^4}{(4\pi)^2} =
      \frac{3}{8}\frac{e^{\frac23}}{(4\pi)^2}\,\LamDInput^4
    \,.
\end{equation}
Inserting
the zero-temperature vacuum energy,
the Hubble rate reads
\begin{align}
    H^2(T) &=
      \frac{1}{3\Mpl^2}\left[\Delta V_0^{}+\rho_\rmii{R}^{}(T)\right]
    \,,&
    \rho_\rmii{R}^{}(T)=
      \frac{\pi^2}{30}\geff^{} (T)T^4
    \,,
\end{align}
with
$\Mpl\simeq2.44 \times 10^{18} \rm{~GeV}$ being the reduced Planck mass and
$\geff$ the effective number of relativistic degrees of freedom.
For strong supercooling,
$\rho_\rmii{R}(T)\ll\Delta V_0$ at the relevant
temperatures, and
the vacuum-dominated limit yields
the constant Hubble rate
\begin{align}
  \label{eq:Hv}
    \Hv^{} &=
      \sqrt{\frac{\Delta V_0^{}}{3\Mpl^2}} =
      \frac{e^{\frac13}}{8\sqrt{2}\,\pi}\frac{\LamDInput^2}{\Mpl^{}}
    \,.
\end{align}
The transition strength $\alpha$ is the ratio of
the released vacuum energy to the radiation energy density,
and
at a reference temperature $\Tstar$ of the transition,
identified with
the leading-order (LO) nucleation temperature $\Tn$
in sec.~\ref{sec:EFTvalidity} and with
the percolation temperature $\Tp$ in sec.~\ref{sec:results},
it reads~\cite{%
  Christiansen:2025xhv,Puchades-Ibanez:2026ksh}
\begin{align}
  \label{eq:alpha}
    \alpha(\Tstar^{}) &=
      \frac{\Delta V_0^{}}{\rho_\rmii{R}^{}(\Tstar^{})} =
      \frac{180\, e^{\frac23}}{(4\pi)^4 \geff^{}(\Tstar^{})}\left(\frac{\LamDInput^{}}{\Tstar^{}}\right)^4 =
      \frac{180}{(4\pi)^4 \geff^{}(\Tstar^{})}g_0^4\left(\frac{v_\rmii{CW}^{}}{\Tstar^{}}\right)^4
    \,.
\end{align}
Here, the zero-temperature $\Delta V_0$ is used since
the broken minimum lies far outside the high-temperature regime,
$\Tstar\ll v_\rmii{CW}$~\cite{Kierkla:2023von,Puchades-Ibanez:2026ksh}.

%
\section{Nucleation framework}
\label{sec:nucleation}

In this section,
we review the semiclassical nucleation
rate~\cite{Langer:1969bc,Coleman:1977py,Callan:1977pt,Affleck:1980ac,Linde:1981zj}
within
the finite-temperature EFT framework of~\cite{Gould:2021ccf,Ekstedt:2021kyx,Hirvonen:2021zej},
focusing on the fluctuation operators that enter it.
Their modifications by higher-dimensional operators are collected in
appendix~\ref{sec:bounce:higherdim}.

%
\subsection{From the free energy to the fluctuation determinant}
\label{sec:freeenergy:to:det}

At finite temperature, the natural starting point is the Euclidean path integral
for the thermal partition function $\mathcal Z(T)$.
Using it, one can obtain
the free energy $F_{\rm therm}$ of the system.
With $\beta=1/T$, we have
\begin{align}
  \label{eq:Zthermal}
    \mathcal Z(T) &=
      \tr\,e^{-\beta\hat H} =
      \int\mathcal D\Psi\,e^{-S_\rmii{E}[\Psi]}
    \,,&
    F_{\rm therm} &=
      - T\ln \mathcal Z(T)
    \,,
\end{align}
where
$\Psi$ collects the full microscopic field content,
$S_\rmii{E}=\int_0^\beta {\rm d}\tau\int_{\vec{x}}\mathcal L_\rmii{E}$
is the Euclidean action on
the thermal circle,
with bosonic (fermionic) fields periodic (anti-periodic) in
$\tau$, and
$\int_{\vec{x}}\equiv\int{\rm d}^dx$ and
$\nabla_d$ denote the integral and gradient
over the $d$ spatial dimensions.

When the relevant degrees of freedom span a wide hierarchy of scales, EFTs resum the associated large corrections and extend the validity of perturbation theory.\footnote{An EFT is therefore not merely a convenient reformulation, but the framework that makes predictions possible when the short-distance contributions affect the low-energy (nucleation) behavior.}
The free energy is then obtained by carrying out the path integral successively,
integrating out the heavier modes down to the scale of interest.
Concretely, we split $\Psi=\Psi_{\rm l}+\Psi_{\rm h}$ into
the light, coarse-grained fields $\Psi_{\rm l}$
that drive the transition and the heavier modes
$\Psi_{\rm h}$.%
\footnote{%
  The collection $\Psi_{\rm l}$ therefore contains
  not only the light scalar field that undergoes
  the transition but, depending on the theory,
  possibly other fields as well
  (e.g.\ the spatial components of gauge fields).
  In general, it comprises all fields with masses at or
  below the nucleation scale $\Lambda_{\rm nucl}$
  somewhere along the bounce.
}
Integrating out the latter defines an effective
free energy
$F_{\rm eff}[\Psi_{\rm l}]$,
\begin{equation}
  \label{eq:Feff:def}
    e^{-\beta F_{\rm eff}[\Psi_{\rm l}]} \equiv
      \int\mathcal D\Psi_{\rm h}\,e^{-S_\rmii{E}[\Psi_{\rm l},\Psi_{\rm h}]}
    \,.
\end{equation}
The heavy fields are thereby absorbed into
$F_{\rm eff}^{ } =
  F_{\rm eff}^{\rmii{LO}}
+ F_{\rm eff}^{\rmii{NLO}}
+ \dots$
to a given order in perturbation theory,
with NLO denoting the next-to-leading order.
The metastable partition function reduces to
a path integral over the light fields alone,
\begin{equation}
  \label{eq:Zmeta}
    \mathcal Z =
      \int\mathcal D\Psi_{\rm l}\,e^{-\beta F_{\rm eff}[\Psi_{\rm l}]}
    \,.
\end{equation}
Given a hierarchy between the light and heavy scales,
$F_{\rm eff}$ is local for light momenta below
the heavy threshold and is organized as
a derivative expansion,
whose Wilson coefficients carry
the short-distance physics.

The non-zero Matsubara modes have thermal masses
$\sim \pi T$,
well above the soft scale $\sim gT$ of the transition.
Integrating them out leaves
a three-dimensional effective theory for
the soft bosonic fields.
The light fields $\Psi_{\rm l}$ contain
the scalar that undergoes the transition,
which we split according to eq.~\eqref{eq:field:decomp} into
a real background $\phibg$ and
the fluctuations $\phiField$ and $\chi$.
For a static background,
the free energy takes the local form
\begin{equation}
  \label{eq:Feff:local}
    F_{\rm eff}[\phibg] =
      \int_{\vec{x}}\left[
          \frac{1}{2}Z(\phibg)(\nabla_d\phibg)^2
        + \Veff(\phibg,T)
        + \dots
      \right]
    \,,
\end{equation}
with $d=3$ after dimensional reduction,
$S_{3,\rmii{EFT}}\equiv\beta F_{\rm eff}$ the corresponding action of the \threedEFT{},
and the dots denoting higher orders in the derivative expansion. 
The constant field-strength normalization from the hard modes is absorbed into
the 3d fields by the matching, whereas within the \threedEFT{} the kinetic term
acquires a field-dependent normalization $Z(\phibg)$,
e.g.\ from higher-dimensional operators or from integrating out heavier 3d modes on the background~\cite{Gould:2021ccf,Ekstedt:2021kyx,Kierkla:2023von}.
At LO, $Z(\phib) =1$ and
the fields are canonically normalized.

The vacuum structure is thus determined by $\phibg$.
The metastable phase is
a homogeneous minimum $\phiF$ of $F_{\rm eff}$,
whereas nucleation proceeds through
the critical bubble $\phib$,
the non-trivial stationary point (the bounce)
that interpolates toward the stable phase (true vacuum),
\begin{align}
  \label{eq:bounce:generic}
    \frac{\delta F_{\rm eff}}{\delta\phibg}\biggr|_{\phib}&=0
    \,,&
    \Delta F_{\rm eff} &=
        F_{\rm eff}[\phib]
      - F_{\rm eff}[\phiF]
    \,,
\end{align}
where
the free-energy barrier $\Delta F_{\rm eff}$
suppresses the rate by $e^{-\Delta F_{\rm eff}/T}$.
The quadratic fluctuations are schematically%
\footnote{%
  For field-dependent $Z$, expanding
  eq.~\eqref{eq:Feff:local} to second order gives
  $F_{\rm eff}''[\phib]=-\nabla_d\cdot Z(\phib)\nabla_d-Z'(\phib)\nabla_d^2\phib-\frac12Z''(\phib)(\nabla_d\phib)^2+\Veff''(\phib,T)$.
  For details see appendix~\ref{sec:app:fluc} and 
  $\fluctO^{(1)}_{\phiField\phiField}$ in eq.~\eqref{eq:app:ops:SA}.
}
governed by
\begin{equation}
  \label{eq:fluc:operator}
    F_{\rm eff}''[\phib^{}] =
      - Z(\phib)\nabla_d^2
      + \Veff''(\phib^{},T)
    \,,
\end{equation}
which, because the bounce is a saddle point rather than a minimum,
has exactly one negative eigenvalue. 
The associated mode is the unstable growth direction of the bubble~\cite{Gould:2021ccf,Gould:2026mtl}.

The leading semiclassical nucleation rate per unit volume and time
factorizes into
a dynamical and
a statistical prefactor.
The statistical prefactor is
a product of determinant ratios,
\begin{align}
  \label{eq:rate:generic}
    \GammaT\equiv\frac{\Gamma}{\Vol_3} &= \Adyn\,\Astat
    \,,\\
  \label{eq:Astat:general}
    - \ln\Astat &=
        \frac{\Delta F_{\rm eff}}{T}
      - \ln\frac{\mathcal{J}}{\mathcal{V}_3}
      + \frac{1}{2}\sum_{i\in\Psi_{\rm l}}
        \ln\frac{
          \big|\det'\big(\fluctO_i(\phib)/(2\pi)\big)\big|}{
          \det\big(\fluctO_i(\phiF)/(2\pi)\big)}
    \,,
\end{align}
where
$\Delta F_{\rm eff}/T=S_{3,\rmii{EFT}}[\phib]-S_{3,\rmii{EFT}}[\phiF]$.
The determinants are expanded in partial waves $\ell$ (for details see appendix~\ref{sec:app:fluc}) and the sum runs over the light fields $i\in \Psi_{\rm l}$,
specified for the Abelian-Higgs model in sec.~\ref{sec:full:det}
and eqs.~\eqref{eq:app:XG:matrix}--\eqref{eq:app:sector:ops}.
These sectors consist of the light fields
$\Psi_{\rm l}$ of eq.~\eqref{eq:Feff:def},
whose fluctuations are not parametrically heavier than the bounce scale
$\Lambda_{\rm nucl}^{}\sim R_{\bnc}^{-1}$, with $R_{\bnc}$ the radius of the critical bubble.
The heavier modes have already been absorbed into
$\Veff$ of eq.~\eqref{eq:Feff:local} and the couplings.
The prime signals that the zero modes of the fluctuation operators
are removed, since a naive Gaussian integration over them diverges.
The zero modes are instead traded for collective coordinates~\cite{Ekstedt:2023sqc, Andreassen:2017rzq, Andreassen:2018ghv, Bhattacharya:2024chz}.
The three translational ones produce
the spatial volume
$\mathcal V_{\rm tr}^3=\Vol_3^{ }$,
which is divided out in $\GammaT$,
and those of spontaneously broken internal symmetries, the group volume
$\Vol_\rmii{G}={\rm vol}(\mathcal G/\mathcal H)$ of the broken
generators~\cite{Buchmuller:1993bq,Ekstedt:2023sqc},
with $\mathcal G$ the internal symmetry group and
$\mathcal H\subset\mathcal G$
the subgroup leaving the bounce invariant.
Together with the Jacobians of the change of variables, the group and translational volumes are collected in
$\mathcal J=\Vol_\rmii{G}\,\Vol_3 \,\mathcal J_\rmi{tr}\mathcal J_\rmii{G}$,
where
$\mathcal J_\rmi{tr}$ and
$\mathcal J_\rmii{G}$ are
the Jacobians of
the translations and of
the $n_\rmii{G}=\dim(\mathcal G/\mathcal H)$ Goldstone modes,
respectively (see appendix~\ref{sec:zero:modes}).
 
The prefactor
$\Astat$ is the equilibrium probability of finding
the critical bubble relative to the false vacuum, which is what
the static EFT computes.
The dynamical prefactor is the real-time flux of configurations across
the saddle~\cite{Langer:1969bc,Hanggi:1990zz,Berera:2019uyp}.
In the no-damping approximation of
sec.~\ref{sec:intro}~\cite{Affleck:1980ac,Ekstedt:2023sqc},
$\Adyn=\sqrt{|\lambda_-|}/(2\pi)$, with
$\lambda_-$ the negative eigenvalue of
the scalar operator $\fluctO_{\rmii{$S$}}(\phib)$.

%
\subsection{Full nucleation rate}
\label{sec:full:det}
 
In the Abelian-Higgs model,
the fields that appear in the determinants are
the radial scalar $\phiField$,
the Goldstone mode $\chi$,
the spatial gauge field $B_i$, and
the temporal scalar $B_0$.
Collecting their contributions in
the scalar determinant $\detS$ of the radial scalar and
the vector determinant $\detV$ of the gauge fields and the Goldstone mode,
both including their zero-mode Jacobians and
given explicitly in
eqs.~\eqref{eq:detS:full}
and~\eqref{eq:detV:full:explicit},
eq.~\eqref{eq:Astat:general}
takes the form 
\begin{align}
  \label{eq:Astat:full}
    \Astat^{} &=
      \left(\frac{\detS^{}}{\Vol_3^{}}\right)\detV^{}\,\exp\left[
        - S^{\rmii{LO}}_{3,\rmii{EFT}}[\phib^{}]
        + C_{\rm sub}^{}[\phib^{}]
        - \int_{\vec{x}}V^{\rmii{NLO}}_{3,\rmii{EFT}}(\phib^{})
      \right]
    \,,\\[1mm]
  \label{eq:Csub}
    C_{\rm sub}^{}[\phib^{}] &\equiv
      - \frac{1}{12\pi}\int_{\vec{x}}\biggl[
          2\mX^3(\phib^{})
        + \mXzero^3(\phib^{})
        - \mXzero^3(\phiF^{})
      \biggr]
    \,,
\end{align}
where
$S_{3,\rmii{EFT}}^{}=S^{\rmii{LO}}_{3,\rmii{EFT}}+\int_{\vec{x}}V^{\rmii{NLO}}_{3,\rmii{EFT}}$ at the order considered, and
the false-vacuum contribution is understood to be subtracted throughout. 
The LO potential $V^{\rmii{LO}}_{3,\rmii{EFT}}$ contains the one-loop gauge
contribution that generates the barrier.
Since the same contribution is also contained in $\detV$, evaluated on the bounce,
it is subtracted from the action through $C_{\rm sub}$ to avoid double
counting~\cite{Kierkla:2025qyz}.
Here, $\mX^2(\phib) = g_3 ^2 \phib^2 $ and $\mXzero^2(\phib) = \mD^2 + h_3^{ } \phib^2$
are the bounce-dependent masses squared, for the spatial and temporal gauge modes, respectively.

Combining with eq.~\eqref{eq:rate:generic} and
restoring the powers of $T$ that render
the logarithm dimensionless,
\begin{align}
  \label{eq:rate:full:log}
    - \ln\frac{\GammaT^{}}{T^4} =&
      - \ln\frac{\Adyn^{}}{T}
      + S^{\rmii{LO}}_{3,\rmii{EFT}}[\phib^{}]
      - C_{\rm sub}^{}[\phib^{}]
      + \int_{\vec{x}}V^{\rmii{NLO}}_{3,\rmii{EFT}}(\phib^{})
      - \ln\frac{\detS^{}}{\Vol_3^{}T^3}
      - \ln\detV^{}
    \,.
\end{align}
The vector contribution factorizes as
\begin{align}
  \label{eq:detV:full}
    \detV &=
      \detXzero\,\detg\,\detXT\,\detXG
    \,,
\end{align}
with $\detXzero$ the contribution of the temporal scalar $B_0$,
$\detg$ the ghost determinant,
$\detXT$ that of the transverse spatial gauge modes, and
$\detXG$ the mixed Goldstone--gauge determinant.
Since the Goldstone and spatial gauge fluctuations mix in the inhomogeneous bounce background,
$\detXG$ is the determinant of the full coupled operator. 
The collective coordinates of the two zero modes produce the volumes $\Vol_3$ and $\Vol_{{\rm U}(1)}=2\pi$, 
while the Jacobians,
combined with the reduced determinants of
the corresponding partial waves,
reduce to
\begin{subequations}
\label{eq:zero:modes}
\begin{align}
  \label{eq:zero:modes:tr}
    \text{translations}\
    &(n_\rmi{tr}^{}=3,\
      \ell=1\ \text{of}\ \fluctO_{\rmii{$S$}}^{}):
    &\mathcal J_\rmi{tr}^{}
      \biggl[
          \frac{
            \det\fluctO^{1}_{\rmii{$S$}}(\phiF^{})}{
            {\rm det}'\fluctO^{1}_{\rmii{$S$}}(\phib^{})}
      \biggr]^{3/2} &=
      \bigl(-2\phibg_\infty^{}\phib''(0)\bigr)^{3/2}
    \,,\\
  \label{eq:zero:modes:G}
    \text{Goldstone}\
    &(n_\rmii{G}^{}=1,\
      \ell=0\ \text{of}\ \fluctO_{\rmii{$XG$}}^{}):
    &\mathcal J_\rmii{G}^{}
      \biggl[
          \frac{
            \det\fluctO^{0}_{\rmii{$XG$}}(\phiF^{})}{
            {\rm det}'\fluctO^{0}_{\rmii{$XG$}}(\phib^{})}
      \biggr]^{1/2} &=
      \bigl(+2\phibg_\infty^{}\phib^{}(0)\bigr)^{1/2}
    \,,
\end{align}
\end{subequations}
with $\phibg_\infty$ the asymptotic amplitude of the bounce,
$\phib(r)\to\phibg_\infty\,e^{-\mF r}/r$, and
$\mF$ the false-vacuum scalar mass.
At $\ell=0$, the mixed Goldstone--gauge and the ghost operators reduce to the Goldstone one.
The factors of eqs.~\eqref{eq:zero:modes:tr} and~\eqref{eq:zero:modes:G},
including the dimension-six corrections,
are derived in appendix~\ref{sec:zero:modes}.
Trading the three $\ell=1$ zero modes for collective coordinates gives
$\detS\propto\Vol_3\,\mathcal J_\rmi{tr}$, and,
after inserting
eq.~\eqref{eq:zero:modes:tr},
the scalar contribution reads
\begin{align}
  \label{eq:detS:full}
    - \ln\frac{\detS^{}}{\Vol_3^{}T^3} =&
      - \frac{3}{2}\ln\left(-2\,\phibg_\infty^{}\,\phib''(0)\right)
      + 3\ln T
    \nn &
      + \frac{1}{2}\ln\frac{
          \big|\det\fluctO_{\rmii{$S$}}^{0}(\phib^{})\big|}{
          \det\fluctO_{\rmii{$S$}}^{0}(\phiF^{})}
      + \frac{1}{2}\sum_{\ell\geq 2}(2\ell+1)\,
        \ln\frac{
          \det\fluctO_{\rmii{$S$}}^{\ell}(\phib^{})}{
          \det\fluctO_{\rmii{$S$}}^{\ell}(\phiF^{})}
    \,.
\end{align}
The $\ell=0$ partial wave contains the single negative eigenvalue, hence the absolute value, while the partial waves $\ell\geq2$ are strictly positive. 
For the vector determinant, one gets
\begin{align}
  \label{eq:detV:full:explicit}
    - \ln\detV^{} =&
      - \ln\detXzero^{}
      - \frac{1}{2}\ln\!\left(2\,\phibg_\infty^{}\,\phib^{}(0)\right)
      -\ln\Vol_{{\rm U}(1)}^{}
    \nn &
      + \frac{1}{2}\sum_{\ell=1}^{\infty}(2\ell+1)\biggl(
          \ln\frac{
            \det\fluctO_{\rmii{$XG$}}^\ell(\phib^{})}{
            \det\fluctO_{\rmii{$XG$}}^\ell(\phiF^{})}
        + \ln\frac{
            \det\fluctO_{\rmii{$X_T$}}^\ell(\phib^{})}{
            \det\fluctO_{\rmii{$X_T$}}^\ell(\phiF^{})}
        - 2\ln\frac{
            \det\fluctO_{\rmi{gh}}^\ell(\phib^{})}{
            \det\fluctO_{\rmi{gh}}^\ell(\phiF^{})}
      \biggr)
    \,,
\end{align}
Each partial-wave determinant in
eqs.~\eqref{eq:detS:full}
and~\eqref{eq:detV:full:explicit}
is evaluated exactly with the Gel'fand--Yaglom method,
such that the full one-loop determinant is obtained
without a derivative expansion.

%
\subsection{Higher-dimensional operators}
\label{sec:higherDimOperators}

The reduction to the nucleation-scale EFT generates
a tower of local, higher-dimensional operators for the light fields $\Psi_{\rm l}$,
with Wilson coefficients set by the hard scale.
In this work,
we propose a framework for including them consistently in the nucleation rate.
Since these operators involve the light fields themselves,
they modify both the bounce and the fluctuations at the nucleation scale
$\Lambda_{\rm nucl}^{}\sim R_{\bnc}^{-1}$, and thus contribute directly to
the fluctuation determinants.
One option is to organize this contribution through a power-counting scheme,
in which the operators enter perturbatively as
an NLO action $S^{\rmii{NLO}}$~\cite{Chala:2024xll}.
Here, we instead keep their full effect by
including them directly in the fluctuation operators of the \threedEFT{}.
This avoids an additional expansion in the Wilson coefficients, and
the size of the resulting correction directly quantifies the truncation error
of the \threedEFT{}, i.e.\ where the dimensional reduction becomes unreliable (cf.\ sec.~\ref{sec:EFTvalidity}).
Since no power counting is assumed,
the method applies to a wide class of models,
for instance, from weak to strong supercooling.%
\footnote{%
  More explicitly,
  the counting $\lambda\simeq g^n$ can differ from
  one scenario to another.
}
More broadly, it offers a generic prescription for including
perturbative effects directly in the fluctuation determinants,
beyond the dimension-six operators considered here.

Up to dimension six,
all operators in which a field carries two or more derivatives,
such as $\Box\phi$,
can be eliminated through integration by parts (IBP)
and field redefinitions based on the equations of motion (EOMs),
in favor of operators in which each field carries at most one derivative.
Schematically,
\begin{equation}
\label{eq:EOM_IBP}
    \phi^n(\Box\phi)^m(\partial\phi)^\ell \xrightarrow[\rm EOM]{\rm IBP}
      \phi^\alpha(\partial\phi)^\beta
    \,.
\end{equation}
This simplification holds for the dimension-six operators relevant for our model. 
At higher dimensions, this is no longer possible in general, and
operators in which a field carries two or more derivatives
can no longer be systematically eliminated.%
\footnote{%
  For simple single-scalar-field theories,
  this even holds up to dimension eight,
  where the operator basis can be written in terms of
  $(\partial\phi)^4$ and a single $\Box$ insertion.
  Within the framework presented here, the new terms then contribute as
  $\fluctO_{\rm fluc}=\partial\bigl(\phi^\alpha(\partial\phi)^{\beta-2}\partial\bigr)$
  or, for a radially symmetric background,
  $\fluctO_{\rm fluc}(r)=\frac{1}{r^2}\partial_r\bigl(r^2\phi^\alpha\phi'(r)^{\beta-2}\partial_r\bigr)$.
  For more general field contents or even higher-dimensional operators,
  this is no longer possible, and one needs to use the generalized version of
  the Gel'fand--Yaglom theorem~\cite{Forman:1987gha} with
  $n$ initial conditions at the origin.
} The question of field-redefinition invariance is discussed in appendix~\ref{sec:app:redef}.
There, we show in a simplified example, and for a restricted class of field redefinitions, that the computation of the fluctuation determinant is invariant under field redefinitions.
This example supports the expectation that invariance holds for the most general field redefinitions as well, which justifies simplifying the interactions by means of eq.~\eqref{eq:EOM_IBP}.
A general proof is, however, beyond the scope of this work, whose main objective is to test the validity of the EFT approach in the supercooled scenario.

After this reduction,
the dimension-six operators modify the quadratic fluctuation Lagrangian,
$\mathcal{L}\supset\frac12\phiField\,\fluctO\,\phiField$,
only through field-dependent corrections to
the kinetic term and to the potential.
Schematically,
\begin{equation}
    \fluctO \equiv
        \fluctO_4
      + \fluctO_6 =
      - \Box
      + V''(\phibg)
      - G(\phibg)\Box
      + F''(\phibg)
    \,,
\end{equation}
where
$G(\phibg)$ parametrizes the corrections to the kinetic term and
$F(\phibg)$ those to the potential.
While $F$ is simply absorbed into the effective potential,
$G$ modifies the differential structure of the fluctuation operator.
Using IBP and the radial symmetry of the bounce,
$G(r)\equiv G(\phib(r))$,
we write the kinetic part of $\fluctO_6$ in terms of
the self-adjoint combinations
\begin{align}
    &\frac{1}{2r^2}\left[\partial_r^{}\left(r^2\,G(r)\right)\right]
    \,,&
    &\frac{1}{r^2}\partial_r^{}\left(r^2\,G(r)\partial_r^{}\right)
    \,,&
    &\frac{1}{2r^2}\left[\partial_r^2\left(r^2\,G(r)\right)\right]
    \,,
\end{align}
which allows us to include them via the Gel'fand--Yaglom theorem.
In appendix~\ref{app:Gel'fand--Yaglom},
we prove explicitly that the theorem still holds for
the renormalized fluctuation determinants with a modified kinetic operator.
As a result, the Gel'fand--Yaglom equations of each partial wave
are modified in the differential operator and in the non-derivative terms,
\begin{align}
    \Box=
      \frac{1}{r^2}\partial_r\left(r^2\partial_r\right) &\to
      \frac{1}{r^2}\partial_r\left(r^2\,(1+G(r))\partial_r\right)
    \,,\\[1mm]
    V''(\phib(r)) &\to
        V''(\phib(r))
      + F''(\phib(r))
      + c_1(r) G(r)
      + c_2(r) G'(r)
      + c_3(r) G''(r)
    \,,
\end{align}
where the coefficients $c_{1,2,3}(r)$ follow from the IBP identities and
the bounce profile.
They differ between
the fluctuation sectors and are given explicitly in
eq.~\eqref{eq:app:ops:SA}.

%
\section{Semi-analytical estimate of EFT validity}
\label{sec:EFTvalidity}

In this section,
we estimate semi-analytically where the \threedEFT{}
description of nucleation remains valid.
From the power counting of
the \threedEFT{} Lagrangian~\cite{%
  Kajantie:1995dw,Braaten:1995cm,Chala:2024xll,Bernardo:2025vkz},
\begin{equation}
  \label{eq:L3d:counting}
    \mathcal{L}_\rmi{3d}^{}\sim
      \widehat{\mathcal{L}}\biggl(
          \frac{g_3^{}\phibg}{4\pi T},
          \frac{D_i^{}}{4\pi T},
          \frac{g_3^2}{(4\pi)^2T}
      \biggr)
    \,,
\end{equation}
the parameter controlling
the suppression of higher-dimensional operators
along the bounce is
\begin{equation}
  \label{eq:Ratio}
    \REFT(r)\equiv
      \frac{g_3\phib(r)}{4\pi T} \sim
      \frac{\mX(\phib(r))}{\Lambda_{\rm h}}
    \,,
\end{equation}
which is
the ratio of the field-dependent gauge-boson mass in the \threedEFT{},
$\mX(\phibg)\equiv g_{3}\phibg$,
evaluated on the bounce,
to the hard scale $\Lambda_{\rm h}\sim\pi T$
set by the non-zero Matsubara modes integrated out in the dimensional reduction.
The smaller $\REFT$, the more reliable the high-temperature expansion and hence
the \threedEFT{}.
Since the bounce is monotonic,
$\REFT(r)$ is largest at the center of the bubble,
$r_{\rm esc}=0$,
where the field reaches its escape point,
$\phib(r_{\rm esc})=\phibg_{\rm esc}$~\cite{Kierkla:2023von,Levi:2022bzt}.
We therefore determine
$\REFT\equiv \REFT(r_{\rm esc})$,
which gives the most conservative estimate along the bounce.

For this estimate,
we approximate the LO potential of the \threedEFT{} by
\begin{equation}
    V(\phibg)=
        \frac{1}{2}m_3^2\,\phibg^2
      - \frac{1}{3}\gamma_3^{}\,\phibg^3
      + \frac{1}{4}\lambda_3^{}\,\phibg^4
    \,,
\end{equation}
where the parameters
$m_3^2$, $\gamma_3$, and $\lambda_3$ are matched to the
four-dimensional couplings as given in appendix~\ref{sec:app:semi-analit}.
For simplicity, we neglect the Debye mass $\mD$,
so that both the spatial gauge modes and the temporal scalar
contribute to $\gamma_3$.
The nucleation rate is then estimated as~\cite{Christiansen:2025xhv}
\begin{equation}
    \GammaT\simeq
      T^4 e^{-\Delta S^{\rmii{LO}}_{3,\rmii{EFT}}}
    \,.
\end{equation}
The nucleation temperature then follows from 
\begin{equation}
  \label{eq:Tn}
    \GammaT(\Tn)\simeq H^4(\Tn)
    \,.
\end{equation}
For this simplified analysis,
we set $\Tstar=\Tn$ throughout this section.
For the potential above, both
the bounce action and
the escape point are
known (see e.g.~\cite{Salvio:2026bco,Levi:2022bzt, Adams:1993zs}) as functions of a single parameter
$\kappa_3=\lambda_3^{}m_3^2/\gamma_3^2$.
The action is obtained analytically and
the escape point from a numerical fit, which is summarized
in appendix~\ref{sec:app:semi-analit}.
Evaluating the action and the escape point at $\Tn$,
we compute $\REFT$ for different values of the gauge coupling $g$
at $\LamDInput=1$~GeV,
together with the corresponding transition strength
$\alpha(\Tn)$ of eq.~\eqref{eq:alpha}.
\begin{figure}[t]
\centering
\includegraphics[width=0.5\textwidth]{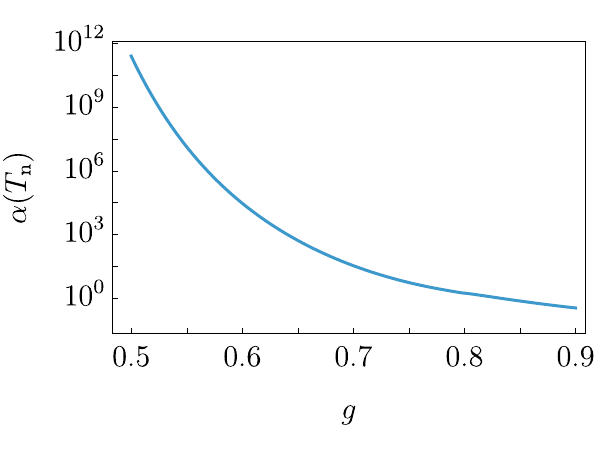}%
\includegraphics[width=0.5\textwidth]{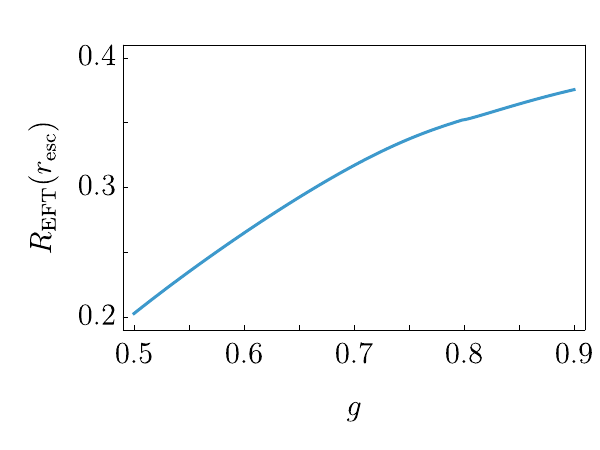}%
\caption{%
    Transition strength $\alpha(\Tn)$ (left) and
    the ratio $\REFT{}=\REFT(r_{\rm esc})$
    in eq.~\eqref{eq:Ratio} (right),
    for different values of the gauge coupling $g$,
    setting $\LamDInput=1$~GeV and
    $\LamDref=\pi T$.
}
\label{fig:plot_eft_Validity}
\end{figure}
As shown in fig.~\ref{fig:plot_eft_Validity},
$\REFT\lesssim1$ throughout, implying
that the gauge-boson mass stays below the hard scale along the bounce.
This supports the validity of our \threedEFT{} construction and shows
that the high-temperature expansion remains valid and
improves for
strongly supercooled transitions,
in agreement
with previous findings in~\cite{Kierkla:2023von,Kierkla:2025qyz}.

Beyond LO, the nucleation temperature shifts.
Compared to daisy-resummed estimates,
it typically increases~\cite{%
  Kierkla:2023von,Kierkla:2025qyz,Kierkla:2026bnm},
whereas relative to the LO \threedEFT{} estimate used here,
the exact gauge determinants can also lower
it~\cite{Kierkla:2025qyz,Puchades-Ibanez:2026ksh}.
In the strict expansion,
NLO corrections are evaluated on the LO bounce,
and
$\phibg_{\rm esc}$ and $\REFT$ change only through $\Tn$.
In practice,
$\REFT$ depends only logarithmically on $\Tn$,
mainly via the running of $\lambda$
(cf.\ eqs.~\eqref{eq:kappadef} and~\eqref{eq:app:phiesc}).
Hence, NLO-induced effects on $\REFT$ are mild and
$\REFT\lesssim1$ persists at NLO,
such that
the high-temperature expansion and
the \threedEFT{} remain valid for nucleation.

%
\section{Results}
\label{sec:results}

%
\subsection{Percolation temperature and bubble separation}

Given the nucleation rate $\GammaT(T)$ computed above,
the transition observables follow from
the false-vacuum fraction
$P_{\rm f}(T)=e^{-I(T)}$,
where
\begin{equation}
    I(T) =
      \frac{4\pi}{3}\int_{T}^{\Tc} \frac{{\rm d} T'}{T'^4}\frac{\GammaT(T')}{H(T')}\biggl[
          \int_{T}^{T'}\!{\rm d} T''\,\frac{\vw}{H(T'')}
      \biggr]^{3}
\end{equation}
is the expected comoving volume converted to the broken phase,
the inner integral being the radius at $T$ of a bubble nucleated at $T'$,
and $\Tc$ is the critical temperature.
We define
the percolation temperature $\Tp$ by the standard criterion
\begin{align}
  \label{eq:percolation-criterion}
    P_{\rm f}(\Tp)\simeq 0.71
    \,,
\end{align}
or equivalently $I(\Tp) \simeq 0.34$.
For a review of percolation in supercooled
phase transitions, see
e.g.~\cite{Athron:2023xlk,Athron:2022mmm}.
The characteristic bubble separation is
the inverse cube root of the physical bubble
number density at percolation,
\begin{equation}
    R_\star^{} =
      \biggl[
          \Tp^{3}\!\int_{\Tp}^{\Tc}\!\frac{{\rm d} T}{T^4}\frac{\GammaT^{}(T)}{H(T)}\,e^{-I(T)}
      \biggr]^{-1/3}
    \,.
\end{equation}
We express $R_\star$ in units of
the Hubble radius at percolation through
the dimensionless product
$R_\star\Hstar$ with
$\Hstar\equiv H(\Tp)$,
which, together with $\Tp$ and $\alpha$, sets the GW spectrum. 

Our calculation relies on two assumptions.
First, the transition is strongly supercooled, so the energy budget is vacuum
dominated, and we use the constant Hubble rate $H(T)\simeq\Hv$ of
eq.~\eqref{eq:Hv} throughout the relevant temperature range.
In particular, $\Hstar=\Hv$.
Second, we assume ultrarelativistic walls, $\vw=1$~\cite{%
  Cline:2021iff,Lewicki:2021pgr,Ellis:2022lft,
  Ekstedt:2024fyq,vandeVis:2025plm,Branchina:2025jou, Branchina:2025adj,
  Ekstedt:2025awx,Branchina:2026nql
  },
appropriate for the very strong transitions, $\alpha\gg1$, considered here.

%
\subsection{Impact of higher-dimensional operators on phase-transition parameters}
\label{sec:results:numerical}

\begin{figure}[t!]
\centering
\includegraphics[width=0.6\textwidth]{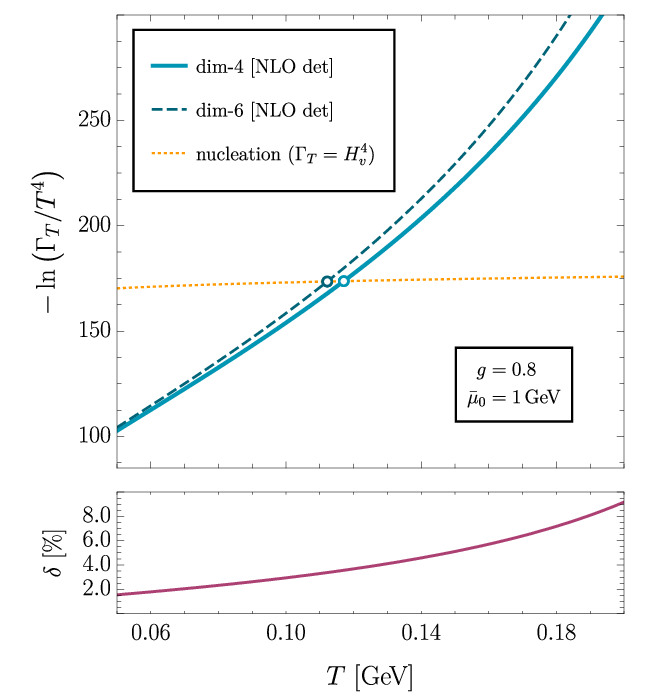}
\caption{%
  Nucleation exponent $-\ln(\GammaT/T^4)$ as a function of temperature
  for the benchmark \hyperref[eq:BM1]{{\tt BM1}}.
  Solid (dashed) lines show the NLO rate with full determinants,
  eq.~\eqref{eq:rate:full:log},
  without (with) dimension-six operators.
  The dotted line is the nucleation condition $\GammaT^{}=\Hv^4$,
  with the vacuum-dominated Hubble rate $\Hv$ of eq.~\eqref{eq:Hv}, while 
  circles mark $\Tn$ (see eq.~\eqref{eq:Tn}).
  The lower panel shows the relative difference $\delta\equiv\delta_{\ln(\GammaT/T^4)}$, eq.~\eqref{eq:Delta:rel},
  of the exponent with and without dimension-six operators.
  }
\label{fig:action_temperature}
\end{figure}

We now quantify the impact of
the dimension-six operators of eq.~\eqref{eq:Lag3d:dim6}
by comparing two computations.
Both use the NLO rate with
the full determinants~\eqref{eq:rate:full:log}
without a derivative expansion,
as required by the scale-shifting gauge modes~\cite{Kierkla:2025qyz}.
The two computations differ only in
the operators kept in the \threedEFT{}:
{\setlength{\leftmargini}{5em}%
\begin{itemize}
  \item[{[dim-4]}:]
    Only the dimension-four operators of eq.~\eqref{eq:Lag3d:dim4} are kept
    for both the bounce and the fluctuation operators.
  \item[{[dim-6]}:]
    The dimension-six operators of eq.~\eqref{eq:Lag3d:dim6} are added
    to the dimension-four ones in the bounce equation~\eqref{eq:app:eom} and
    in all fluctuation operators of appendix~\ref{sec:app:fluc}.
\end{itemize}}
For a quantity $X$,
we define the relative difference
\begin{equation}
  \label{eq:Delta:rel}
    \delta^{\text{[dim-6]}}_X \equiv
      \frac{
        X_\text{[dim-6]}-X_\text{[dim-4]}}{
        X_\text{[dim-4]}}
    \,.
\end{equation} 
As a benchmark point, we take
\begin{align}
  \label{eq:BM1}
    g_0 &= 0.8
    \,,&
    \LamDInput &= 1~\text{GeV}
    \,,
  \tag{{\tt BM1}}
\end{align}
with the input conditions of eq.~\eqref{eq:RG:prescription}
and the central matching scale $\LamDref=\pi T$.
Figure~\ref{fig:action_temperature} shows
the nucleation exponent $-\ln(\GammaT/T^4)$ for \hyperref[eq:BM1]{{\tt BM1}}.
Its intersection with
$\GammaT^{}=\Hv^4$
determines the nucleation temperature via eq.~\eqref{eq:Tn}.
The dimension-six operators increase the exponent at
all temperatures that are shown,
and $\delta$ grows
from $1.5\%$ at the lowest
to $9\%$ at the highest temperatures.
At $\Tn$, with
$\delta\simeq 3.6\%$,
the rate is suppressed by
$\Gamma_\text{[dim-6]}/\Gamma_\text{[dim-4]}\simeq 10^{-3}$,
but, the exponent being steep in $T$,
$\Tn$ reduces by only $4\%$.

The exponent increases because the only operators
non-vanishing on the real bounce,
$(\PhiField^\dagger\PhiField)^3$ and
$(D_i\PhiField)^\dagger (D_i\PhiField)(\PhiField^\dagger\PhiField)$,
have positive Wilson coefficients for $|\lambda|\ll g^2$
(appendix~\ref{sec:dim6:matching})
and thus raise the bounce action while
suppressing the rate.
We expect this to dominate the total shift,
although the dimension-six terms in
the fluctuation operators
can in principle
come with both signs.

\begin{figure}[t!]
\centering
\includegraphics[width=0.5\textwidth]{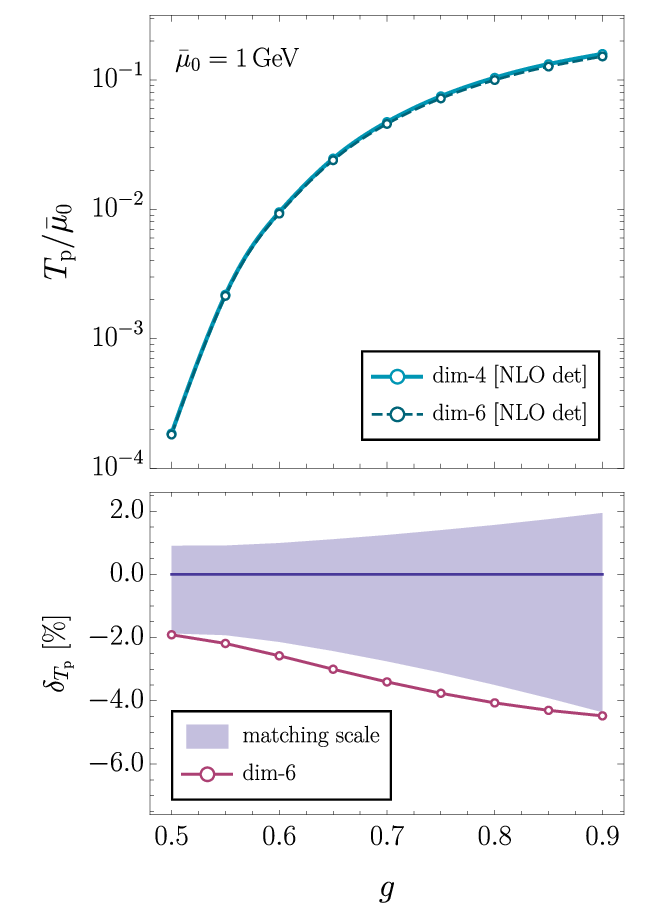}%
\includegraphics[width=0.5\textwidth]{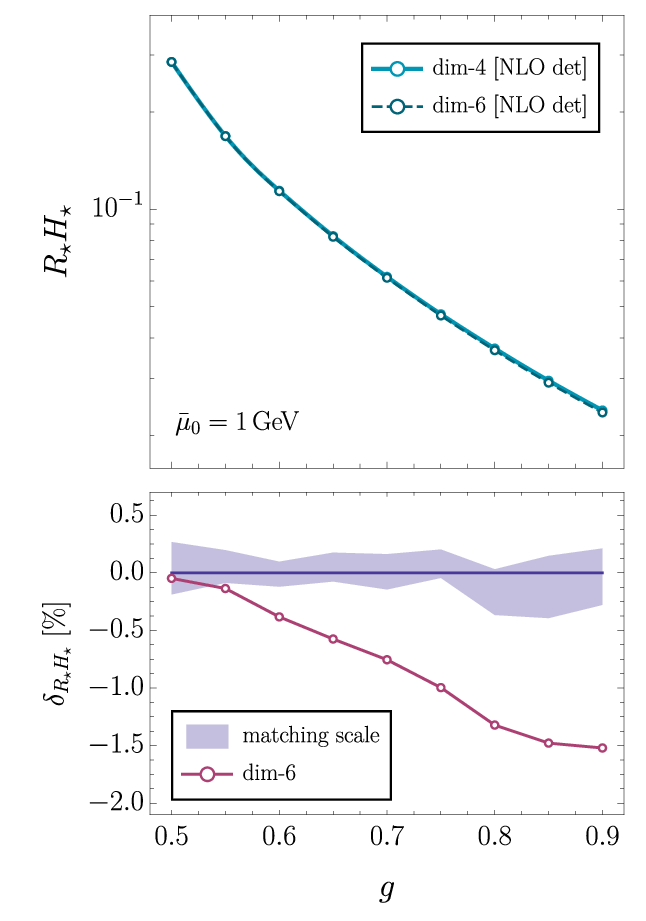}
\caption{%
  Percolation temperature $\Tp/\LamDInput$ (left) and
  bubble separation $R_\star\Hstar$ (right)
  as functions of the gauge coupling
  $g=g_0$ of eq.~\eqref{eq:RG:prescription},
  computed with the NLO determinants
  without (dim-4) and
  with (dim-6)
  dimension-six operators.
  The lower panels show the relative differences
  $\delta_{\Tp}$ and $\delta_{R_\star\Hstar}$ of eq.~\eqref{eq:Delta:rel}.
  The dim-6 shifts are shown in magenta, while the variation with matching scale, $\LamD = \LamDref \times [0.5,2]$, for the dim-4 result is shown as a purple band.
}
\label{fig:Tp_RHstar_comparison}
\end{figure}

Figure~\ref{fig:Tp_RHstar_comparison} (left) shows
the percolation temperature of eq.~\eqref{eq:percolation-criterion}
as a function of the coupling.
Toward smaller $g$,
the supercooling strengthens rapidly, and
$\Tp/\LamDInput$ falls by almost three orders of magnitude,
from $0.16$ at $g=0.9$ to
$2\times10^{-4}$ at $g=0.5$.
On this scale, the results with and without dimension-six operators
are indiscernible.
Their relative difference $\delta_{\Tp}$
is negative throughout,
as expected from the larger exponent,
and its magnitude grows monotonically with $g$,
from
$1.9\%$ at $g=0.5$ to
$4.5\%$ at $g=0.9$.
Since $\alpha\propto\Tp^{-4}$ at fixed $\geff$, eq.~\eqref{eq:alpha},
the transition strength increases by $8\%$--$20\%$.
The GW spectrum depends on $\alpha$ only through
the kinetic-energy fraction
$K\propto\alpha/(1+\alpha)$~\cite{Steinhardt:1981ct,Espinosa:2010hh},
which saturates at $K\simeq1$ for
$\alpha\gg1$.
The relative shifts of $\alpha$
change $K$
by less than $10^{-12}$ at $g=0.5$ and
by less than $0.5\%$ for $g\leq0.7$.
Only at the largest couplings, where
$\alpha=\mathcal{O}(1)$
(cf.\ fig.~\ref{fig:plot_eft_Validity}, left),
a shift of $\alpha$ starts to matter.
There, $\rho_\rmii{R}(\Tp)$ is also no longer negligible
in the Hubble rate compared with $\Delta V_0$.

The bubble separation $R_\star\Hstar$ in the right panels of
fig.~\ref{fig:Tp_RHstar_comparison} is even less affected.
Its relative change is
below $0.5\%$ in magnitude for $g\leq 0.6$ and
about $-1.6\%$ at $g=0.9$.
Since $\Hstar=\Hv$ is fixed by
$\Delta V_0$ via eq.~\eqref{eq:Hv},
$R_\star$ is fully responsible for this shift.
$R_\star$ shrinks because
the dimension-six operators
increase the rate faster as the Universe cools,
producing more and hence smaller bubbles.
Correspondingly~\cite{Kierkla:2026bnm,Megevand:2016lpr,Matuszak:2026xsz},
\begin{equation}
    \frac{\beta}{\Hstar}\simeq\frac{\vw(8\pi)^{1/3}}{R_\star\Hstar(1-P_{\rm f}(T_{\rm p}))^{1/3}}
\end{equation}
increases, however, only by at most $1.6\%$.

The dimension-six shifts of the nucleation exponent, $\Tp$, and $R_\star\Hstar$
in figs.~\ref{fig:action_temperature} and~\ref{fig:Tp_RHstar_comparison}
quantify the truncation error of the \threedEFT{}
anticipated in sec.~\ref{sec:higherDimOperators}.
By the power counting of eq.~\eqref{eq:L3d:counting},
dimension-six operators are suppressed relative to dimension-four ones by
$\REFT^2$ via eq.~\eqref{eq:Ratio}.
With $\REFT\simeq0.2$--$0.4$ for $g=0.5$--$0.9$
(fig.~\ref{fig:plot_eft_Validity}, right),
corrections of at most $\mathcal{O}(10\%)$ are expected.
The shifts of a few percent found with the full NLO rate,
growing with $g$ together with $\REFT$,
confirm the estimate of sec.~\ref{sec:EFTvalidity}.
They are of the same order as the effect of the $\phibg^6$ operator alone,
which changes the exponent at $\Tp$ by $0.2\%$--$1.3\%$
for $g=0.5$--$0.7$~\cite{Puchades-Ibanez:2026ksh}.
In contrast to equilibrium observables near $\Tc$,
for which dimension-six operators can dominate over higher-loop
corrections~\cite{Bernardo:2025vkz,Bernardo:2026whs},
their impact on nucleation is small,
since for strong supercooling
the bounce stays within the high-temperature regime,
far from the broken minimum~\cite{Kierkla:2023von,Puchades-Ibanez:2026ksh}.
Consistently,
the effect grows toward larger $g$ and higher $T$,
i.e.\ toward weaker supercooling.

The shifts are also small compared with
other theoretical uncertainties of supercooled transitions.
In the classically conformal ${\rm SU}(2)$ model,
the NLO corrections shift $\Tp$ by $50\%$--$100\%$ relative to the LO result,
and at NLO a factor-two variation of the four-dimensional matching scale
still changes $\Tp$ by up to $10\%$ and
$R_\star\Hstar$ by up to $2\%$~\cite{Kierkla:2023von, Kierkla:2025qyz}.
In the conformal ${\rm U}(1)$ model with a Higgs portal,
daisy resummation and the NLO rate with determinants
lead to gauge couplings reconstructed from a
Laser Interferometer Space Antenna (LISA) signal
that differ by $\mathcal{O}(10\%)$,
whereas the scale dependence amounts to $\mathcal{O}(1\%)$ for daisy
and is negligible at NLO~\cite{Kierkla:2026bnm}.

Despite the RG improvement of sec.~\ref{sec:RG:improvement},
the residual matching-scale dependence
is comparable to the dimension-six corrections for $\Tp$
and somewhat smaller for $R_\star\Hstar$.
Denoting by
$\delta_X^{(\LamD)}$
the relative change of
$X\in\{\Tp,R_\star\Hstar\}$
by varying
the reference scale
in the range of
$\LamD = \LamDref \times [0.5,2]$
(bands in fig.~\ref{fig:Tp_RHstar_comparison}) and
by $\delta_X^{\rmii{NLO}}$
the shift from the lower-order (LO or daisy) to the NLO rate,
the uncertainties thus follow the hierarchy
\begin{align}
  \label{eq:uncertainty:hierarchy}
    \delta_{\Tp}^{(\LamD)} & \simeq 
      \delta_{\Tp}^{\text{[dim-6]}}\simeq \mathcal{O}(1\%) \ll
      \delta_{\Tp}^{\rmii{NLO}}
    \,,\nn
    \delta_{R_\star\Hstar}^{(\LamD)} & \lesssim
      \delta_{R_\star\Hstar}^{\text{[dim-6]}}\simeq \mathcal{O}(1\%) \ll
      \delta_{R_\star\Hstar}^{\rmii{NLO}}
    \,,
\end{align}
where $\delta_{\Tp}^{\rmii{NLO}}$ and $\delta_{R_\star\Hstar}^{\rmii{NLO}}$ are taken from~\cite{Kierkla:2023von,Kierkla:2026bnm,Puchades-Ibanez:2026ksh}. 
Dimension-six operators are thus subleading to the NLO corrections,
but of the same order as the residual scale dependence at NLO.

For $\vw=1$ and $K\simeq1$,
the GW spectrum depends only on $\beta/\Hstar$ and
on the reheating temperature, which is fixed by $\Delta V_0$
except at the largest couplings~\cite{Kierkla:2026bnm,Puchades-Ibanez:2026ksh},
so that the shifts found here correspond to a shift of the reconstructed
gauge coupling comparable to the LISA reconstruction uncertainty
for the strongest signals~\cite{Kierkla:2026bnm}.
\vfill
%
\section{Conclusions}
\label{sec:conclusion}

We developed a framework
for including small perturbations of
the effective action in the full one-loop bubble-nucleation rate.
In this work,
we specifically studied thermal higher-dimensional operators
in supercooled first-order phase transitions.
The modifications considered enter both
the bounce and
the fluctuation determinants
as field-dependent corrections
to the potential and
to the kinetic operator.
Casting the bilinear modifications into a self-adjoint form,
we showed that the Gel'fand--Yaglom theorem continues to hold,
and all determinants can be evaluated exactly
without a derivative expansion.
Without assuming a concrete power counting,
the approach applies
to both weak and strong supercooling regimes.
The framework is also not restricted to
the operator bases considered in this work (see sec.~\ref{sec:higherDimOperators} for the details).

As an application,
we included
the dimension-six operators generated by
the hard-to-soft matching of dimensional reduction
in the \threedEFT{} of the classically conformal Abelian-Higgs model,
including the mixed Goldstone--gauge sector.
Restricting to a pure scalar sector,
we verified explicitly that the result is
invariant under field redefinitions (see appendix~\ref{sec:app:redef}).
For the strongly supercooled transitions considered,
the dimension-six operators modify
the nucleation exponent,
the percolation temperature $\Tp$, and
$R_\star\Hstar$ at $\mathcal{O}(1\%)$.
These modifications grow with the coupling,
stay below typical NLO corrections~\cite{Kierkla:2023von,Puchades-Ibanez:2026ksh},
and are of the same order as the residual matching-scale dependence
of our RG-improved NLO computation
(cf.~the hierarchy of eq.~\eqref{eq:uncertainty:hierarchy}).
However, these shifts are comparable to
the LISA reconstruction uncertainty
for the strongest signals~\cite{Kierkla:2026bnm}.

As a byproduct,
we estimated the validity of the high-temperature EFT along the bounce.
The ratio of
the gauge-boson mass at the escape point to
the hard scale
remains below unity over the parameter range studied,
as estimated with the LO action.
The $\mathcal{O}(1\%)$ dimension-six corrections
found with the one-loop determinants
confirm this trend beyond LO
and the claims of~\cite{Kierkla:2023von}.
As a result,
using dimensional reduction for the nucleation rate is justified
even in strongly supercooled transitions.
While our numerical results are restricted to a single model,
our framework applies directly to
all gauge-Higgs theories and beyond-the-Standard-Model extensions as well as
to general corrections to the effective action,
such as those from higher loop orders.

%
\section*{Acknowledgments}

We are grateful to
Luis Gil,
Miha Nemevšek,
Cristina Puchades Ibáñez,
Daniel Schmitt,
Bogumi{\l}a {\'S}wie{\.z}ewska,
Jorinde van de Vis, and
Felix Yu
for enlightening discussions during the development of this project.
F.~Bernardo and P.~Schicho are supported by
the Swiss National Science Foundation (SNSF) under grant
\href{https://data.snf.ch/grants/grant/215997}{\tt PZ00P2-215997}.
M.~Kierkla is supported by the Carl Trygger Foundation through
grant no.~{\tt CTS~24:3412}.
N.~Leister and P.~Schwaller acknowledge support by
the Cluster of Excellence ``PRISMA$^{++}$'' funded by the German Research Foundation
(DFG) within the German Excellence Strategy (project no.~390831469).
N.~Leister is grateful to the German Academic Scholarship Foundation for the award of a
PhD fellowship.
\\

\noindent
{\bf Data availability statement.}
The matching relations of the higher-dimensional operators
are given in appendix~\ref{sec:dim6:matching}
and~\cite{Bernardo:2025vkz,Bernardo:2026nyq}.
\\

\noindent
{\bf AI disclosure statement.}
During the preparation of this manuscript,
the authors used Claude Opus 5.5 and GPT-5.6 Sol for coding assistance,
language editing and rephrasing, and
checking the consistency of notation and
of statements about the cited literature.
All AI-assisted output was reviewed,
verified, and
edited by the authors,
who take full responsibility for the content of this publication.

%
\appendix
\addtocontents{toc}{\protect\setcounter{tocdepth}{1}}
\renewcommand{\thesection}{\Alph{section}}
\renewcommand{\thesubsection}{\Alph{section}.\arabic{subsection}}
\renewcommand{\theequation}{\Alph{section}.\arabic{equation}}

%
\section{Higher-dimensional operators}
\label{sec:dim6:matching}

This appendix collects the dimension-six operators generated by
dimensional reduction and their Wilson coefficients.
Only operators that contribute at quadratic order in the fluctuations
enter the one-loop nucleation rate.
Operators that start at cubic
order affect only higher-loop corrections and are therefore omitted
from the determinant analysis.
The operators listed in tab.~\ref{tab:softop:dim6}
constitute
the off-shell dimension-six operator basis of
the \threedEFT,
with hatted Wilson coefficients $\alphaF_i$~\cite{Bernardo:2025vkz}.
\begin{table}[t]
\centering
\renewcommand{\arraystretch}{1.2}
\begin{tabular}[t]{|Sl|Sl|}
\hline
\multicolumn{2}{|c|}{Dimension-six operator basis}\\
\hline
\hline
  $F_{ij}^{}F_{ij}^{} B_0^2$ &
  $Z_{B_0} Z_{B_i}\alphaF_{B_0^2 F^2}$
  \\ \hline
  $F_{ij} F_{ij} \PhiField^\dagger \PhiField$ &
  $Z_{B_i} Z_\phi\alphaF_{\phi^2 F^2}$
  \\ \hline
  $(D_i \PhiField)^\dagger (D_i \PhiField) (\PhiField^\dagger \PhiField)$ &
  $Z_\phi^2  \alphaF_{D^2\phi^4, 1}^{}$
  \\ \hline
  $(D_i^{}\PhiField)^\dagger (D_i^{} \PhiField) B_0^2$ &
  $Z_\phi Z_{B_0}\alphaF_{D^2\phi^2 B_0^2, 3}$
  \\ \hline
  $B_0^6$ & $Z_{B_0}^3\alphaF_{B_0^6}^{}$
  \\ \hline
  $B_0^4 (\PhiField^\dagger\PhiField) $ &
  $Z_\phi^{} Z_{B_0}^2\alphaF_{\phi^2 B_0^4}^{}$
  \\ \hline
  $B_0^2 (\PhiField^\dagger\PhiField)^2 $ &
  $Z_\phi^2 Z_{B_0}^{}\alphaF_{\phi^4 B_0^2}^{}$
  \\ \hline
  $(\PhiField^\dagger\PhiField)^3$ &
  $Z_\phi^3\alphaF_{\phi^6 }^{}$
  \\ \hline
\end{tabular}
\hspace{1cm}
\begin{tabular}[t]{|Sl|Sl|}
\hline
\multicolumn{2}{|c|}{Redundant operators}\\
\hline
\hline
  $(\partial_i F_{ij})^2$ &
  $Z_{B_i}\alphaF_{D^2 F^2}$
  \\ \hline
  $B_0 \Box^2 B_0$ &
  $Z_{B_0}\alphaF_{D^4B_0^2}$
  \\ \hline
  $B_0^3 \Box B_0^{}$ &
  $Z_{B_0}^2\alphaF_{D^2B_0^4}^{}$
  \\ \hline
  $(D^2 \PhiField)^\dagger (D^2 \PhiField)$ &
  $Z_\phi\alphaF_{D^4\phi^2}$
  \\ \hline
  $(\PhiField^\dagger \PhiField) (\PhiField^\dagger D^2 \PhiField + \textit{h.c.})$ &
  $Z_\phi^2 \alphaF_{D^2\phi^4,2}^{}$
  \\ \hline
  $(\partial_i F_{ij}) i \PhiField^\dagger (D_j \PhiField)$ &
  $Z_\phi^{} Z_{B_i}^{1/2}\alphaF_{D^2\phi^2 F}^{}$
  \\ \hline
  $(\PhiField^\dagger \PhiField) B_0 \Box B_0$ &
  $Z_\phi Z_{B_0}\alphaF_{D^2\phi^2 B_0^2, 1}$
  \\ \hline
  $\PhiField^\dagger (D^2\PhiField) B_0^2 + \textit{h.c.}$ &
  $Z_\phi Z_{B_0}\alphaF_{D^2\phi^2 B_0^2, 2}$
  \\ \hline
\end{tabular}
\caption{%
  The off-shell basis of dimension-six operators in the \threedEFT{}, including the
  redundant operators that can be removed by field redefinitions.
  We refer to ref.~\cite{Bernardo:2025vkz} for details.
  }
\label{tab:softop:dim6}
\end{table}
After field redefinitions, only the non-redundant operators remain,
with on-shell Wilson coefficients $\alphaFR_i$ given by
\begin{eqnarray}
    \alphaFR_{B_0^6} &=&
      \mathcal{O}(g^8)
    \,, \\[1mm]
    \alphaFR_{\phi^2B_0^4} &=&
      - \frac{\zeta_3}{288\,\pi^4}g^{6}
      + \mathcal{O}(g^8)
    \,, \\[1mm]
    \alphaFR_{\phi^4B_0^2} &=&
      - \frac{\zeta_3}{960\,\pi^4}\Big(
          31 g^{6}-260 g^{4}\lambda
        + 280 g^{2}\lambda^2
      \Big)
      + \mathcal{O}(g^8)
    \,, \\[1mm]
    \alphaFR_{\phi^6} &=&
      \frac{\zeta_3}{480\,\pi^4}\Big(
          15 g^{6}-18g^{4}\lambda
        + 75 g^{2}\lambda^2
        + 100 \lambda^3
      \Big)
      + \mathcal{O}(g^8)
    \,, \\[1mm]
    \alphaFR_{D^2\phi^2B_0^2} &=&
      - \frac{\zeta_3}{384\,\pi^4T}\Big(
          7 g^{4}-4 g^{2}\lambda
      \Big)
      + \mathcal{O}(g^6)
    \,, \\[1mm]
    \alphaFR_{B_0^2F^2} &=&
        \frac{\zeta_3}{384\,\pi^4T}g^{4}
      + \mathcal{O}(g^6)
    \,, \\[1mm]
    \alphaFR_{\phi^2F^2} &=&
      \frac{\zeta_3}{768\,\pi^4T}\Big(
          7 g^{4}-4 g^{2}\lambda
      \Big)
      + \mathcal{O}(g^6)
    \,, \\[1mm]
    \alphaFR_{D^2\phi^4} &=&
      \frac{\zeta_3}{960\,\pi^4T}\Big(
          51 g^{4}+200 g^{2}\lambda
        - 20 \lambda^2
      \Big)
      + \mathcal{O}(g^6)
    \,.
\end{eqnarray}
    
%
\section{Bounce equation and fluctuation operators}
\label{sec:bounce:higherdim}

This appendix collects the technical ingredients needed to evaluate
the one-loop nucleation rate of sec.~\ref{sec:nucleation} once
the dimension-six operators of eq.~\eqref{eq:Lag3d:dim6} are included.
We proceed in three steps: sec.~\ref{sec:app:bounce} derives
the EOM for the bounce,
sec.~\ref{sec:app:fluc} constructs the quadratic fluctuation operators in
each sector and casts them in the self-adjoint form required by 
the Gel'fand--Yaglom method, and appendix~\ref{sec:app:WKB} works out
their large angular-momentum behavior
in the Wentzel--Kramers--Brillouin (WKB) approximation,
which controls the subtraction of the partial-wave sums.
Throughout,
$\phib(r)$ denotes the radial bounce profile and
$\phib'\equiv\partial_r^{}\phib^{}$.
To keep track of the new hard contributions, all contributions generated by
the dimension-six operators are printed in \newDimSix{this color}.
Note that setting the Wilson coefficients $\alpha_i\to0$
recovers the super-renormalizable \threedEFT{} and
the usual form of the fluctuation determinants.

%
\subsection{Background bounce equation}
\label{sec:app:bounce}

We expand the fields around a real,
static background $\phib(x)$, using
the decomposition~\eqref{eq:field:decomp},
\begin{align}
  \label{eq:app:fluc:param}
    \PhiField(x) &=
        \frac{1}{\sqrt{2}}\phib(x)
      + \frac{1}{\sqrt{2}}\bigl(\phiField(x)+i\chi(x)\bigr)
    \,,&
    B_i(x) &= a_i(x)
    \,,
\end{align}
so that
$\phiField$ is the radial fluctuation,
$\chi$ the Goldstone mode, and
$a_i$ the gauge fluctuation
in three spatial dimensions.
In the classically conformal model,
after locally expanding the action around
a vanishing temporal scalar background $B_0 = 0$,%
\footnote{%
  Such an expansion corresponds
  to assuming the trivial holonomy in the action.
  However, it was recently shown that
  the general case of a non-trivial holonomy can have a sizable impact
  on the equilibrium thermodynamics of the phase
  transition~\cite{Bandyopadhyay:2026nrv,Biermann:2026qju}.
}
the bounce is carried entirely by $\phib$.
Varying the action~\eqref{eq:Lag3d:dim4}--\eqref{eq:Lag3d:dim6} with respect to
$\phiField$ gives the EOM for the background%
\footnote{%
  Throughout,
  $V(\phib)\equiv V(\PhiField)\big|_{\PhiField=\phib/\sqrt2}$,
  and primes on $V$ denote derivatives with respect to $\phib$.
}
\begin{align}
  \label{eq:app:eom}
    - \Box \phib^{}
    + V'\!\left(\phib^{}\right)
    \newDimSix{
      - \frac{\alpha_{D^2\phi^4}^{}}{2}\Bigl(
          \phib^2\Box \phib^{}
        + (\partial_i^{}\phib^{})^2\phib^{}
      \Bigr)
      + \frac{3}{4}\alpha_{\phi^6}^{}\phib^5}
    &= 0
    \,,
\end{align}
which, for a rotationally invariant profile $\phib\equiv\phib(r)$, reduces to the radial equation
\begin{align}
  \label{eq:app:eom:radial}
    - \phib''
    - \frac{2}{r}\phib'
    + V'\!(\phib^{})
    \newDimSix{
      - \frac{\alpha_{D^2\phi^4}^{}}{2}\Bigl(
          \phib^2\phib''
        + \frac{2}{r}\phib^2\phib'
        + \phib'^2\phib^{}
      \Bigr)
      + \frac{3}{4}\alpha_{\phi^6}^{}\phib^5}
    &= 0
    \,,
\end{align}
with boundary conditions
\begin{align}
    \phib'(0)&=0
    \,,&
    \phib^{}(\infty)&=0
    \,.
\end{align}
Moreover, collecting the two-derivative terms, eq.~\eqref{eq:app:eom:radial} can be
written in the manifestly self-adjoint form
$-r^{-2}\partial_r(r^2K\phib')+\tfrac12K'(\phib)\phib'^2+\dots=0$ with
$K(\phib^{})=1+\tfrac12\alpha_{D^2\phi^4}^{}\phib^2$.
For more details, we refer to the 
zero-mode analysis of appendix~\ref{sec:zero:modes}.

%
\subsection{Quadratic fluctuation operators}
\label{sec:app:fluc}

Expanding the action to second order in the fluctuations,
$(\phiField,\chi,a_i,B_0)$,
defines the fluctuation operators that enter
the determinants in sec.~\ref{sec:full:det}.
They hold for a generic static background $\phibg$
and are evaluated on the bounce, $\phibg=\phib$,
and in the false vacuum, $\phibg=\phiF$.
At leading (super-renormalizable) order, and in Feynman gauge $\xi=1$,
the quadratic Lagrangian is
\begin{align}
  \label{eq:app:L2:LO}
    \mathcal{L}^{(2)}_0 &=
      \frac{1}{2}B_0^{}\Bigl(
        - \Box
        + \mD^2
        + h_3^{}\phibg^2
      \Bigr)B_0^{}
      + \frac{1}{2}\phiField\Bigl(
        - \Box
        + V''(\phibg)
      \Bigr)\phiField
    \nn &
      + \frac{1}{2}a_i^{}\Bigl(
        - \Box
        + g^2\phibg^2
      \Bigr)\delta^{ij}a_j^{}
      + \frac{1}{2}\chi\Bigl(
        - \Box
        + \frac{V'(\phibg)}{\phibg}
        + g^2\phibg^2
      \Bigr)\chi
      - \frac{1}{2}\chi\,4g\,(\partial_i^{} \phibg)\,a_i^{}
    \,.
\end{align}
The dimension-six operators of eq.~\eqref{eq:Lag3d:dim6} add further quadratic
terms.
Rather than listing the Lagrangian, we give the resulting operators directly below.
The functional determinant is gauge independent~\cite{Baacke:1999sc}, which we
exploit by choosing the gauge-fixing function $f(\phibg)=f_0(\phibg)$ of
eq.~\eqref{eq:gauge-fixing}.
This choice cancels
the dimension-six Goldstone--gauge cross terms
and simplifies the mixing sector.

The temporal scalar $B_0$ and the radial scalar $\phiField$ decouple from the
other fluctuations, whereas the Goldstone $\chi$ mixes
with the gauge modes in the inhomogeneous background.
Working in a basis of spherical harmonics
$Y_\ell^m$,
we denote
\begin{align}
  \label{eq:app:dell}
    \partial_\ell^2 &\equiv
        \partial_r^2
      + \frac{2}{r}\partial_r^{}
      - \frac{L^2}{r^2}
    \,,&
    L^2 &\equiv \ell(\ell+1)
    \,,
\end{align}
and decompose the gauge field into
longitudinal, radial, and transverse parts,
\begin{align}
  \label{eq:app:gauge:decomp}
    a_i^{}(x)&=
      \sum_{\ell m}\left(
          \frac{f^{\ell}_a(r)}{L}\,r\,\partial_i^{}
        + f_b^\ell(r)\frac{x_i^{}}{r}
        + \frac{f_c^\ell(r)}{L}\,\epsilon_{ijk}^{}x_j^{}\partial_k^{}
      \right)Y_\ell^m
    \,,&
    \chi(x)&=
      \sum_{\ell m} f^\ell_\chi(r)\,Y_\ell^m
    \,,
\end{align}
where the longitudinal and radial modes $(f_a,f_b)$ couple to the Goldstone $f_\chi$,
while the
transverse mode $f_c$ decouples.

%
\subsection{Self-adjoint form}

We write all fluctuation operators in a form that is manifestly self-adjoint with respect to
the radial measure
$\int_r \equiv \int_0^\infty {\rm d}r\, r^2$.
There are two reasons for this. First, only
the symmetric part of an operator contributes to
the quadratic form in the path integral,
since any antisymmetric part drops out.
Second, a well-defined evaluation of
the fluctuation determinants requires
the operators to be manifestly self-adjoint
(cf.~\cite{Kirsten:2004qv} and
appendix~\ref{app:Gel'fand--Yaglom}).

The off-diagonal entries are made manifestly
symmetric by the gauge choice $f=f_0$,
while the diagonal derivative terms are
symmetrized with the IBP identities
\begin{align}
    \int_{r} f\,G(r)\,\partial_r^{} f &\stackrel{\text{IBP}}{=}
      - \int_{r} f\,\frac{1}{2r^2}\left[\partial_r^{}\!\left( r^2\,G\right)\right]\,f
    \,,\nn
    \int_{r} f\,G(r)\,\partial_r^2 f &\stackrel{\text{IBP}}{=}
        \int_{r} f\,\frac{1}{r^2}\partial_r^{}\!\left(r^2\,G\,\partial_r^{}\right)f
      + \int_{r} f\,\frac{1}{2r^2}\left[\partial_r^2 \!\left(r^2\,G\right)\right]f
    \,,
\end{align}
valid for any smooth $G(r)$. Consequently 
the three combinations
\begin{align}
    &\frac{1}{2r^2}\left[\partial_r^{}\!\left(r^2\,G\right)\right]
    \,,&
    &\frac{1}{r^2}\partial_r^{}\!\left(r^2\,G\,\partial_r^{}\right)
    \,,&
    &\frac{1}{2r^2}\left[\partial_r^2\!\left(r^2\,G\right)\right]
    \,,
\end{align}
are self-adjoint.
It is also convenient to introduce the anti-self-adjoint
first-order operator
\begin{align}
  \label{eq:app:DG}
    \mathcal{D}_G &\equiv
        \frac{G(r)}{r}\partial_r
      + \frac{1}{2r^2}\left[\partial_r\!\left(r\,G\right)\right]
    \,,&
    \int_{r} \psi\,\mathcal{D}_G\phi &=
      - \int_{r} \phi\,\mathcal{D}_G\psi
    \,,
\end{align}
which pairs the $(a,b)$ off-diagonal entries into a symmetric block.

In this basis,
the fluctuation operators,
divided into a leading (super-renormalizable)
part $\fluctO^{(0)}$ and
a dimension-six correction $\newDimSix{\fluctO^{(1)}}$,
take the self-adjoint form
\begin{align}
  \label{eq:app:ops:SA}
    \fluctO^{(0)}_{B_0B_0}&=
      - \partial_\ell^2
      + \mD^2
      + h_3^{}\phibg^2
    \,,\nn
    \newDimSix{\fluctO^{(1)}_{B_0B_0}}&=\newDimSix{
        \frac{\alpha_{\phi^4B_0^2}^{}}{2}\phibg^4
      + \alpha_{D^2\phi^2B_0^2}^{}\phibg'^2}
    \,,\nn[1mm]
    \fluctO^{(0)}_{\phiField\phiField}&=
      - \partial_\ell^2
      + V''(\phibg)
    \,,\nn
    \newDimSix{\fluctO^{(1)}_{\phiField\phiField}}&=\newDimSix{
      - \frac{\alpha_{D^2\phi^4}^{}}{2r^2}\partial_r^{}\!\left(r^2\,\phibg^2\partial_r^{}\right)
      + \alpha_{D^2\phi^4}^{}\left(
          \frac{L^2\,\phibg^2}{2r^2}
        - \frac{2\phibg\phibg'}{r}
        - \frac{\phibg'^2}{2}
        - \phibg\phibg''
      \right)
      + \frac{15}{4}\alpha_{\phi^6}^{}\phibg^4}
    \,,\nn[1mm]
    \fluctO^{(0)}_{\chi\chi}&=
      - \partial_\ell^2
      + \frac{V'(\phibg)}{\phibg}
      + g^2\phibg^2
    \,,\nn
    \newDimSix{\fluctO^{(1)}_{\chi\chi}}&=\newDimSix{
      - \frac{\alpha_{D^2\phi^4}^{}}{2r^2}\partial_r^{}\!\left(r^2\phibg^2\partial_r^{}\right)
      + \alpha_{D^2\phi^4}^{}\left(
          \frac{L^2\,\phibg^2}{2r^2}
        + g^2\,\phibg^4
        + \frac{1}{2}\phibg'^2
      \right)
      + \frac{3}{4}\alpha_{\phi^6}^{}\phibg^4}
    \,,\nn[1mm]
    \fluctO^{(0)}_{a\chi}=\fluctO^{(0)}_{\chi a}&=0
    \,,\nn
    \newDimSix{\fluctO^{(1)}_{a\chi}=\fluctO^{(1)}_{\chi a}}&=\newDimSix{0}
    \,,\nn
    \fluctO^{(0)}_{b\chi}=\fluctO^{(0)}_{\chi b}&=-2g\phibg'
    \,,\nn
    \newDimSix{\fluctO^{(1)}_{b\chi}=\fluctO^{(1)}_{\chi b}}&=\newDimSix{
      - 2g\,\alpha_{D^2\phi^4}^{}\phibg^2\phibg'}
    \,,\nn[1mm]
    \fluctO^{(0)}_{aa}&=
      - \partial_\ell^2
      + g^2\phibg^2
    \,,\nn
    \newDimSix{\fluctO^{(1)}_{aa}}&=\newDimSix{
      - \frac{2\alpha_{\phi^2F^2}^{}}{r^2}\partial_r^{}\!\left(r^2\phibg^2\partial_r^{}\right)
      + \frac{1}{2}g^2\alpha_{D^2\phi^4}^{}\phibg^4
      - \frac{4\phibg\phibg'}{r}\alpha_{\phi^2F^2}^{}}
    \,,\nn[1mm]
    \fluctO^{(0)}_{bb}&=
      - \partial_\ell^2
      + \frac{2}{r^2}
      + g^2\phibg^2
    \,,\nn
    \newDimSix{\fluctO^{(1)}_{bb}}&=\newDimSix{
        2\alpha_{\phi^2F^2}^{}L^2\frac{\phibg^2}{r^2}
      + \frac{1}{2}g^2\alpha_{D^2\phi^4}^{}\phibg^4}
    \,,\nn[1mm]
    \fluctO^{(0)}_{ab}=\fluctO^{(0)}_{ba}&=-\frac{2L}{r^2}
    \,,\nn
    \newDimSix{\fluctO^{(1)}_{ab}}&=\newDimSix{
      \alpha_{\phi^2F^2}^{}\left(
        - L\frac{\phibg^2}{r^2}
        + 2L\frac{\phibg\phibg'}{r}
        + 2L\,\mathcal{D}_{\phibg^2}^{}
      \right) =
      \alpha_{\phi^2F^2}^{}\left(
          4L\frac{\phibg\phibg'}{r}
        + 2L\frac{\phibg^2}{r}\partial_r^{}
      \right)}
    \,,\nn
    \newDimSix{\fluctO^{(1)}_{ba}}&=\newDimSix{
      \alpha_{\phi^2F^2}^{}\left(
        - L\frac{\phibg^2}{r^2}
        + 2L\frac{\phibg\phibg'}{r}
        - 2L\,\mathcal{D}_{\phibg^2}^{}
      \right) =
      \alpha_{\phi^2F^2}^{}\left(
        - 2L\frac{\phibg^2}{r^2}
        - 2L\frac{\phibg^2}{r}\partial_r^{}
      \right)}
    \,,\nn
    \fluctO^{(0)}_{cc}&=
      - \partial_\ell^2
      + g^2\,\phibg^2
    \,,\nn
    \newDimSix{\fluctO^{(1)}_{cc}}&=\newDimSix{
      - \frac{2\alpha_{\phi^2F^2}^{}}{r^2}\partial_r^{}\!\left(r^2\,\phibg^2\partial_r^{}\right)
      + \frac{1}{2}g^2\alpha_{D^2\phi^4}^{}\phibg^4
      + \alpha_{\phi^2F^2}^{}\left(
        - \frac{4\phibg\phibg'}{r}
        + \frac{2L^2\phibg^2}{r^2}
      \right)}
    \,,\nn
    \fluctO^{(0)}_{\rmi{gh}}&=
      - \partial_\ell^2
      + g^2\phibg^2
    \,,\nn[1mm]
    \newDimSix{\fluctO^{(1)}_{\rmi{gh}}}&=\newDimSix{
      \frac{1}{2}g^2\alpha_{D^2\phi^4}^{}\phibg^4}
    \,,
\end{align}
where
the first of the two equivalent expressions for
$\newDimSix{\fluctO^{(1)}_{ab}}$ and
$\newDimSix{\fluctO^{(1)}_{ba}}$ displays
the manifestly self-adjoint form (with
$\mathcal{D}_G$ and $G=\phibg^2$ in eq.~\eqref{eq:app:DG}),
and the second the expanded one.
Assembling these entries,
the coupled Goldstone--gauge fluctuations are
governed by the self-adjoint $3\times3$ operator
\begin{align}
    \fluctO^{\ell}_{\rmii{$XG$}} &=
      \begin{pmatrix}
          \fluctO_{aa}^{} & \fluctO_{ab}^{} & \fluctO_{a\chi}^{}\\[1mm]
          \fluctO_{ba}^{} & \fluctO_{bb}^{} & \fluctO_{b\chi}^{}\\[1mm]
          \fluctO_{\chi a}^{} & \fluctO_{\chi b}^{} & \fluctO_{\chi\chi}^{}
      \end{pmatrix}
    \,,&
  \label{eq:app:XG:matrix}
    \fluctO_{nm}^{} &\equiv
        \fluctO^{(0)}_{nm}
      + \newDimSix{\fluctO^{(1)}_{nm}}
    \,,
\end{align}
while the remaining sectors are diagonal,
\begin{align}
  \label{eq:app:sector:ops}
    \fluctO^{\ell}_{\rmii{$X_0$}} &=
        \fluctO^{(0)}_{B_0B_0}
      + \newDimSix{\fluctO^{(1)}_{B_0B_0}}
    \,, &
    \fluctO^{\ell}_{\rmii{$S$}} &=
        \fluctO^{(0)}_{\phiField\phiField}
      + \newDimSix{\fluctO^{(1)}_{\phiField\phiField}}
    \,, \nonumber\\
    \fluctO^{\ell}_{\rmii{$X_T$}} &=
        \fluctO^{(0)}_{cc}
      + \newDimSix{\fluctO^{(1)}_{cc}}
    \,, &
    \fluctO^{\ell}_{\rmii{gh}} &=
        \fluctO^{(0)}_{\rmii{gh}}
      + \newDimSix{\fluctO^{(1)}_{\rmii{gh}}}
    \,.
\end{align}
Altogether, these are exactly the operators whose determinants
$\detXzero$, $\detS$,
$\detg$, $\detXT$, and $\detXG$ appear in the statistical part of the nucleation
rate,
eqs.~\eqref{eq:detS:full}
and~\eqref{eq:detV:full:explicit}.
Each partial-wave determinant is evaluated exactly with
the Gel'fand--Yaglom method,
and no derivative expansion is needed.

%
\section{Gel'fand--Yaglom theorem}
\label{app:Gel'fand--Yaglom}

Each partial-wave determinant can be related to an initial-value problem by
the Gel'fand--Yaglom~(Forman) theorem~\cite{%
Gelfand:1959nq,Forman:1987gha,Dunne:2006ct,Ekstedt:2023sqc}
\begin{equation}
  \label{eq:GY}
    \frac{
      \det\fluctO_1}{
      \det\fluctO_2} =
      \frac{\psi_1}{\psi_2}\Bigg|_{\partial\Omega}
    \,,
\end{equation}
where $\psi_i$ are
the solutions of the equations
$\fluctO_i\,\psi_i=0$, and
$\partial\Omega$ is the endpoint $r=R\to\infty$
of the radial interval $\Omega=[0,R]$.

In this section,
we investigate the effects of
higher-dimensional operators in
a fully decoupled sector.
We refer the reader to appendix~\ref{sec:app:WKB}
for the more general case involving coupled operators.
Higher-dimensional operators can modify
the kinetic part of the fluctuation operator,
which is therefore of
the more general Sturm--Liouville form~\cite{everitt2005catalogue,Kirsten:2004qv}
\begin{equation}
  \label{eq:sturm-liouville}
    \fluctO_\ell =
      - \frac{1}{r^2}\partial_r\!\left[r^2 K(r)\,\partial_r\right]
      + K(r)\,\frac{\ell(\ell+1)}{r^2}
      + U(r)
    \,,
\end{equation}
where the homogeneous solution $\psi(r)$ satisfies
\begin{equation}
    \fluctO_\ell\psi(r)=0
    \,.
\end{equation}
In the false vacuum the kinetic function is trivial, $K_{\fv}=1$,
whereas on the bounce $K(r)$ is non-trivial but switches off at large radius,
$K(r)\to1$ as $r\to\infty$.
In general, this turns out to be an issue since the Gel'fand--Yaglom theorem
applies when $\fluctO_1,\,\fluctO_2$ have the same kinetic factor (both for the
boundary and the bulk)~\cite{Kirsten:2004qv}. 
In the following, we will show, after a WKB subtraction,
that the Gel'fand--Yaglom theorem remains applicable even for non-trivial $K(r)$.

%
\subsection{Liouville transformation}

First, the operator is brought to the Schr\"odinger form by a change of variable
and rescaling~\cite{everitt2005catalogue}
\begin{align}
    s(r) &=
      \int_0^r\frac{{\rm d}r'}{\sqrt{K(r')}}
    \,,&
  \label{eq:liouville-map}
    X(s) &=
      r\,K(r)^{1/4}\,\psi(r)
    \,.
\end{align}
This map preserves the radial norm,
\begin{equation}
    \int_0^\infty {\rm d}r\,r^2\,|\psi(r)|^2 =
      \int_0^\infty {\rm d}s\,|X(s)|^2
    \,,
\end{equation}
so it is unitary~\cite{Everitt1982} and leaves the spectrum (and hence the determinant) unchanged.
The new homogeneous solution $X$ satisfies the Schr\"odinger equation
\begin{equation}
  \label{eq:schrodinger-form}
    \widetilde{\fluctO}_\ell^{}\,X =
      \left[-\partial_s^2+Q_\ell^{}(s)\right]X=0
    \,,
\end{equation}
to which the usual Gel'fand--Yaglom argument applies.

For the homogeneous equation,
the Liouville map is equivalent to the simpler
transformation
\begin{equation}
    u(r)=
      r\;K(r)^{1/2}\;\psi(r)
    \,,
\end{equation}
used in the following sections.
It keeps $r$ as the variable and only removes
the first-derivative term.
The two are related by 
\begin{align}
  \label{eq:liouvill_trafo_simple}
    \big[-\partial_s^2+Q_\ell^{}(s)\big]X =
      K(r)^{3/4}\big[-\partial_r^2+\mathcal V(r)\big]u
    \qquad {\rm with} \qquad 
    Q_\ell^{} =
        K\,\mathcal V
      - K^{3/4}\big(K^{1/4}\big)''\!\!.
\end{align}
Since $K(r)\to1$,
$X\stackrel{r\to\infty}{=}u\stackrel{r\to\infty}{=}r\psi$,
and
the large-$r$ behavior of
the Gel'fand--Yaglom ratio~\eqref{eq:GY}
is unchanged.

%
\subsection{Normalization and the finite factor}

The Gel'fand--Yaglom theorem expresses the determinant ratio
in terms of the \emph{regular} solutions of
eq.~\eqref{eq:schrodinger-form},
which must be normalized in the same way at the origin
for the bounce ($\bnc$) and the false vacuum ($\fv$). The standard normalization condition reads
\begin{equation}
\label{eq:standard_norm}
    \psi_\ell(r)\sim r^\ell \qquad (\text{standard normalization for}\ r\rightarrow0).
\end{equation} 
Using $s\simeq r/\sqrt{K(0)}$ near the origin, we obtain
\begin{equation}
    X_\ell(s)=
      r\,K^{1/4}\psi_\ell \stackrel{s\to0}{\simeq}
      K(0)^{(2\ell+3)/4}\,s^{\ell+1}
    \,.
\end{equation}
To apply the Gel'fand--Yaglom theorem,
we introduce the canonically normalized solution
\begin{align}
  \label{eq:xhat-norm}
    \widehat X_\ell(s) &\stackrel{s\to0}{\sim}s^{\ell+1}
    &&\text{(same normalization on false vacuum and bounce)}
    \,,
\end{align}
which is related to the Liouville-transformed solution via
\begin{equation}
  \label{eq:xhat-def}
    \widehat X_\ell(s)=
      K(0)^{-(2\ell+3)/4}\,X_\ell(s) =
      K(0)^{-(2\ell+3)/4}\,r\,K^{1/4}\,\psi_\ell(r)
    \,.
\end{equation}
The normalizations of the false-vacuum and bounce configurations differ, because in general $1=K_{\fv}(0)\neq K_{\bnc}(0)$ and fixing $\psi_\ell\sim r^\ell$
does not fix $\widehat X_\ell\sim s^{\ell+1}$.
In this work,
we adopt the standard normalization
in radial coordinates~\eqref{eq:standard_norm}
on both branches.
This is a matter of convention,
as it shifts the finite factors of
$K(0)$ between $\widehat X$ and $\psi$.
With our convention,
the boundary value of eq.~\eqref{eq:GY}
remains unchanged.

Applying the Gel'fand--Yaglom theorem to the canonically normalized solutions,
we find that the determinant ratio is the ratio of the $\widehat X_\ell$ at the outer boundary.
Each branch is evaluated at the endpoint
$s(R)$, where $R\to\infty$.
This means the operator domains and the boundaries of the two branches differ.
We use that~\cite{Forman:1987gha,Forman:1992abc}
\begin{equation}
    \det \widetilde{\fluctO}_\ell =
      \mathcal{C}\times X(\rm boundary)
    \,,
\end{equation}
where $\mathcal{C}$ does not depend on $X$ (but potentially on the boundary conditions).
Using that $K_{\bnc}(R)\to1$ and
$X_\ell\to r\psi_\ell$ on both branches, we find
\begin{equation}
  \label{eq:det-ratio}
    \frac{
      \det\fluctO_\ell^{\bnc}}{
      \det\fluctO_\ell^{\fv}} =
      \frac{\mathcal{C}^{\bnc}}{\mathcal{C}^{\fv}}\lim_{r\to\infty}\frac{
        \widehat X_\ell^{\bnc}(s(r))}{
        \widehat X_\ell^{\fv}(r)} =
      \frac{\mathcal{C}^{\bnc}}{\mathcal{C}^{\fv}}K(0)^{-(2\ell+3)/4}\,\lim_{r\to\infty}\frac{
        \psi_\ell^{\bnc}(r)}{
        \psi_\ell^{\fv}(r)} \equiv
      \mathcal{K}_\ell^{}\,T_\ell^{}(\infty)
    \,.
\end{equation}
Here, we employ the notation
$T_\ell^{}(\infty) \equiv \lim\limits_{r\to\infty}\frac{\psi_\ell^{\bnc}(r)}{\psi_\ell^{\fv}(r)}$,
and we explicitly keep potential corrections from the different boundaries
$\mathcal{C}^{\bnc}/\mathcal{C}^{\fv}$,
which cancel for equal boundary conditions.
Combined with the non-trivial kinetic coefficient at the origin, $r=0$,
this gives the finite factor $\mathcal K_\ell$ per partial wave.

%
\subsection{Renormalization}

We renormalize the fluctuation determinants by WKB subtraction.
Because $K(0)^{-(2\ell+3)/4}$ is a fixed normalization factor,
independent of the potential and of the wavefunction,
it appears identically in the full and the WKB determinants,
both normalized as in eq.~\eqref{eq:xhat-norm}, and cancels \emph{term by term} in
$\ln\Gamma$.
For each partial wave,
\begin{align}
  \label{eq:app:lnT:renorm}
      \ln\!\Bigl[\mathcal{K}_\ell^{}\,T_\ell^{}(\infty)\Bigr]\Big|_{\text{full}}
    - \ln\!\Bigl[\mathcal{K}_\ell^{}\,T_\ell^{}(\infty)\Bigr]\Big|_{\rmii{WKB}} &=
        \ln T_\ell^{\text{full}}(\infty)
      - \ln T_\ell^{\rmii{WKB}}(\infty)
    \,.
\end{align}
The boundary factor therefore never enters the renormalized determinant.
Its cancellation is also necessary.
Otherwise, the sum over partial waves would generate
\begin{equation}
    \sum_\ell (2\ell+1)\,\frac{2\ell+3}{4}\,\ln K(0)\ \sim
      \ L_{\rm cut}^{3}\,\ln K(0)
    \,,
\end{equation}
a spurious, cutoff-dependent contribution with no associated counterterm.

%
\subsection{Different viewpoint}

Another perspective on the validity of the Gel'fand--Yaglom theorem follows from computing
the following ratios of functional determinants instead
\begin{equation}
  \label{eq_:ratios_shuffled}
    \frac{
      \det\fluctO_\ell^{\bnc}}{
      \det\fluctO_\ell^{\fv}}\Bigg|_{\text{full}}
    \times \frac{
      \det\fluctO_\ell^{\fv}}{
      \det\fluctO_\ell^{\bnc}}\Bigg|_{\rmii{WKB}} =
      \frac{
        \det\fluctO_\ell^{\bnc,\text{full}}}{
        \det\fluctO_\ell^{\bnc,\rmii{WKB}}}
      \times \frac{
        \det\fluctO_\ell^{\fv,\rmii{WKB}}}{
        \det\fluctO_\ell^{\fv,\text{full}}}
    \,,
\end{equation}
with $\fluctO_\ell^{\rmii{WKB}}$ the corresponding operator from the WKB expansion,
which has the same kinetic factor as $\fluctO_\ell^{\text{full}}$.
For generic operators in the Sturm--Liouville form
\begin{equation}
    \fluctO_i=
        -\frac{{\rm d}}{{\rm d}x}\left(K_i(x)\frac{{\rm d}}{{\rm d}x}\right)
      + U_i(x)
    \,,
\end{equation}
it was shown~\cite{Kirsten:2004qv} that
\begin{equation}
    \frac{
      \det\fluctO_1}{
      \det\fluctO_2} =
      \frac{\psi_1(r)}{\psi_2(r)}
    \,,
\end{equation}
as long as $K_1=K_2$.
When comparing the false-vacuum to the bounce operator, $K_1\neq K_2$
and the theorem cannot be applied.
We are, however, interested in the renormalized fluctuation determinants.
For that, one needs to subtract the leading-order large-$\ell$ behavior to cancel
ultraviolet (UV) divergences and render them finite.
This large-$\ell$ limit is described by the WKB approximation of the wavefunction,
but can also be written in terms of operators directly.
Generic radial operators considered here can be written as%
\footnote{%
  Following~\cite{Langer:1937qr}, we set
  $\bar L \equiv \ell+\tfrac12$ and
  $L^2\to \bar L^2$,
  which improves the WKB expansion near the origin.
}
\begin{align}
    \fluctO_i^\ell &=
      - \frac{1}{r^2}\partial_r^{}\!\left[r^2 K_i^{}(r)\partial_r^{}\right]
      + K_i^{}(r)\frac{L^2}{r^2}
      + U_i^{}(r)
  \label{eq:app:KU}
    \,.
\end{align}
Rotational invariance of all operators forces the same $K_i(r)$ to multiply the full
Laplacian
$-\Box=-\tfrac{1}{r^2}\partial_r(r^2\partial_r)+\tfrac{\bar L^2}{r^2}$.
Using the transformation of eq.~\eqref{eq:liouvill_trafo_simple},
the equation then takes the standard form
\begin{equation}
    - u_i''(r)
    + \mathcal V_i^{}(r)u_i^{}(r)
    = 0
    \,,
\end{equation}
with
\begin{equation}
    \mathcal V_i^{}(r) =
        \frac{L^2}{r^2}
      + \frac{U_i^{}(r)}{K_i^{}(r)}
      + \frac{K_i''(r)}{2K_i^{}(r)}
      + \frac{K_i'(r)}{rK_i^{}(r)}
      - \frac{K_i'(r)^2}{4K_i^{}(r)^2} \equiv
        \frac{\bar L^2}{r^2}
      + \mathcal{M}(r)
    \,.
  \label{eq:VeffWKB}
\end{equation}
The leading WKB solutions are
\begin{equation}
  \label{eq:WKBwavefunction}
    \psi_{i,\rmii{WKB}}^\pm(r) =
      \frac{
        \sqrt{K_i^{}(0)}\,\sqrt{\bar L}}{
        r\sqrt{K_i^{}(r)}\,\mathcal V_i^{}(r)^{1/4}}
      \exp\left[\pm\int^r {\rm d}s\,\sqrt{\mathcal V_i^{}(s)}\right]
    \,,
\end{equation}
normalized such that
$\psi_{\rmii{WKB}}\stackrel{r\to 0}{=} r^\ell$.
More generally, for any prescribed non-zero radial wavefunction $\psi(r)$, an operator
with the same kinetic structure for which $\psi$ is an exact
zero mode can be written as (here for $\psi=\psi_{\rmii{WKB}}$)
\begin{equation}
    \widetilde{\fluctO}_{i,\rmii{WKB}}^\ell =
      - \frac{1}{r^2}\partial_r^{}\!\left[r^2K_i^{}(r)\partial_r^{}\right]
      + \frac{1}{r^2\psi_{\rmii{WKB}}^{}(r)}\partial_r^{}\!\left[r^2K_i^{}(r)\psi_{\rmii{WKB}}'(r)\right]
    \,.
\end{equation}
This proves that now both ratios in eq.~\eqref{eq_:ratios_shuffled}
have exactly the same kinetic modification
$K_1=K_2$.
The generalized
Sturm--Liouville Gel'fand--Yaglom theorem~\cite{Kirsten:2004qv}
is therefore applicable and yields
\begin{equation}
    \frac{
      \psi_\ell^{\bnc,\text{full}}(\infty)}{
      \psi_\ell^{\bnc,\rmii{WKB}}(\infty)}
    \times \frac{
      \psi_\ell^{\fv,\rmii{WKB}}(\infty)}{
      \psi_\ell^{\fv,\text{full}}(\infty)} =
      T_\ell^{\text{full}}(\infty)/T_\ell^{\rmii{WKB}}(\infty)
    \,,
\end{equation}
in agreement with eq.~\eqref{eq:app:lnT:renorm}.

The reorganization used in
this proof agrees with the intuitive picture presented in~\cite{Fadakar:2026tfo}.
For operators with the same asymptotic behavior,
the UV dependence of the normalization constant cancels in the determinant ratio, leaving only the low-energy sensitivity.
Here, however, the higher-dimensional operators change the UV and hence the asymptotic behavior.
This UV sensitivity is canceled by the respective WKB operator,
which intuitively explains the reshuffling in eq.~\eqref{eq_:ratios_shuffled}.

%
\section{WKB approximation}
\label{sec:app:WKB}

In the previous section,
we found that 
\begin{align}
  \label{eq:app:lnpsi}
    \ln \psi^{\rmii{WKB}}_\ell(\infty) =
        \frac{1}{2}\ln K(0)
      + \int_0^\infty\!{\rm d}r\, \sqrt{\mathcal{V}(r)}
    \,,
\end{align}
as $K(r)\stackrel{r\rightarrow\infty}{=} 1 $
and up to normalization terms
that cancel in the bounce/false-vacuum ratio.
The determinant ratio is controlled by
\begin{align}
    \Omega_\ell^{} \equiv
      \ln \frac{
        \psi_\ell^{\bnc}(\infty)}{
        \psi_\ell^{\fv}(\infty)}
    \,.
\end{align}

%
\subsection{Series expansion}

Expanding eq.~\eqref{eq:VeffWKB} at large $\bar L$,
\begin{align}
    \sqrt{\mathcal{V}(r)} =
        \frac{\bar L}{r}
      + \frac{r\,\mathcal{M}(r)}{2\bar L}
      - \frac{r^3\mathcal{M}(r)^2}{8\bar L^3}
      + \frac{r^5\mathcal{M}(r)^3}{16\bar L^5}
      + \mathcal{O}(\bar L^{-7})
    \,,
\end{align}
and subtracting the false vacuum, the first coefficients of $\Omega_\ell$ are
\begin{align}
  \label{eq:app:Omega:series}
    \Omega_\ell^{} \simeq
        \frac{1}{2}\ln{\frac{K_{\bnc}(0)}{K_{\fv}(0)}}+\frac{1}{2\bar L}\int_0^\infty {\rm d}r\, r\,\left[\mathcal{M}_{\bnc}^{}(r)-\mathcal{M}_{\fv}^{}(r)\right]
      - \frac{1}{8\bar L^3}\int_0^\infty {\rm d}r\, r^3\!\left[\mathcal{M}_{\bnc}^2-\mathcal{M}_{\fv}^2\right]
    \,.
\end{align}
The first two terms are subtracted to cancel UV divergences.
The subleading terms collect higher orders and can be used to estimate the tail contribution in
the partial-wave expansion beyond a given angular-momentum cutoff $L_{\rm cut}$.

%
\subsection{Corrected WKB masses}

Treating the dimension-six corrections as perturbations, the NLO correction of the effective WKB mass of eq.~\eqref{eq:VeffWKB} becomes
(evaluated on the bounce, $\phibg=\phib$, for $\mathcal M_{\bnc}$
and in the false vacuum, $\phibg=\phiF=0$, for $\mathcal M_{\fv}$,
where all dimension-six corrections vanish)
\begin{align}
  \label{eq:app:MWKB}
    \mathcal{M}_i^\rmii{WKB} &=
        \mathcal{M}_{i}^{(0)}
      + \delta\mathcal{M}_{i}^{ }
    \,,&
    \delta\mathcal{M}_i^{} &=
        \delta U_i^{}
      - \delta K_i^{}\,\mathcal{M}_i^{(0)}
      + \frac{(\delta K_i^{})'}{r}
      + \frac{(\delta K_i^{})''}{2}
    \,.
\end{align}
The zeroth-order masses are
\begin{align}
    \mathcal{M}_{\rmii{$X_0$}}^{(0)} &=
        \mD^2
      + h_3^{}\phibg^2
    \,,&
    \mathcal{M}_{\rmii{$S$}}^{(0)} &=
      V''\!\left(\phibg\right)
    \,,\nn
    \mathcal{M}_\chi^{(0)} &=
        \frac{V'(\phibg)}{\phibg}
      + g^2\phibg^2
    \,,&
    \mathcal{M}_{\rmii{$A$}}^{(0)} = \mathcal{M}_{\rmii{$B$}}^{(0)} = 
    \mathcal{M}_c^{(0)} =
    \mathcal{M}_{\rmii{gh}}^{(0)}&= g^2\phibg^2
    \,,
\end{align}
and the dimension-six kinetic terms $\delta K_i$ are
\begin{align}
    K_{\rmii{$X_0$}}^{}
    =K_{\rmii{gh}}^{}
    &= 1
    \,,&
    K_{\rmii{$S$}}^{} = K_\chi^{} &=
        1
      + \frac12 \alpha_{D^2\phi^4}^{}\phibg^2
    \,,&
    K_c^{} &=
        1
      + 2\alpha_{\phi^2F^2}^{}\phibg^2
    \,,
\end{align}
and $\delta U_i$,
\begin{align}
    \delta U_{\rmii{$X_0$}}^{} &=
        \frac12 \alpha_{\phi^4B_0^2}^{} \phibg^4
      + \alpha_{D^2\phi^2B_0^2}^{} \phibg'^2
    \,,&
    \delta U_c^{} &=
        \frac12 g^2\alpha_{D^2\phi^4}^{}\phibg^4
      - 4\alpha_{\phi^2F^2}^{}\frac{\phibg\phibg'}{r}
    \,,\nn
    \delta U_\chi^{} &=
      \alpha_{D^2\phi^4}^{}\Bigl[
          g^2\phibg^4
        + \frac12\phibg'^2
      \Bigr]
      + \frac34 \alpha_{\phi^6}^{}\phibg^4
    \,,&
    \delta U_\rmii{gh}^{}&=
      \frac{1}{2}g^2\alpha_{D^2\phi^4}^{}\phibg^4
    \,,\nn
    \delta U_{\rmii{$S$}}^{} &=
      \alpha_{D^2\phi^4}^{}\Bigl[
        - \frac{2\phibg\phibg'}{r}
        - \frac12\phibg'^2
        - \phibg\phibg''
      \Bigr]
      + \frac{15}{4}\alpha_{\phi^6}^{}\phibg^4
    \,.
\end{align}
Thus,
the overall dimension-six contributions to the WKB masses
take the following form:
\begin{align}
    \delta\mathcal M_{\rmii{$X_0$}}^{} &=
        \frac12 \alpha_{\phi^4B_0^2}^{} \phibg^4
      + \alpha_{D^2\phi^2B_0^2}^{} \phibg'^2
    \,,\nn
    \delta\mathcal M_{\rmii{$S$}}^{} &=
        \frac{15}{4}\alpha_{\phi^6}^{}\phibg^4
      - \alpha_{D^2\phi^4}^{}\left[
          \frac{\phibg\phibg'}{r}
        + \frac12\phibg\phibg''
        + \frac12\phibg^2\mathcal M_{\rmii{$S$}}^{(0)}
      \right]
    \,,\nn
    \delta\mathcal M_\chi^{} &=
        \frac34\alpha_{\phi^6}^{}\phibg^4
      + \alpha_{D^2\phi^4}^{}\left[
          g^2\phibg^4
        - \frac12\phibg^2\mathcal M_\chi^{(0)}
        + \frac{\phibg\phibg'}{r}
        + \phibg'^2
        + \frac12\phibg\phibg''
      \right]
    \,,\nn
    \delta\mathcal M_c^{} &=
        \frac12 g^2\alpha_{D^2\phi^4}^{}\phibg^4
      - 2\alpha_{\phi^2F^2}^{}g^2\phibg^4
      + 2\alpha_{\phi^2F^2}^{}\phibg'^2
      + 2\alpha_{\phi^2F^2}^{}\phibg\phibg''
    \,,\nn[1mm]
    \delta\mathcal M_\rmii{gh}^{}&=
      \frac{1}{2} g^2\alpha_{D^2\phi^4}^{}\phibg^4
    \,.
\end{align}

%
\subsection{Vector block and rotated WKB basis}

For the mixed determinant,
the radial $(a,b)$ gauge modes are not WKB
eigenchannels by themselves and must be diagonalized first.
At zeroth order in
the Wilson coefficients, and
before the Goldstone mixing,
the $(a,b)$ block is
the vector Laplacian with mass $\WX^{}=g^2\phibg^2$.
At leading order in $L$ (cf.\ eq.~\eqref{eq:app:dell}),
the WKB solution is
$\psi_{\rmii{WKB}}(r)=r^\ell + \mathcal{O}(r^n/L)$,
and the block becomes
\begin{align}
    \fluctO^{(0)}_{(a,b)}=
      \frac{1}{r^2}\begin{pmatrix}
          L^2 & -2L\\
          -2L & L^2+2
      \end{pmatrix}
      + \mathcal{O}(r^n/L)
    \,.
\end{align}
The corresponding eigenvectors are
\begin{align}
    e_{\rmii{$A$}}&=
      \begin{pmatrix}\sqrt{w_{\rmii{$A$}}}\\ \sqrt{w_{\rmii{$B$}}}\end{pmatrix}
    \,,&
    e_{\rmii{$B$}}&=
      \begin{pmatrix}\hphantom{+}\sqrt{w_{\rmii{$B$}}}\\ -\sqrt{w_{\rmii{$A$}}}\end{pmatrix}
    \,,&
    w_{\rmii{$A$}}&=
      \frac{\ell+1}{2\ell+1}
    \,,&
    w_{\rmii{$B$}}&=
      \frac{\ell}{2\ell+1}
    \,,
\end{align}
with eigenvalues $L_i^2/r^2$, where
$L_i^2\equiv\ell_i(\ell_i+1)$ with
$\ell_{\rmii{$A$}}=\ell-1$ and $\ell_{\rmii{$B$}}=\ell+1$.
The block thus splits into two channels with
angular momenta $\ell_{\rmii{$A$}}$ and $\ell_{\rmii{$B$}}$.%
\footnote{%
  This modification induces
  a new term in the WKB expansion at NLO.
  The splitting of the respective piece in
  eq.~\eqref{eq:app:Omega:series},
  $2\bar L\bigl[\frac{1}{2(\bar L-1)}+\frac{1}{2(\bar L+1)}\bigr]=2+\tfrac{2}{\bar L^2}+\mathcal{O}(\bar L^{-4})$,
  acquires subleading corrections in the partial-wave expansion of the tail. For the coupled block, eq.~\eqref{eq:app:Omega:series} holds with traces over $(2\ell+1)\Omega_\ell\simeq\int{\rm d}r\,r\,{\rm tr}\,\Delta\mathcal M-\frac{1}{4\bar L^2}\int{\rm d}r\,r^3\,{\rm tr}\,\Delta\mathcal M^2$. The Goldstone--gauge mixing enters through the off-diagonal elements in ${\rm tr}\,\mathcal M^2$ and thus contributes to the tail at $\mathcal{O}(\bar L^{-2})$.

}

The dimension-six terms are treated in standard first-order perturbation
theory (for the explicit expressions, see eq.~\eqref{eq:app:ops:SA}):
\begin{align}
    \fluctO^{(1)}_n\equiv
      \langle n|\fluctO^{(1)}|n\rangle=
      e_n^T\,\fluctO^{(1)}\,e_n^{}\,,\qquad n=A,B
    \,.
\end{align}
The antisymmetric couplings
$\pm 2L\alpha_{\phi^2F^2}\,\mathcal D_{\phibg^2}$,
with the anti-self-adjoint operator $\mathcal D_G$ of eq.~\eqref{eq:app:DG} for
$G=\phibg^2$,
cancel in the symmetric trace,
resulting in 
\begin{align}
    K_{\rmii{$A$}}^{}&=
        1
      + 2\alpha_{\phi^2F^2}^{}\,w_{\rmii{$A$}}^{}\phibg^2
    \,,&
    \delta U_{\rmii{$A$}}^{}&=
        \tfrac12g^2\alpha_{D^2\phi^4}^{}\phibg^4
      + 4\alpha_{\phi^2F^2}^{}\,w_{\rmii{$A$}}^{}(\ell-1)\frac{\phibg\phibg'}{r}
    \,,\\
    K_{\rmii{$B$}}^{}&=
        1
      + 2\alpha_{\phi^2F^2}^{}\,w_{\rmii{$B$}}^{}\phibg^2
    \,,&
    \delta U_{\rmii{$B$}}^{}&=
        \tfrac12g^2\alpha_{D^2\phi^4}^{}\phibg^4
      - 4\alpha_{\phi^2F^2}^{}\,w_{\rmii{$B$}}^{}(\ell+2)\frac{\phibg\phibg'}{r}
    \,.
\end{align}
This gives
\begin{align}
  \label{eq:app:Mchannels}
    \delta\mathcal M_{\rmii{$A$}}^{}&=
        \tfrac12g^2\alpha_{D^2\phi^4}^{}\phibg^4
      + 2\alpha_{\phi^2F^2}^{}\biggl[
          w_{\rmii{$A$}}^{}\bigl(
            - \WX^{}\phibg^2
            + \phibg'^2
            + \phibg\phibg''
          \bigr)
        + \frac{2\ell(\ell+1)}{2\ell+1}\,\frac{\phibg\phibg'}{r}
      \biggr]
    \,,\\
    \delta\mathcal M_{\rmii{$B$}}^{}&=
        \tfrac12g^2\alpha_{D^2\phi^4}^{}\phibg^4
      + 2\alpha_{\phi^2F^2}^{}\biggl[
          w_{\rmii{$B$}}^{}\bigl(
            - \WX^{}\phibg^2
            + \phibg'^2
            + \phibg\phibg''
          \bigr)
        - \frac{2\ell(\ell+1)}{2\ell+1}\,\frac{\phibg\phibg'}{r}
      \biggr]
    \,.
\end{align}
The last terms of $\delta\mathcal M_{\rmii{$A$}}$ and $\delta\mathcal M_{\rmii{$B$}}$ are
$\mathcal O(L)$ with opposite signs, so their sum is $\mathcal O(L^0)$. This surviving
piece must be kept in the leading-order WKB subtraction in order to fully cancel all
emerging UV divergences in the full solution. Beyond it, the operator shifts the
already-convergent $\mathcal O(L^{-2})$ and $\mathcal O(L^{-3})$ terms, by a relative
amount set by $L_{\rm cut}\,\alpha_{\phi^2F^2}\phib^2(0)\ll1$. We checked that our code reproduces these findings and converges for increasing
$L_{\rm cut}\in[0,50]$.

%
\section{Zero modes}
\label{sec:zero:modes}

The bounce breaks the translations and the global ${\rm U}(1)$ of the false vacuum.
This gives three translational zero modes in the $\ell=1$ Higgs sector,
$\psi_0^{ }\propto\phib'$, and one phase zero mode in the $\ell=0$ Goldstone sector,
$\psi_0^{ }\propto\phib$.%
\footnote{%
  This is only true up to gauge-fixing terms.
  These, however, are unphysical and
  get canceled by
  the ghost contribution,
  such that $\phib$ is an exact zero mode.
  Another way to see this
  is to treat all partial waves independently and
  to go to Landau gauge for the zero-mode contribution,
  which eliminates all gauge-fixing contributions trivially.
}
The corresponding operators are
$\fluctO^{1}_{\rmii{$S$}}$ of eq.~\eqref{eq:app:sector:ops} and
the Goldstone operator, i.e.\
$\fluctO_{\chi\chi}$ at $\ell=0$ without the
gauge-fixing terms (which are canceled by the ghost determinant),
\begin{equation}
 \mathcal O_{\rmii{G}}
  =-\frac{1}{r^2}\partial_r\!\left[r^2\newDimSix{K}\partial_r\right]
  +\frac{V'(\phibg)}{\phibg}
  +\newDimSix{\frac34\alpha_{\phi^6}\phibg^4+\frac{\alpha_{D^2\phi^4}}{2}\phibg'^2}
  \,,\qquad
  \newDimSix{K}=1+\newDimSix{\tfrac12\alpha_{D^2\phi^4}\phibg^2}
  \,.
\end{equation}
Both are of the form $\fluctO=-r^{-2}\partial_r(r^2K\partial_r)+U(r)$, and the bounce
equation~\eqref{eq:app:eom:radial} and its derivative give
$\mathcal O_{\rmii{G}}\phib=\fluctO^{1}_{\rmii{$S$}}\phib'=0$.

The zero eigenvalue is removed as
${\rm det}'\fluctO=\lim_{\varepsilon\to0}\varepsilon^{-1}\det(\fluctO+\varepsilon)$.
Projecting $(\fluctO+\varepsilon)y^\varepsilon=0$ onto $\psi_0$, with both normalized
equally at the origin, and integrating by parts gives~\cite{Ekstedt:2021kyx,Andreassen:2017rzq}
\begin{align}
  \frac{{\rm det}'\fluctO(\phib)}{\det\fluctO(\phiF)} &=
  - \frac{\int_0^\infty{\rm d}r\,r^2\psi_0^2}
  {R^2\newDimSix{K(R)}\bigl[\psi_0'y_{\fv}-\psi_0y_{\fv}'\bigr]_{r=R}}
  \,,&
  R\to\infty
  \,.
\end{align}
The dimension-six operators enter only through
$\newDimSix{K(R)}\to1$ and through the
bounce itself.
The divergent normalization factors
$\mathcal{C}^{\bnc/\fv}$ cancel against
the respective WKB piece and
are therefore not written explicitly.
With $\phib\simeq\phibg_\infty e^{-\mF r}/r$
one recovers the standard
results~\cite{Ekstedt:2023sqc},
\begin{align}
  \frac{{\rm det}'\mathcal O_{\rmii{G}}(\phib)}{\det\mathcal O_{\rmii{G}}(\phiF)}
  &=\frac{\int_0^\infty{\rm d}r\,r^2\phib^2}{\phi_\infty\phib(0)}
  \,,&
  \frac{{\rm det}'\fluctO^{1}_{\rmii{$S$}}(\phib)}{\det \fluctO^{1}_{\rmii{$S$}}(\phiF)}
  &=\frac{\int_0^\infty{\rm d}r\,r^2\phib'^2}{-3\phi_\infty\phib''(0)}
  \,.
\end{align}
Combined with the Jacobian of the collective coordinates, this yields the factors
$\sqrt{2\phibg_\infty\phib(0)}$ and $\bigl(-2\phibg_\infty\phib''(0)\bigr)^{3/2}$ of
eqs.~\eqref{eq:zero:modes:G} and~\eqref{eq:zero:modes:tr},
which are structurally equivalent to the dimension-four case,
but evaluated on the full bounce. Like all other partial waves, the zero modes are renormalized with the respective WKB subtraction of eq.~\eqref{eq:app:Omega:series}, including the finite $K(0)$ factors.

%
\section{Field-redefinition invariance}
\label{sec:app:redef}

%
\subsection{General idea}

In this appendix, we illustrate the invariance of the tunneling rate under field redefinitions with a simple example.
For simplicity, we restrict ourselves to redefinitions that involve only the fields and not their derivatives, namely $\phi\to\mathcal{F}(\phi)$. We expect the argument to extend to more general redefinitions of the form $\phi\to\mathcal{F}(\phi,\partial\phi,\partial^2\phi,\dots)$.
Field-redefinition invariance holds order by order in a strict perturbative expansion, analogously to what was shown for the bounce action in ref.~\cite{Chala:2024xll}. Here we work to first order. We stress that the computation in the main text does not rely on such a strict expansion, but instead solves for the bounce directly with the higher-dimensional operators included. The two approaches differ only by terms that are formally of higher order in the strict perturbative expansion, and these lie within the theoretical uncertainties of our calculation.

We start from the Lagrangian
\begin{equation}
  \label{eq:lag1_fri}
    \mathcal{L} =
      - \frac{1}{2}K(\PhiField_1)\,\PhiField_1\Box\PhiField_1
      + V(\PhiField_1)
    \,,
\end{equation}
where $K(\PhiField)=1+\epsilon\,\delta K(\PhiField)$.
To $\mathcal{O}(\epsilon)$, the non-canonical kinetic term can be brought into canonical
form by performing the field redefinition
$\PhiField_1= \PhiField_2- \frac{1}{2}\epsilon\,\delta K(\PhiField_2)\,\PhiField_2$.
Substituting this relation into eq.~\eqref{eq:lag1_fri}
and expanding to first order in $\epsilon$, we obtain
\begin{equation}
    \mathcal{L} =
      - \frac{1}{2}\PhiField_2\Box\PhiField_2
      + \underbrace{
          V(\PhiField_2)
        - \frac{1}{2}\epsilon\,
          \delta K(\PhiField_2)\PhiField_2 V'(\PhiField_2)
      }_{\displaystyle \widetilde{V}(\PhiField_2)}
      + \mathcal{O}(\epsilon^2)
    \,.
\end{equation}
The fluctuation operator of this Lagrangian,
expanded around the background,
$\PhiField_2^{}=\phibg_{\bnc,2}^{}+\phiField$, is
$\fluctO_2=-\Box+\widetilde{V}''(\phibg_{\bnc,2})$.
On the other hand, expanding the original Lagrangian in
eq.~\eqref{eq:lag1_fri} around the background $\phibg_{\bnc,1}$
gives the following fluctuation operator:
\begin{equation}
    \fluctO_1^{}=
      - (K_1^{}+\phibg_{\bnc,1}^{}\,K'_1)\Box
      + V_1''
      - K'_1\Box\phibg_{\bnc,1}^{}
      - \frac{1}{2}K''_1\phibg_{\bnc,1}^{}\Box\phibg_{\bnc,1}^{}
    \,,
\end{equation}
where we have defined for simplicity
$K_1\equiv K(\phibg_{\bnc,1}),\,V_1\equiv V(\phibg_{\bnc,1})$.
Using the EOM of the background
$\Box\phibg_{\bnc,1}^{}=V'_1+\mathcal{O}(\epsilon)$ together with
$\phibg_{\bnc,1}=\phibg_{\bnc,2}-\frac{1}{2}\epsilon\,\delta K_2\,\phibg_{\bnc,2}$, and
defining $Z\equiv K(\PhiField)+\PhiField\,K'(\PhiField)$, one finds
\begin{equation}
    \fluctO_1^{}=
        Z_2^{}\,(-\Box+\widetilde{V}''_2)
      + \mathcal{O}(\epsilon^2)=
        Z_2^{}\,\fluctO_2^{}
      + \mathcal{O}(\epsilon^2)
    \,.
\end{equation}
Therefore,
$\det\,\fluctO_1=\det\,Z_2\,\det\,\fluctO_2$.
The $\det\,Z_2$ term accounts for the determinant of
the Jacobian in the path-integral measure since
\begin{equation}
    J(x,y)\equiv
      \frac{
        \delta\PhiField_1^{}(x)}{
        \delta\PhiField_2^{}(y)}=
      \delta(x-y)\left(
          1
        - \frac{1}{2}\epsilon\,\delta K'_2\PhiField_2^{}
        - \frac{1}{2}\epsilon\,\delta K_2^{}
      \right)
    \,,
\end{equation}
therefore 
\begin{eqnarray}
    \det\,J&=&
      \exp\left[
          \delta^d(0)\int_{\vec{x}}\ln\left(
              1
            - \frac{1}{2}\epsilon\,\delta K'_2\PhiField_2^{}
            - \frac{1}{2}\epsilon\,\delta K_2^{}
          \right)
      \right]
    \nn
    &=&
      \exp\left[\delta^d(0)\int_{\vec{x}}\ln\left(1-\frac{1}{2}\delta Z_2^{}\right)\right]=
        \frac{1}{\sqrt{\det\,Z_2^{}}}
      + \mathcal{O}(\epsilon^2)
    \,.
\end{eqnarray}
Indeed, consider the path integral
\begin{equation}
    \mathcal{Z}_{\PhiField_1} =
      \int \mathcal{D}\PhiField_1\,
      \exp\left[-S_1[\PhiField_1]\right] \simeq
      \exp\left[-S[\phibg_{\bnc,1}]\right]
      \frac{1}{\sqrt{\det\,\fluctO_1}}
    \,,
\end{equation}
where in the last step we have expanded the action around
the saddle point $\phibg_{\bnc,1}$.
Upon performing the field redefinition $\PhiField_1=\PhiField_1(\PhiField_2)$,
the path integral becomes
\begin{align}
    \mathcal{Z}_{\PhiField_1}=
      \int \mathcal{D}\PhiField_2\,
      \det\,J\,
      \exp\left[-S_2[\PhiField_2]\right] &\simeq
      \exp\left[-S_2[\phibg_{\bnc,2}]\right]
      \frac{
        \det\,J}{
        \sqrt{\det\,\fluctO_2}}
    \nn &=
      \exp\left[-S_1[\phibg_{\bnc,1}]\right]
      \frac{1}{\sqrt{\det\,\fluctO_1}}
    \,,
\end{align}
where we have used the invariance of the action evaluated on the background under field
redefinitions~\cite{Chala:2024xll}.
Since the tunneling rate is proportional to the imaginary part of $\mathcal{Z}_{\PhiField}$,
and since the latter is invariant under field redefinitions, we conclude
that the nucleation rate is also invariant under such redefinitions, as expected. 
The computation turns out to be even simpler
in dimensional regularization
with $d=3-2\varepsilon$ spatial dimensions,
in which the scaleless integral
$\delta^d(0)=\int\!\frac{{\rm d}^dp}{(2\pi)^d}$ vanishes,
$\delta^d(0)\stackrel{d=3-2\varepsilon}{=}0$, yielding
$\det\,J\stackrel{d=3-2\varepsilon}{=}1$ and
$\det\,Z_2\stackrel{d=3-2\varepsilon}{=}1$.

%
\subsection{Explicit example}

As a consistency check on the fluctuation-determinant setup proposed in this
work, we show explicitly how field-redefinition invariance is realized within the
framework of the Gel'fand--Yaglom theorem.
To keep
the argument transparent we strip away all gauge and fermionic fields and work
purely in the scalar sector,
treating $\PhiField$ as a single real field, with
the leading Lagrangian
\begin{equation}
  \label{eq:L_toy}
    \mathcal L =
        \frac{1}{2}(\partial\PhiField)^2
      + V(\PhiField)
    \,.
\end{equation}
The redundancy of the scalar dimension-six operators can be exposed by the simplest
field redefinition,
\begin{equation}
  \label{eq:app:fieldredef}
    \tilde\PhiField =
        \PhiField
      + \epsilon\,\PhiField^3
    \,,
\end{equation}
where $\PhiField$ ($\tilde\PhiField$) denotes the field in
the \textit{new} (\textit{old}) basis.
This redefinition generates Wilson coefficients of the order $\mathcal{O}(\epsilon)$.
Substituting back into the Lagrangian and
keeping only terms linear in
$\epsilon$ generates the two scalar operators of
eq.~\eqref{eq:Lag3d:dim6},
\begin{equation}
  \label{eq:app:redef:ops}
    \delta\mathcal{L} =
        3\epsilon\,\PhiField^2(\partial\PhiField)^2
      + \epsilon\,\PhiField^3 V'(\PhiField)
    \,.
\end{equation}
No operators with additional derivatives are produced,
yielding the Sturm--Liouville form of
eq.~\eqref{eq:sturm-liouville}.
The redefinition~\eqref{eq:app:fieldredef}
also moves the classical solution.
With $\tilde\phib$ denoting
the bounce of the original potential
of eq.~\eqref{eq:L_toy},
the bounce in the new basis reads,
at leading order in $\epsilon$ ($\epsilon\,\tilde\PhiField^3\simeq\epsilon\,\PhiField^3$),
\begin{equation}
  \label{eq:app:bounceshift}
    \phib =
        \tilde\phib
      - \epsilon\,\tilde\phib^3
    \,.
\end{equation}
Writing $\PhiField=\phib+\phiField$ and expanding to quadratic order, one obtains
\begin{equation}
  \label{eq:app:redef:quad}
    \delta\mathcal{L}^{(2)} =
        3\epsilon\,\phib^2\,(\partial\phiField)^2
      + 3\epsilon\,\phib'^2\,\phiField^2
      + 12\epsilon\,\phib^{}\phib'\,\phiField\,\partial\phiField
      + \tfrac12\epsilon\,(\phib^3V')''\,\phiField^2
    \,.
\end{equation}
The first term is a kinetic modification with 
$\delta K_{\rmii{$S$}}=6\epsilon\,\phib^2$, while
the second and third terms act as corrections to the mass term.
The latter follows from the IBP reduction
$12\epsilon\phib^{}\phib'\,\phiField\,\partial\phiField\xrightarrow{\rm IBP}
-6\epsilon(\phib'^2+\phib^{}\Box\phib^{})\,\phiField^2$.
Combining this with the potential term and
with the bounce-induced shift of
the leading-order mass term, and using
the bounce equation ($\Box \phib = V'(\phib)$),
yields
\begin{align}
    \delta U_{\rmii{$S$}}^{} &=
      - 6\epsilon\phib'^2
      - 12\epsilon\phib^{}\Box\phib^{}
      + \epsilon\big(6\phib^{} V'+6\phib^2V''+\phib^3V'''\big)
      + \big(V''(\phib^{})-V''(\tilde\phib^{})\big)
    \nn &=
      6\epsilon\,\big(\phib^2 V'' - \phib^{} V' - \phib'^2\big)
    \,.
\end{align}
Inserting into the WKB mass~\eqref{eq:app:MWKB} gives
\begin{equation}
  \label{eq:app:redef:dM}
    \delta\mathcal{M}_{\rmii{$S$}}^{} =
        \delta U_{\rmii{$S$}}^{}
      - \delta K_{\rmii{$S$}}^{}\,\mathcal{M}_{\rmii{$S$}}^{(0)}
      + \frac{(\delta K_{\rmii{$S$}}^{})'}{r}
      + \frac{(\delta K_{\rmii{$S$}}^{})''}{2} =
      6\epsilon\,\phib^{}\!\left(
          \phib''
        + \frac{2}{r}\phib'
        - V'(\phib^{})
      \right) =
      0
    \,,
\end{equation}
which vanishes again under the bounce equation.
Thus $\mathcal{V}_{\rmii{$S$}}(r)$, and
with it the fluctuation determinant
(see the canonically normalized fluctuation operator in eq.~\eqref{eq:liouvill_trafo_simple}),
is invariant
under the field redefinition~\eqref{eq:app:fieldredef}
at linear order in $\epsilon$.
It only shifts
$\alpha_{D^2\phi^4}$ and $\alpha_{\phi^6}$
along the direction generated by eq.~\eqref{eq:app:redef:ops},
leaving the physical WKB potential unchanged.

%
\section{Semi-analytical computation of the bounce action and escape point}
\label{sec:app:semi-analit}

This appendix details the semi-analytical computation of
the bounce action and
the escape point of sec.~\ref{sec:EFTvalidity}.

For a potential of the form
\begin{equation}
    V(\phibg) =
        \frac{1}{2}m_3^2\,\phibg^2
      - \frac{1}{3}\gamma_3^{}\,\phibg^3
      + \frac{1}{4}\lambda_3^{}\,\phibg^4
    \,,
\end{equation}
the bounce action is given by~\cite{Salvio:2026bco,Levi:2022bzt}
\begin{equation}
    S^{\rmii{LO}}_{3,\rmii{EFT}} =
      \frac{m_3^3}{\gamma_3^2}\times
      \begin{cases}
        \dfrac{2\pi}{3(\kappa_3-\kappa_c^{})^2}\,
        \overline{B}_3^{}(\kappa_3)
        \,,
        & \lambda_3^{}>0
        \,,\\[3mm]
        \dfrac{27\pi}{2}\,
        \dfrac{
          1+e^{-1/\sqrt{|\kappa_3|}}}{
          1+\frac{2}{9}|\kappa_3|}
        \,,
        & \lambda_3^{}<0
        \,,
      \end{cases}
\end{equation}
where we have defined
\begin{align}
  \label{eq:kappadef}
    \kappa_3 &\equiv
      \frac{\lambda_3^{} m_3^2}{\gamma_3^2}
    \,,&
    \kappa_c^{} &\equiv \frac{2}{9}
    \,.
\end{align}
For $\lambda_3>0$, the function $\overline{B}_3(\kappa_3)$ is given by~\cite{Levi:2022bzt}
\begin{align}
    \overline{B}_3(\kappa_3) \equiv
      \frac{16}{243}\Big[
          1
        - 38.23\left(\kappa_3-\kappa_c\right)
        &
        + 115.26\left(\kappa_3-\kappa_c\right)^2
    \nn &
        + 58.07\sqrt{\kappa_3}\left(\kappa_3-\kappa_c\right)^2
        + 229.07\kappa_3\left(\kappa_3-\kappa_c\right)^2
      \Big]
    \,.
\end{align}
The corresponding escape point can be expressed in a dimensionless form as
\begin{equation}
  \label{eq:app:phiesc}
    \phibg_{\rm esc}^{} =
      \frac{m_3^2}{\gamma_3^{}}\,\mathcal{F}(\kappa_3)
    \,.
\end{equation}
To determine the function $\mathcal{F}(\kappa_3)$,
we numerically computed the escape point for the rescaled potential~\cite{Levi:2022bzt}
\begin{equation}
    \widetilde{V}(\hat\phibg) =
        \frac{1}{2}\hat\phibg^2
      - \frac{1}{3}\hat\phibg^3
      + \frac{\kappa_3}{4}\hat\phibg^4
    \,,
\end{equation}
for a range of values of $\kappa_3$, and used the resulting numerical data to construct an
interpolation for $\mathcal{F}(\kappa_3)$.

In the model under consideration, the parameters entering the three-dimensional
effective potential are related to the underlying four-dimensional couplings through the
matching relations~\cite{Bernardo:2026whs}
\begin{align}
    m_3^2 &=
        \frac{1}{4}g^2 T^2
      + \frac{1}{3}\lambda T^2
    \,,\nn
    \lambda_3^{} &=
        \lambda T
      + \frac{T}{(4\pi)^2}\left[
          (2-3\Lb^{})g^4
        + 6\Lb^{}\,g^2\lambda
        - 10\Lb^{}\,\lambda^2
      \right]
    \,,\nn
    \gamma_3^{} &=
      \frac{3}{4\pi}g^3 T^{3/2}
    \,,
\end{align}
where
$\Lb \equiv 2\ln\frac{\LamDref}{T} - 2(\ln(4\pi) - \gammaE)$.

{\small
%
\bibliographystyle{utphys}
\bibliography{ref}

@article{Ekstedt:2023sqc,
    author = "Ekstedt, Andreas and Gould, Oliver and Hirvonen, Joonas",
    title = "{BubbleDet: a Python package to compute functional determinants for bubble nucleation}",
    eprint = "2308.15652",
    archivePrefix = "arXiv",
    primaryClass = "hep-ph",
    doi = "10.1007/JHEP12(2023)056",
    journal = "JHEP",
    volume = "12",
    pages = "056",
    year = "2023"
}

@article{Athron:2022mmm,
    author = "Athron, Peter and Bal{\'a}zs, Csaba and Morris, Lachlan",
    title = "{Supercool subtleties of cosmological phase transitions}",
    eprint = "2212.07559",
    archivePrefix = "arXiv",
    primaryClass = "hep-ph",
    doi = "10.1088/1475-7516/2023/03/006",
    journal = "JCAP",
    volume = "03",
    pages = "006",
    year = "2023"
}

@article{Salvio:2026bco,
    author = "Salvio, Alberto",
    title = "{Supercooled phase transitions with radiative symmetry breaking}",
    eprint = "2602.20246",
    archivePrefix = "arXiv",
    primaryClass = "hep-ph",
    doi = "10.1142/s0217751x26460048",
    journal = "Int. J. Mod. Phys. A",
    volume = "41",
    number = "24n25",
    pages = "2646004",
    year = "2026"
}

@article{Levi:2022bzt,
    author = "Levi, Noam and Opferkuch, Toby and Redigolo, Diego",
    title = "{The supercooling window at weak and strong coupling}",
    eprint = "2212.08085",
    archivePrefix = "arXiv",
    primaryClass = "hep-ph",
    doi = "10.1007/JHEP02(2023)125",
    journal = "JHEP",
    volume = "02",
    pages = "125",
    year = "2023"
}

@article{Ekstedt:2022ceo,
    author = "Ekstedt, Andreas",
    title = "{Convergence of the nucleation rate for first-order phase transitions}",
    eprint = "2205.05145",
    archivePrefix = "arXiv",
    primaryClass = "hep-ph",
    doi = "10.1103/PhysRevD.106.095026",
    journal = "Phys. Rev. D",
    volume = "106",
    number = "9",
    pages = "095026",
    year = "2022"
}

@article{Christiansen:2025xhv,
    author = "Christiansen, Martin and Madge, Eric and Puchades-Ib{\'a}{\~n}ez, Cristina and Ramirez-Quezada, Maura E. and Schwaller, Pedro",
    title = "{Beyond the daisy chain: running and the 3D EFT view of supercooled phase transitions}",
    eprint = "2511.02910",
    archivePrefix = "arXiv",
    primaryClass = "hep-ph",
    reportNumber = "IFT-UAM/CSIC 25-139 and MITP-25-071",
    doi = "10.1007/JHEP05(2026)014",
    journal = "JHEP",
    volume = "05",
    pages = "014",
    year = "2026"
}

@article{Gould:2021oba,
    author = "Gould, Oliver and Tenkanen, Tuomas V. I.",
    title = "{On the perturbative expansion at high temperature and implications for cosmological phase transitions}",
    eprint = "2104.04399",
    archivePrefix = "arXiv",
    primaryClass = "hep-ph",
    reportNumber = "NORDITA 2021-010",
    doi = "10.1007/JHEP06(2021)069",
    journal = "JHEP",
    volume = "06",
    pages = "069",
    year = "2021"
}

@article{Affleck:1980ac,
  author        = {Affleck, Ian},
  title         = {{Quantum Statistical Metastability}},
  reportnumber  = {HUTP-80/A062},
  doi           = {10.1103/PhysRevLett.46.388},
  journal       = {Phys. Rev. Lett.},
  volume        = {46},
  pages         = {388},
  year          = {1981},
  inspire       = {https://arxiv.org/abs/}
}

@article{NANOGrav:2023gor,
    author = "Agazie, Gabriella and others",
    collaboration = "NANOGrav",
    title = "{The NANOGrav 15 yr Data Set: Evidence for a Gravitational-wave Background}",
    eprint = "2306.16213",
    archivePrefix = "arXiv",
    primaryClass = "astro-ph.HE",
    doi = "10.3847/2041-8213/acdac6",
    journal = "Astrophys. J. Lett.",
    volume = "951",
    number = "1",
    pages = "L8",
    year = "2023"
}

@article{Andreassen:2017rzq,
  author        = {Andreassen, Anders and Frost, William and Schwartz, Matthew
                  D.},
  title         = {{Scale Invariant Instantons and the Complete Lifetime of the
                  Standard Model}},
  eprint        = {1707.08124},
  archiveprefix = {arXiv},
  primaryclass  = {hep-ph},
  doi           = {10.1103/PhysRevD.97.056006},
  journal       = {Phys. Rev. D},
  volume        = {97},
  number        = {5},
  pages         = {056006},
  year          = {2018},
  inspire       = {https://arxiv.org/abs/1707.08124}
}

@phdthesis{Andreassen:2018ghv,
    author = "Andreassen, Anders Johan",
    title = "{Precision Tunneling Rate Calculations in Quantum Field Theory and the Ultimate Fate of Our Universe}",
    school = "Harvard U.",
    url = "https://dash.harvard.edu/handle/1/40050090",
    year = "2018"
}

@article{Kirsten:2004qv,
    author = "Kirsten, Klaus and McKane, Alan J.",
    title = "{Functional determinants for general Sturm-Liouville problems}",
    eprint = "math-ph/0403050",
    archivePrefix = "arXiv",
    doi = "10.1088/0305-4470/37/16/014",
    journal = "J. Phys. A",
    volume = "37",
    pages = "4649--4670",
    year = "2004"
}

@article{Forman:1992abc,
    author = "Forman, Robin",
    title = "{Determinants, finite-difference operators and boundary value problems}",
    doi = "10.1007/BF02097240",
    journal = "Commun. Math. Phys.",
    volume = "147",
    number = "3",
    pages = "485--526",
    year = "1992"
}

@article{Fadakar:2026tfo,
    author = "Fadakar, Ipak and Pimentel, Guilherme L. and Tafreshi, Behrang",
    title = "{Feynman Tree Theorem and the Gelfand--Yaglom Formula}",
    eprint = "2607.24923",
    archivePrefix = "arXiv",
    primaryClass = "hep-th",
    month = "7",
    year = "2026"
}

@article{Elahi:2026ldn,
    author = "Elahi, Fatemeh and Fortini, Diego and Gouttenoire, Yann and Leister, Nicholas and Schwaller, Pedro",
    title = "{Big Bang Nucleosynthesis Confronts Domain Walls and First-Order Phase Transitions as the Pulsar Timing Array Explanations}",
    eprint = "2609.25950",
    archivePrefix = "arXiv",
    primaryClass = "hep-ph",
    reportNumber = "MITP-26-047",
    month = "9",
    year = "2026"
}

@incollection{everitt2005catalogue,
  title={{A catalogue of Sturm-Liouville differential equations}},
  author={Everitt, W. Norrie},
  booktitle={Sturm-Liouville Theory: Past and Present},
  editor={Amrein, Werner O. and Hinz, Andreas M. and Pearson, David B.},
  pages={271--331},
  year={2005},
  publisher={Birkh{\"a}user},
  address={Basel},
  doi={10.1007/3-7643-7359-8_12}
}

@article{Everitt1982,
  author = {Everitt, William Norrie},
  title = {{On the transformation theory of ordinary second-order linear symmetric differential expressions}},
  journal = {Czechoslovak Math. J.},
  volume = {32},
  number = {2},
  pages = {275--306},
  year = {1982},
  doi = {10.21136/CMJ.1982.101802}
}

@article{Berera:2019uyp,
  author        = {Berera, Arjun and Mabillard, Jo\"el and Mintz, Bruno W. and
                  Ramos, Rudnei O.},
  title         = {{Formulating the Kramers problem in field theory}},
  eprint        = {1906.08684},
  archiveprefix = {arXiv},
  primaryclass  = {hep-ph},
  doi           = {10.1103/PhysRevD.100.076005},
  journal       = {Phys. Rev. D},
  volume        = {100},
  number        = {7},
  pages         = {076005},
  year          = {2019},
  inspire       = {https://arxiv.org/abs/1906.08684}
}

@article{Buchmuller:1993bq,
    author = "Buchmuller, W. and Fodor, Z. and Helbig, T. and Walliser, D.",
    title = "{The weak electroweak phase transition}",
    eprint = "hep-ph/9303251",
    archivePrefix = "arXiv",
    reportNumber = "DESY-93-021",
    doi = "10.1006/aphy.1994.1082",
    journal = "Annals Phys.",
    volume = "234",
    pages = "260--299",
    year = "1994"
}

@article{Callan:1977pt,
  author        = {Callan, Jr., Curtis G. and Coleman, Sidney R.},
  title         = {{The Fate of the False Vacuum. 2. First Quantum
                  Corrections}},
  reportnumber  = {HUTP-77-A032},
  doi           = {10.1103/PhysRevD.16.1762},
  journal       = {Phys. Rev. D},
  volume        = {16},
  pages         = {1762--1768},
  year          = {1977},
  inspire       = {https://arxiv.org/abs/}
}

@article{EPTA:2023fyk,
    author = "Antoniadis, J. and others",
    collaboration = "EPTA, InPTA:",
    title = "{The second data release from the European Pulsar Timing Array - III. Search for gravitational wave signals}",
    eprint = "2306.16214",
    archivePrefix = "arXiv",
    primaryClass = "astro-ph.HE",
    doi = "10.1051/0004-6361/202346844",
    journal = "Astron. Astrophys.",
    volume = "678",
    pages = "A50",
    year = "2023"
}

@article{Reardon:2023gzh,
    author = "Reardon, Daniel J. and others",
    title = "{Search for an Isotropic Gravitational-wave Background with the Parkes Pulsar Timing Array}",
    eprint = "2306.16215",
    archivePrefix = "arXiv",
    primaryClass = "astro-ph.HE",
    doi = "10.3847/2041-8213/acdd02",
    journal = "Astrophys. J. Lett.",
    volume = "951",
    number = "1",
    pages = "L6",
    year = "2023"
}

@article{Xu:2023wog,
    author = "Xu, Heng and others",
    title = "{Searching for the Nano-Hertz Stochastic Gravitational Wave Background with the Chinese Pulsar Timing Array Data Release I}",
    eprint = "2306.16216",
    archivePrefix = "arXiv",
    primaryClass = "astro-ph.HE",
    doi = "10.1088/1674-4527/acdfa5",
    journal = "Res. Astron. Astrophys.",
    volume = "23",
    number = "7",
    pages = "075024",
    year = "2023"
}

@article{Coleman:1977py,
    author = "Coleman, Sidney R.",
    title = "{The Fate of the False Vacuum. 1. Semiclassical Theory}",
    reportNumber = "HUTP-77-A004",
    doi = "10.1103/PhysRevD.15.2929",
    journal = "Phys. Rev. D",
    volume = "15",
    pages = "2929--2936",
    year = "1977",
    note = "[Erratum: Phys.Rev.D 16, 1248 (1977)]"
}

@article{Coleman:1973jx,
  archiveprefix = {arXiv},
  arxivid       = {hep-th/0507214},
  author        = {Coleman, Sidney and Weinberg, Erick},
  doi           = {10.1103/PhysRevD.7.1888},
  eprint        = {hep-th/0507214},
  issn          = {0556-2821},
  journal       = {Phys. Rev. D},
  month         = {mar},
  number        = {6},
  pages         = {1888--1910},
  primaryclass  = {hep-th},
  title         = {{Radiative Corrections as the Origin of Spontaneous Symmetry
                  Breaking}},
  volume        = {7},
  year          = {1973},
  inspire       = {https://arxiv.org/abs/0507214}
}

@article{Croon:2020cgk,
  author        = {Croon, Djuna and Gould, Oliver and Schicho, Philipp and
                  Tenkanen, Tuomas V. I. and White, Graham},
  title         = {{Theoretical uncertainties for cosmological first-order phase
                  transitions}},
  eprint        = {2009.10080},
  archiveprefix = {arXiv},
  primaryclass  = {hep-ph},
  reportnumber  = {HIP-2020-26/TH},
  doi           = {10.1007/JHEP04(2021)055},
  journal       = {JHEP},
  volume        = {04},
  pages         = {055},
  url           = {http://arxiv.org/abs/2009.10080
                  https://link.springer.com/10.1007/JHEP04(2021)055},
  month         = {4},
  year          = {2021},
  inspire       = {https://arxiv.org/abs/2009.10080}
}

@article{Ekstedt:2022tqk,
  author        = {Ekstedt, Andreas},
  title         = {{Bubble nucleation to all orders}},
  eprint        = {2201.07331},
  archiveprefix = {arXiv},
  primaryclass  = {hep-ph},
  doi           = {10.1007/JHEP08(2022)115},
  journal       = {JHEP},
  volume        = {08},
  pages         = {115},
  year          = {2022},
  inspire       = {https://arxiv.org/abs/2201.07331}
}

@article{Ekstedt:2021kyx,
  author        = {Ekstedt, Andreas},
  title         = {{Higher-order corrections to the bubble-nucleation rate at
                  finite temperature}},
  eprint        = {2104.11804},
  archiveprefix = {arXiv},
  primaryclass  = {hep-ph},
  doi           = {10.1140/epjc/s10052-022-10130-5},
  journal       = {Eur. Phys. J. C},
  volume        = {82},
  number        = {2},
  pages         = {173},
  year          = {2022},
  inspire       = {https://arxiv.org/abs/2104.11804}
}

@article{Ellis:2022lft,
  author        = {Ellis, John and Lewicki, Marek and Merchand, Marco and No,
                  Jos\'e Miguel and Zych, Mateusz},
  title         = {{The Scalar Singlet Extension of the Standard Model:
                  Gravitational Waves versus Baryogenesis}},
  eprint        = {2210.16305},
  archiveprefix = {arXiv},
  primaryclass  = {hep-ph},
  month         = {10},
  year          = {2022},
  inspire       = {https://arxiv.org/abs/2210.16305}
}

@article{Farakos:1994kx,
  author        = {Farakos, K. and Kajantie, K. and Rummukainen, K. and
                  Shaposhnikov, Mikhail E.},
  title         = {{3-D physics and the electroweak phase transition:
                  Perturbation theory}},
  eprint        = {hep-ph/9404201},
  archiveprefix = {arXiv},
  reportnumber  = {CERN-TH-6973-94, IUHET-273},
  doi           = {10.1016/0550-3213(94)90173-2},
  journal       = {Nucl. Phys. B},
  volume        = {425},
  pages         = {67--109},
  year          = {1994},
  inspire       = {https://arxiv.org/abs/hep-ph/9404201}
}

@article{Gelfand:1959nq,
  author        = {Gelfand, I. M. and Yaglom, A. M.},
  title         = {{Integration in functional spaces and its applications in
                  quantum physics}},
  doi           = {10.1063/1.1703636},
  journal       = {J. Math. Phys.},
  volume        = {1},
  pages         = {48},
  year          = {1960},
  inspire       = {https://arxiv.org/abs/}
}

@article{Gould:2021ccf,
  author        = {Gould, Oliver and Hirvonen, Joonas},
  title         = {{Effective field theory approach to thermal bubble
                  nucleation}},
  eprint        = {2108.04377},
  archiveprefix = {arXiv},
  primaryclass  = {hep-ph},
  reportnumber  = {HIP-2020-19/TH},
  doi           = {10.1103/PhysRevD.104.096015},
  journal       = {Phys. Rev. D},
  volume        = {104},
  number        = {9},
  pages         = {096015},
  year          = {2021},
  inspire       = {https://arxiv.org/abs/2108.04377}
}

@article{Hirvonen:2021zej,
  author        = {Hirvonen, Joonas and L\"ofgren, Johan and Ramsey-Musolf,
                  Michael J. and Schicho, Philipp and Tenkanen, Tuomas V. I.},
  title         = {{Computing the gauge-invariant bubble nucleation rate in
                  finite temperature effective field theory}},
  eprint        = {2112.08912},
  archiveprefix = {arXiv},
  primaryclass  = {hep-ph},
  reportnumber  = {ACFI-T21-16, HIP-2021-45/TH, NORDITA 2021-111},
  doi           = {10.1007/JHEP07(2022)135},
  journal       = {JHEP},
  volume        = {07},
  pages         = {135},
  url           = {http://arxiv.org/abs/2112.08912
                  https://link.springer.com/10.1007/JHEP07(2022)135},
  month         = {7},
  year          = {2022},
  inspire       = {https://arxiv.org/abs/2112.08912}
}

@article{Kajantie:1995dw,
  author        = {Kajantie, K. and Laine, M. and Rummukainen, K. and
                  Shaposhnikov, Mikhail E.},
  title         = {{Generic rules for high temperature dimensional reduction and
                  their application to the standard model}},
  eprint        = {hep-ph/9508379},
  archiveprefix = {arXiv},
  reportnumber  = {CERN-TH-95-226, HU-TFT-95-50, IUHET-312},
  doi           = {10.1016/0550-3213(95)00549-8},
  journal       = {Nucl. Phys. B},
  volume        = {458},
  pages         = {90--136},
  year          = {1996},
  inspire       = {https://arxiv.org/abs/hep-ph/9508379}
}

@article{Kierkla:2022odc,
    author = "Kierkla, Maciej and Karam, Alexandros and Swiezewska, Bogumila",
    title = "{Conformal model for gravitational waves and dark matter: a status update}",
    eprint = "2210.07075",
    archivePrefix = "arXiv",
    primaryClass = "astro-ph.CO",
    doi = "10.1007/JHEP03(2023)007",
    journal = "JHEP",
    volume = "03",
    pages = "007",
    year = "2023"
}

@article{Langer:1967ax,
    author = "Langer, J. S.",
    title = "{Theory of the condensation point}",
    doi = "10.1016/0003-4916(67)90200-X",
    journal = "Annals Phys.",
    volume = "41",
    pages = "108--157",
    year = "1967"
}

@article{Langer:1969bc,
    author = "Langer, J. S.",
    title = "{Statistical theory of the decay of metastable states}",
    doi = "10.1016/0003-4916(69)90153-5",
    journal = "Annals Phys.",
    volume = "54",
    pages = "258--275",
    year = "1969"
}

@article{Lewicki:2021pgr,
  author        = {Lewicki, Marek and Merchand, Marco and Zych, Mateusz},
  title         = {{Electroweak bubble wall expansion: gravitational waves and
                  baryogenesis in Standard Model-like thermal plasma}},
  eprint        = {2111.02393},
  archiveprefix = {arXiv},
  primaryclass  = {astro-ph.CO},
  month         = {11},
  year          = {2021},
  inspire       = {https://arxiv.org/abs/2111.02393}
}

@article{Ekstedt:2025awx,
    author = "Ekstedt, Andreas and Konstandin, Thomas and van de Vis, Jorinde",
    title = "{Scalar damping in cosmological phase transitions}",
    eprint = "2512.16663",
    archivePrefix = "arXiv",
    primaryClass = "hep-ph",
    reportNumber = "DESY-25-195, CERN-TH-2025-262",
    doi = "10.1088/1475-7516/2026/06/068",
    journal = "JCAP",
    volume = "06",
    pages = "068",
    year = "2026"
}

@article{Linde:1981zj,
  author        = {Linde, Andrei D.},
  title         = {{Decay of the False Vacuum at Finite Temperature}},
  reportnumber  = {LEBEDEV-81-265},
  doi           = {10.1016/0550-3213(83)90072-X},
  journal       = {Nucl. Phys. B},
  volume        = {216},
  pages         = {421},
  year          = {1983},
  note          = {[Erratum: Nucl.Phys.B 223, 544 (1983)]},
  inspire       = {https://arxiv.org/abs/}
}

@article{Martin:2018emo,
  archiveprefix = {arXiv},
  arxivid       = {1808.07615},
  author        = {Martin, Stephen P. and Patel, Hiren H.},
  doi           = {10.1103/PhysRevD.98.076008},
  eprint        = {1808.07615},
  issn          = {2470-0010},
  journal       = {Phys. Rev. D},
  month         = {oct},
  number        = {7},
  pages         = {076008},
  title         = {{Two-loop effective potential for generalized gauge fixing}},
  volume        = {98},
  year          = {2018},
  inspire       = {https://arxiv.org/abs/1808.07615}
}

@article{Ekstedt:2024etx,
    author = "Ekstedt, Andreas and Schicho, Philipp and Tenkanen, Tuomas V. I.",
    title = "{Cosmological phase transitions at three loops: The final verdict on perturbation theory}",
    eprint = "2405.18349",
    archivePrefix = "arXiv",
    primaryClass = "hep-ph",
    reportNumber = "HIP-2024-15/TH",
    doi = "10.1103/PhysRevD.110.096006",
    journal = "Phys. Rev. D",
    volume = "110",
    number = "9",
    pages = "096006",
    year = "2024"
}

@article{Bernardo:2025vkz,
    author = "Bernardo, Fabio and Klose, Philipp and Schicho, Philipp and Tenkanen, Tuomas V. I.",
    title = "{Higher-dimensional operators at finite temperature affect gravitational-wave predictions}",
    eprint = "2503.18904",
    archivePrefix = "arXiv",
    primaryClass = "hep-ph",
    reportNumber = "HIP-2025-6/TH",
    doi = "10.1007/JHEP08(2025)109",
    journal = "JHEP",
    volume = "08",
    pages = "109",
    year = "2025"
}

@article{Megevand:2016lpr,
    author = "Megevand, Ariel and Ramirez, Santiago",
    title = "{Bubble nucleation and growth in very strong cosmological phase transitions}",
    eprint = "1611.05853",
    archivePrefix = "arXiv",
    primaryClass = "astro-ph.CO",
    doi = "10.1016/j.nuclphysb.2017.03.009",
    journal = "Nucl. Phys. B",
    volume = "919",
    pages = "74--109",
    year = "2017"
}

@article{Braaten:1995cm,
    author = "Braaten, Eric and Nieto, Agustin",
    title = "{Effective field theory approach to high temperature thermodynamics}",
    eprint = "hep-ph/9501375",
    archivePrefix = "arXiv",
    reportNumber = "NUHEP-TH-95-2",
    doi = "10.1103/PhysRevD.51.6990",
    journal = "Phys. Rev. D",
    volume = "51",
    pages = "6990--7006",
    year = "1995"
}

@article{Navarrete:2025yxy,
    author = {Navarrete, Pablo and Paatelainen, Risto and Sepp{\"a}nen, Kaapo and Tenkanen, Tuomas V. I.},
    title = "{Cosmological phase transitions without high-temperature expansions}",
    eprint = "2507.07014",
    archivePrefix = "arXiv",
    primaryClass = "hep-ph",
    reportNumber = "HIP-2025-20/TH",
    doi = "10.1007/JHEP01(2026)113",
    journal = "JHEP",
    volume = "01",
    pages = "113",
    year = "2026"
}

@article{Branchina:2026nql,
    author = "Branchina, Carlo and De Curtis, Stefania and Delle Rose, Luigi and Notari, Alessio and Panico, Giuliano and Starbuck, Matthew",
    title = "{Thermal Masses and Bubble-Wall Friction in Cosmological Phase Transitions}",
    eprint = "2607.14867",
    archivePrefix = "arXiv",
    primaryClass = "hep-ph",
    reportNumber = "KA-TP-15-2026, P3H-26-048",
    month = "7",
    year = "2026"
}

@article{Branchina:2025jou,
    author = "Branchina, Carlo and Conaci, Angela and Delle Rose, Luigi and De Curtis, Stefania",
    title = "{Electroweak phase transition and bubble wall velocity in local thermal equilibrium}",
    eprint = "2504.21213",
    archivePrefix = "arXiv",
    primaryClass = "hep-ph",
    doi = "10.1103/lkht-1nv1",
    journal = "Phys. Rev. D",
    volume = "112",
    number = "9",
    pages = "095008",
    year = "2025"
}

@article{Branchina:2025adj,
    author = "Branchina, Carlo and Conaci, Angela and Delle Rose, Luigi and De Curtis, Stefania",
    title = "{Bubble wall velocity with out-of-equilibrium corrections}",
    eprint = "2510.21942",
    archivePrefix = "arXiv",
    primaryClass = "hep-ph",
    doi = "10.1103/nmkw-7kgk",
    journal = "Phys. Rev. D",
    volume = "113",
    number = "3",
    pages = "035024",
    year = "2026"
}

@article{Cline:2021iff,
    author = "Cline, James M. and Friedlander, Avi and He, Dong-Ming and Kainulainen, Kimmo and Laurent, Benoit and Tucker-Smith, David",
    title = "{Baryogenesis and gravity waves from a UV-completed electroweak phase transition}",
    eprint = "2102.12490",
    archivePrefix = "arXiv",
    primaryClass = "hep-ph",
    doi = "10.1103/PhysRevD.103.123529",
    journal = "Phys. Rev. D",
    volume = "103",
    number = "12",
    pages = "123529",
    year = "2021"
}

@article{Bernardo:2026whs,
    author = "Bernardo, Fabio and Chala, Mikael and Gil, Luis and Schicho, Philipp",
    title = "{Hard thermal contributions to phase transition observables at NNLO}",
    eprint = "2602.06962",
    archivePrefix = "arXiv",
    primaryClass = "hep-ph",
    doi = "10.1007/JHEP06(2026)281",
    journal = "JHEP",
    volume = "06",
    pages = "281",
    year = "2026"
}

@article{Steinhardt:1981ct,
    author = "Steinhardt, Paul Joseph",
    title = "{Relativistic Detonation Waves and Bubble Growth in False Vacuum Decay}",
    reportNumber = "UPR-0181T",
    doi = "10.1103/PhysRevD.25.2074",
    journal = "Phys. Rev. D",
    volume = "25",
    pages = "2074",
    year = "1982"
}

@article{Espinosa:2010hh,
    author = "Espinosa, Jose R. and Konstandin, Thomas and No, Jose M. and Servant, Geraldine",
    title = "{Energy Budget of Cosmological First-order Phase Transitions}",
    eprint = "1004.4187",
    archivePrefix = "arXiv",
    primaryClass = "hep-ph",
    reportNumber = "CERN-PH-TH-2010-027",
    doi = "10.1088/1475-7516/2010/06/028",
    journal = "JCAP",
    volume = "06",
    pages = "028",
    year = "2010"
}

@article{Bernardo:2026nyq,
    author = "Bernardo, Fabio and Reinle, Romain Guillermo and Schicho, Philipp",
    title = "{Matching higher-dimensional operators at finite temperature for general models}",
    eprint = "2605.15176",
    archivePrefix = "arXiv",
    primaryClass = "hep-ph",
    month = "5",
    year = "2026"
}

@article{Chala:2024xll,
    author = "Chala, Mikael and Criado, Juan Carlos and Gil, Luis and Miras, Javier L{\'o}pez",
    title = "{Higher-order-operator corrections to phase-transition parameters in dimensional reduction}",
    eprint = "2406.02667",
    archivePrefix = "arXiv",
    primaryClass = "hep-ph",
    doi = "10.1007/JHEP10(2024)025",
    journal = "JHEP",
    volume = "10",
    pages = "025",
    year = "2024"
}

@article{Matuszak:2026xsz,
    author = "Matuszak, Jonas and Tasillo, Carlo",
    title = "{TransitionListener v2.0 -- Robust gravitational wave predictions for cosmological phase transitions}",
    eprint = "2605.15259",
    archivePrefix = "arXiv",
    primaryClass = "hep-ph",
    month = "5",
    year = "2026"
}

@article{Kierkla:2025qyz,
    author = "Kierkla, Maciej and Schicho, Philipp and Swiezewska, Bogumila and Tenkanen, Tuomas V. I. and van de Vis, Jorinde",
    title = "{Finite-temperature bubble nucleation with shifting scale hierarchies}",
    eprint = "2503.13597",
    archivePrefix = "arXiv",
    primaryClass = "hep-ph",
    reportNumber = "CERN-TH-2025-046, HIP-2024-27/TH",
    doi = "10.1007/JHEP07(2025)153",
    journal = "JHEP",
    volume = "07",
    pages = "153",
    year = "2025"
}

@article{Kierkla:2025vwp,
    author = "Kierkla, Maciej and Ramberg, Nicklas and Schicho, Philipp and Schmitt, Daniel",
    title = "{Thermodynamical uncertainties for primordial black holes from cosmological phase transitions}",
    eprint = "2506.15496",
    archivePrefix = "arXiv",
    primaryClass = "hep-ph",
    reportNumber = "SISSA 07/2025/FISI",
    doi = "10.1103/nj27-ltyg",
    journal = "Phys. Rev. D",
    volume = "113",
    number = "9",
    pages = "095024",
    year = "2026"
}

@article{Kierkla:2023von,
    author = "Kierkla, Maciej and Swiezewska, Bogumila and Tenkanen, Tuomas V. I. and van de Vis, Jorinde",
    title = "{Gravitational waves from supercooled phase transitions: dimensional transmutation meets dimensional reduction}",
    eprint = "2312.12413",
    archivePrefix = "arXiv",
    primaryClass = "hep-ph",
    doi = "10.1007/JHEP02(2024)234",
    journal = "JHEP",
    volume = "02",
    pages = "234",
    year = "2024"
}

@article{Forman:1987gha,
    author = "Forman, Robin",
    title = "{Functional determinants and geometry}",
    doi = "10.1007/BF01391828",
    journal = "Invent. Math.",
    volume = "88",
    number = "3",
    pages = "447--493",
    year = "1987"
}

@article{Kirsten:2010eg,
    author = "Kirsten, Klaus",
    title = "{Functional determinants in higher dimensions using contour integrals}",
    eprint = "1005.2595",
    archivePrefix = "arXiv",
    primaryClass = "hep-th",
    month = "5",
    year = "2010"
}

@article{Athron:2023xlk,
    author = "Athron, Peter and Bal{\'a}zs, Csaba and Fowlie, Andrew and Morris, Lachlan and Wu, Lei",
    title = "{Cosmological phase transitions: From perturbative particle physics to gravitational waves}",
    eprint = "2305.02357",
    archivePrefix = "arXiv",
    primaryClass = "hep-ph",
    doi = "10.1016/j.ppnp.2023.104094",
    journal = "Prog. Part. Nucl. Phys.",
    volume = "135",
    pages = "104094",
    year = "2024"
}

@article{Ellis:2018mja,
    author = "Ellis, John and Lewicki, Marek and No, Jos{\'e} Miguel",
    title = "{On the Maximal Strength of a First-Order Electroweak Phase Transition and its Gravitational Wave Signal}",
    eprint = "1809.08242",
    archivePrefix = "arXiv",
    primaryClass = "hep-ph",
    reportNumber = "KCL-PH-TH/2018-46, CERN-TH/2018-197, IFT-UAM/CSIC-18-94, CERN-TH-2018-197",
    doi = "10.1088/1475-7516/2019/04/003",
    journal = "JCAP",
    volume = "04",
    pages = "003",
    year = "2019"
}

@article{vandeVis:2025plm,
    author = "van de Vis, Jorinde and Schicho, Philipp and Niemi, Lauri and Laurent, Benoit and Hirvonen, Joonas and Gould, Oliver",
    title = "{WallGo investigates: Theoretical uncertainties in the bubble wall velocity}",
    eprint = "2510.27691",
    archivePrefix = "arXiv",
    primaryClass = "hep-ph",
    reportNumber = "CERN-TH-2025-221",
    doi = "10.1007/JHEP04(2026)041",
    journal = "JHEP",
    volume = "04",
    pages = "041",
    year = "2026"
}

@article{Ekstedt:2024fyq,
    author = "Ekstedt, Andreas and Gould, Oliver and Hirvonen, Joonas and Laurent, Benoit and Niemi, Lauri and Schicho, Philipp and van de Vis, Jorinde",
    title = "{How fast does the WallGo? A package for computing wall velocities in first-order phase transitions}",
    eprint = "2411.04970",
    archivePrefix = "arXiv",
    primaryClass = "hep-ph",
    reportNumber = "CERN-TH-2024-174, DESY-24-162, HIP-2024-21/TH",
    doi = "10.1007/JHEP04(2025)101",
    journal = "JHEP",
    volume = "04",
    pages = "101",
    year = "2025"
}

@article{Baratella:2025dum,
    author = "Baratella, Pietro and Nemev{\v{s}}ek, Miha and Shoji, Yutaro and Trailovi{\'c}, Katarina and Ubaldi, Lorenzo",
    title = "{Functional determinants for false vacuum decay}",
    eprint = "2502.15878",
    archivePrefix = "arXiv",
    primaryClass = "hep-ph",
    doi = "10.1007/JHEP05(2025)100",
    journal = "JHEP",
    volume = "05",
    pages = "100",
    year = "2025"
}

@article{Baacke:1999sc,
    author = "Baacke, Jurgen and Heitmann, Katrin",
    title = "{Gauge invariance of the one loop effective action of the Higgs field in the SU(2) Higgs model}",
    eprint = "hep-th/9905201",
    archivePrefix = "arXiv",
    reportNumber = "DO-TH-99-05A, DO-TH-99-08",
    doi = "10.1103/PhysRevD.60.105037",
    journal = "Phys. Rev. D",
    volume = "60",
    pages = "105037",
    year = "1999"
}

@article{Hanggi:1990zz,
    author = "Hanggi, Peter and Talkner, Peter and Borkovec, Michal",
    title = "{Reaction-Rate Theory: Fifty Years After Kramers}",
    doi = "10.1103/RevModPhys.62.251",
    journal = "Rev. Mod. Phys.",
    volume = "62",
    pages = "251--341",
    year = "1990"
}

@article{Hirvonen:2024rfg,
    author = "Hirvonen, Joonas",
    title = "{Real-time nucleation and off-equilibrium effects in high-temperature quantum field theories}",
    eprint = "2403.07987",
    archivePrefix = "arXiv",
    primaryClass = "hep-ph",
    reportNumber = "HIP-2024-6/TH",
    doi = "10.1103/4p7n-nyln",
    journal = "Phys. Rev. D",
    volume = "111",
    number = "11",
    pages = "116020",
    year = "2025"
}

@article{Hirvonen:2025hqn,
    author = "Hirvonen, Joonas and Gould, Oliver",
    title = "{Langer{\textquoteright}s Nucleation Rate Reproduced on the Lattice}",
    eprint = "2505.22732",
    archivePrefix = "arXiv",
    primaryClass = "hep-ph",
    doi = "10.1103/c75g-xbw4",
    journal = "Phys. Rev. Lett.",
    volume = "136",
    number = "8",
    pages = "081601",
    year = "2026"
}

@article{Hirvonen:2026zaq,
    author = "Hirvonen, Joonas",
    title = "{Quantum field nucleating and Wigner functions}",
    eprint = "2607.09233",
    archivePrefix = "arXiv",
    primaryClass = "hep-th",
    month = "7",
    year = "2026"
}

@article{Biermann:2026qju,
    author = "Biermann, Lisa and Chakrabortty, Joydeep and Englert, Christoph",
    title = "{Polyakov loops tame phase transitions}",
    eprint = "2607.06203",
    archivePrefix = "arXiv",
    primaryClass = "hep-ph",
    doi = "10.1103/5cbz-vt5j",
    journal = "Phys. Rev. D",
    volume = "114",
    number = "5",
    pages = "056031",
    year = "2026"
}

@article{Bandyopadhyay:2026nrv,
    author = "Bandyopadhyay, Siddhartha and Chakrabortty, Joydeep and Dey, Debmalya and Schicho, Philipp and Tushar",
    title = "{Higher-dimensional operators and Polyakov loop in hot Scalar QED from the heat kernel}",
    eprint = "2606.09779",
    archivePrefix = "arXiv",
    primaryClass = "hep-ph",
    month = "6",
    year = "2026"
}

@article{Gould:2026mtl,
    author = "Gould, Oliver and Hirvonen, Joonas and Shkerin, Andrey and Sibiryakov, Sergey",
    title = "{Dynamics of nucleation in thermal phase transitions}",
    eprint = "2608.00169",
    archivePrefix = "arXiv",
    primaryClass = "hep-th",
    month = "7",
    year = "2026"
}

@article{Puchades-Ibanez:2026ksh,
    author = "Puchades-Ib{\'a}{\~n}ez, Cristina and Schwaller, Pedro",
    title = "{The devil in the transition: NLO nucleation and the particle physics behind the PTA signal}",
    eprint = "2609.12098",
    archivePrefix = "arXiv",
    primaryClass = "hep-ph",
    month = "9",
    year = "2026"
}

@article{Bezuglov:2018qpq,
    author = "Bezuglov, M. A. and Onishchenko, A. I.",
    title = "{Two-loop corrections to false vacuum decay in scalar field theory}",
    eprint = "1805.06482",
    archivePrefix = "arXiv",
    primaryClass = "hep-ph",
    doi = "10.1016/j.physletb.2018.11.005",
    journal = "Phys. Lett. B",
    volume = "788",
    pages = "122--130",
    year = "2019"
}

@article{Bezuglov:2019uxg,
    author = "Bezuglov, M. A. and Onishchenko, A. I.",
    title = "{NNLO corrections to false vacuum decay rate in thin-wall approximation}",
    eprint = "1903.10864",
    archivePrefix = "arXiv",
    primaryClass = "hep-ph",
    month = "3",
    year = "2019"
}

@article{Kierkla:2026bnm,
    author = "Kierkla, Maciej and Lewicki, Marek and Schicho, Philipp and Schmitt, Daniel and Swiezewska, Bogumila",
    title = "{Theoretical uncertainties in reconstructing model parameters with gravitational waves from supercooled phase transitions}",
    eprint = "2607.18233",
    archivePrefix = "arXiv",
    primaryClass = "hep-ph",
    reportNumber = "KA-TP-17-2026",
    month = "7",
    year = "2026"
}

@article{Li:2025kyo,
    author = "Li, Xu-Xiang and Ramsey-Musolf, Michael J. and Tenkanen, Tuomas V. I. and Wu, Yanda",
    title = "{An Effective Sphaleron Awakens}",
    eprint = "2506.01585",
    archivePrefix = "arXiv",
    primaryClass = "hep-ph",
    reportNumber = "HIP-2025-8/TH",
    month = "6",
    year = "2025"
}

@article{Dunne:2006ct,
    author = "Dunne, Gerald V. and Kirsten, Klaus",
    title = "{Functional determinants for radial operators}",
    eprint = "hep-th/0607066",
    archivePrefix = "arXiv",
    doi = "10.1088/0305-4470/39/38/017",
    journal = "J. Phys. A",
    volume = "39",
    pages = "11915--11928",
    year = "2006"
}

@article{Langer:1937qr,
    author = "Langer, Rudolph E.",
    title = "{On the Connection Formulas and the Solutions of the Wave Equation}",
    doi = "10.1103/PhysRev.51.669",
    journal = "Phys. Rev.",
    volume = "51",
    pages = "669--676",
    year = "1937"
}

@article{Chakrabortty:2026swu,
    author = "Chakrabortty, Joydeep and Siqueira Eduardo, Bruno and Karmakar, Siddhartha and Schicho, Philipp",
    title = "{Finite-temperature operator basis on {\ensuremath{\mathbb{R}}}$^{3}$ {\texttimes} S$^{1}$ for SMEFT}",
    eprint = "2605.02878",
    archivePrefix = "arXiv",
    primaryClass = "hep-th",
    doi = "10.1007/JHEP09(2026)241",
    journal = "JHEP",
    volume = "09",
    pages = "241",
    year = "2026"
}

@article{Girmohanta:2026ogq,
    author = "Girmohanta, Sudhakantha and Kamada, Kohei and Nakai, Yuichiro and Uchida, Fumio",
    title = "{Dark Transport Erases Baryon Inhomogeneity from Supercooled Phase Transitions}",
    eprint = "2609.27978",
    archivePrefix = "arXiv",
    primaryClass = "hep-ph",
    reportNumber = "CTPU-PTC-26-24",
    month = "9",
    year = "2026"
}

@article{Tanaka:2026ssu,
    author = "Tanaka, Masanori",
    title = "{Big Bang Nucleosynthesis as a Probe of First-Order Phase Transitions}",
    eprint = "2609.28348",
    archivePrefix = "arXiv",
    primaryClass = "hep-ph",
    month = "9",
    year = "2026"
}

@article{Baratella:2024hju,
    author = "Baratella, Pietro and Nemev{\v{s}}ek, Miha and Shoji, Yutaro and Trailovi{\'c}, Katarina and Ubaldi, Lorenzo",
    title = "{Revising the Full One-Loop Gauge Prefactor in Electroweak Vacuum Stability}",
    eprint = "2406.05180",
    archivePrefix = "arXiv",
    primaryClass = "hep-ph",
    doi = "10.1103/PhysRevLett.134.011601",
    journal = "Phys. Rev. Lett.",
    volume = "134",
    number = "1",
    pages = "011601",
    year = "2025"
}

@article{Gildener:1976ih,
    author = "Gildener, Eldad and Weinberg, Steven",
    title = "{Symmetry Breaking and Scalar Bosons}",
    reportNumber = "PRINT-76-0068 (HARVARD)",
    doi = "10.1103/PhysRevD.13.3333",
    journal = "Phys. Rev. D",
    volume = "13",
    pages = "3333",
    year = "1976"
}

@article{Ellis:2020nnr,
    author = "Ellis, John and Lewicki, Marek and Vaskonen, Ville",
    title = "{Updated predictions for gravitational waves produced in a strongly supercooled phase transition}",
    eprint = "2007.15586",
    archivePrefix = "arXiv",
    primaryClass = "astro-ph.CO",
    reportNumber = "KCL-PH-TH/2020-40, CERN-TH-2020-129",
    doi = "10.1088/1475-7516/2020/11/020",
    journal = "JCAP",
    volume = "11",
    pages = "020",
    year = "2020"
}

@article{Bhattacharya:2024chz,
    author = "Bhattacharya, Arindam and Cotler, Jordan and Dersy, Aur{\'e}lien and Schwartz, Matthew D.",
    title = "{Collective coordinate fix in the path integral}",
    eprint = "2402.18633",
    archivePrefix = "arXiv",
    primaryClass = "hep-th",
    doi = "10.1103/PhysRevD.110.116023",
    journal = "Phys. Rev. D",
    volume = "110",
    number = "11",
    pages = "116023",
    year = "2024"
}

@article{Klose:2026pux,
    author = "Klose, Philipp",
    title = "{Thermal Resummation for Very Strong First-order Phase Transitions}",
    eprint = "2607.21252",
    archivePrefix = "arXiv",
    primaryClass = "hep-ph",
    month = "7",
    year = "2026"
}

@article{Berthiaume:2026dkt,
    author = {Berthiaume, Rapha{\"e}l and Curtin, David and Luke, Michael and Rasovic, Andrija and Roy, Jyotirmoy},
    title = "{Hybrid Partial Dressing: Correct Effective Potentials at All Temperatures}",
    eprint = "2609.30366",
    archivePrefix = "arXiv",
    primaryClass = "hep-ph",
    month = "9",
    year = "2026"
}

@article{Carosi:2024lop,
    author = {Carosi, Matthias and Garbrecht, Bj{\"o}rn},
    title = "{False vacuum decay beyond the quadratic approximation: Summation of nonlocal self-energies}",
    eprint = "2411.18421",
    archivePrefix = "arXiv",
    primaryClass = "hep-th",
    reportNumber = "TUM-HEP-1540/24",
    doi = "10.1103/PhysRevD.111.085002",
    journal = "Phys. Rev. D",
    volume = "111",
    number = "8",
    pages = "085002",
    year = "2025"
}

@article{Carosi:2026gpi,
    author = "Carosi, Matthias",
    title = "{On equivalent methods for functional determinants}",
    eprint = "2601.08686",
    archivePrefix = "arXiv",
    primaryClass = "hep-th",
    reportNumber = "TUM-1590/26",
    doi = "10.1007/JHEP06(2026)073",
    journal = "JHEP",
    volume = "06",
    pages = "073",
    year = "2026"
}

@article{Adams:1993zs,
    author = "Adams, Fred C.",
    title = "{General solutions for tunneling of scalar fields with quartic potentials}",
    eprint = "hep-ph/9302321",
    archivePrefix = "arXiv",
    reportNumber = "UM-AC-93-4",
    doi = "10.1103/PhysRevD.48.2800",
    journal = "Phys. Rev. D",
    volume = "48",
    pages = "2800--2805",
    year = "1993"
}

}
\end{document}